\documentclass[journal]{IEEEtran}

\usepackage{setspace}
\usepackage{cite,graphicx,amsmath,amssymb}
\usepackage{hyperref}
\usepackage{subfigure}
\usepackage{url}
\usepackage{fancyhdr}
\usepackage{mdwmath}
\usepackage{mdwtab}
\usepackage{balance}
\usepackage{xcolor}
\usepackage{bm}
\usepackage{amsthm}
\usepackage{threeparttable}
\usepackage{algorithm}
\usepackage{algorithmic}
\usepackage{multirow}
\usepackage{flafter}
\usepackage{makecell}
\usepackage{diagbox}
\usepackage{tcolorbox}
\theoremstyle{definition}

\newtheorem{theorem}{Theorem}

\newtheorem{lemma}{Lemma}

\newtheorem{corollary}{Corollary}

\usepackage{xurl}

\providecommand{\url}[1]{#1}
\allowdisplaybreaks
\begin{document}

\title{Simultaneously Transmitting and Reflecting Surfaces for Ubiquitous Next Generation Multiple Access in 6G and Beyond}

\author{Xidong Mu, Jiaqi Xu, Zhaolin Wang, and Naofal Al-Dhahir,~\IEEEmembership{Fellow,~IEEE}\\
\vspace{0.2cm}

\thanks{Xidong Mu and Zhaolin Wang are with the School of Electronic Engineering and Computer Science, Queen Mary University of London, London E1 4NS, U.K. (e-mail: xidong.mu@qmul.ac.uk; zhaolin.wang@qmul.ac.uk).}
\thanks{Jiaqi Xu is with the Department of Electrical Engineering and Computer Science, University of California, Irvine, CA 92697, USA (email: xu.jiaqi@uci.edu).}
\thanks{Naofal Al-Dhahir is with the Department of Electrical and Computer Engineering, The University of Texas at Dallas, Richardson, TX 75080 USA (e-mail: aldhahir@utdallas.edu).}
}

\maketitle
\begin{abstract}
The ultimate goal of next generation multiple access (NGMA) is to support massive terminals and facilitate multiple functionalities over the limited radio resources of wireless networks in the most efficient manner possible. However, the random and uncontrollable wireless radio environment is a major obstacle to realizing this NGMA vision. Given the prominent feature of achieving 360° smart radio environment, simultaneously transmitting and reflecting surfaces (STARS) are emerging as one key enabling technology among the family of reconfigurable intelligent surfaces for NGMA. This paper provides a comprehensive overview of the recent research progress of STARS, focusing on fundamentals, performance analysis, and full-space beamforming design, as well as promising employments of STARS in NGMA. In particular, we first introduce the basics of STARS by elaborating on the foundational principles and operating protocols as well as discussing different STARS categories and prototypes. Moreover, we systematically survey the existing performance analysis and beamforming design for STARS-aided wireless communications in terms of diverse objectives and different mathematical approaches. Given the superiority of STARS, we further discuss advanced STARS applications as well as the attractive interplay between STARS and other emerging techniques to motivate future works for realizing efficient NGMA.

\end{abstract}

\section{Introduction} \label{sec:intro}
Since the feasibility of wireless communications was demonstrated at the end of the 19th century, there has been a rapid development of wireless communication technologies, which totally changed human life and workstyle. Although the fifth generation (5G) wireless network has started to be globally deployed, the communication requirements are far from being satisfied given the emergence of new revolutionary applications, such as augmented reality, virtual reality, autonomous driving, and Metaverse. Therefore, the research community has begun to investigate next generation (NG) wireless networks (e.g., sixth generation (6G) and beyond)~\cite{8766143,8869705,10054381}. Compared to 5G, on the one hand, NG wireless networks have to support extremely higher data rates and ultra massive connectivity in a ubiquitous space-air-ground-sea range~\cite{10054381,9628162}. On the other hand, multiple beyond communication functionalities, including but not limited to radio frequency (RF) sensing, imaging, computing, and positioning, have to be facilitated and even integrated in a harmonious manner in NG wireless networks~\cite{8016573,9737357,10024901}. 

To satisfy the aforementioned stringent and new requirements imposed by NG wireless networks, advanced wireless technologies should be conceived. Given the fact that the available radio resource is always limited, multiple access (MA) will continue to play an unprecedentedly important role in supporting NG wireless networks~\cite{8085125,9205230,9693417}. Nevertheless, in contrast to previous orthogonal communication-oriented MA technologies employed from first generation (1G) to 5G (i.e., frequency/time/code division multiple access (FDMA/TDMA/CDMA) and orthogonal frequency division multiple access (OFDMA)), new advanced MA technologies, namely next generation multiple access (NGMA), need to be developed~\cite{9693417}. Generally speaking, the key idea of NGMA is to intelligently accommodate multiple terminals and multiple functionalities in the allotted resource blocks in the most efficient manner possible. To achieve this exciting NGMA vision, there are many promising candidates, such as non-orthogonal multiple access (NOMA)~\cite{8114722}, space division multiple access (SDMA)~\cite{9113273}, and rate splitting multiple access (RSMA)~\cite{9831440}, which share the same \emph{non-othogonal} principle in different resource domains for enhancing the resource efficiency and improving the connectivity capability in NG wireless networks. Note that the non-orthogonal use of radio resources in NGMA leads to complicated interference management, access scheduling, and resource allocation tasks among different terminals and functionalities. Despite numerous approaches that have been proposed in the literature to address these issues in NGMA, the maximum performance achieved is highly determined by the given wireless propagation environment.
\begin{figure}[t!]
\begin{center}
    \includegraphics[width=3in]{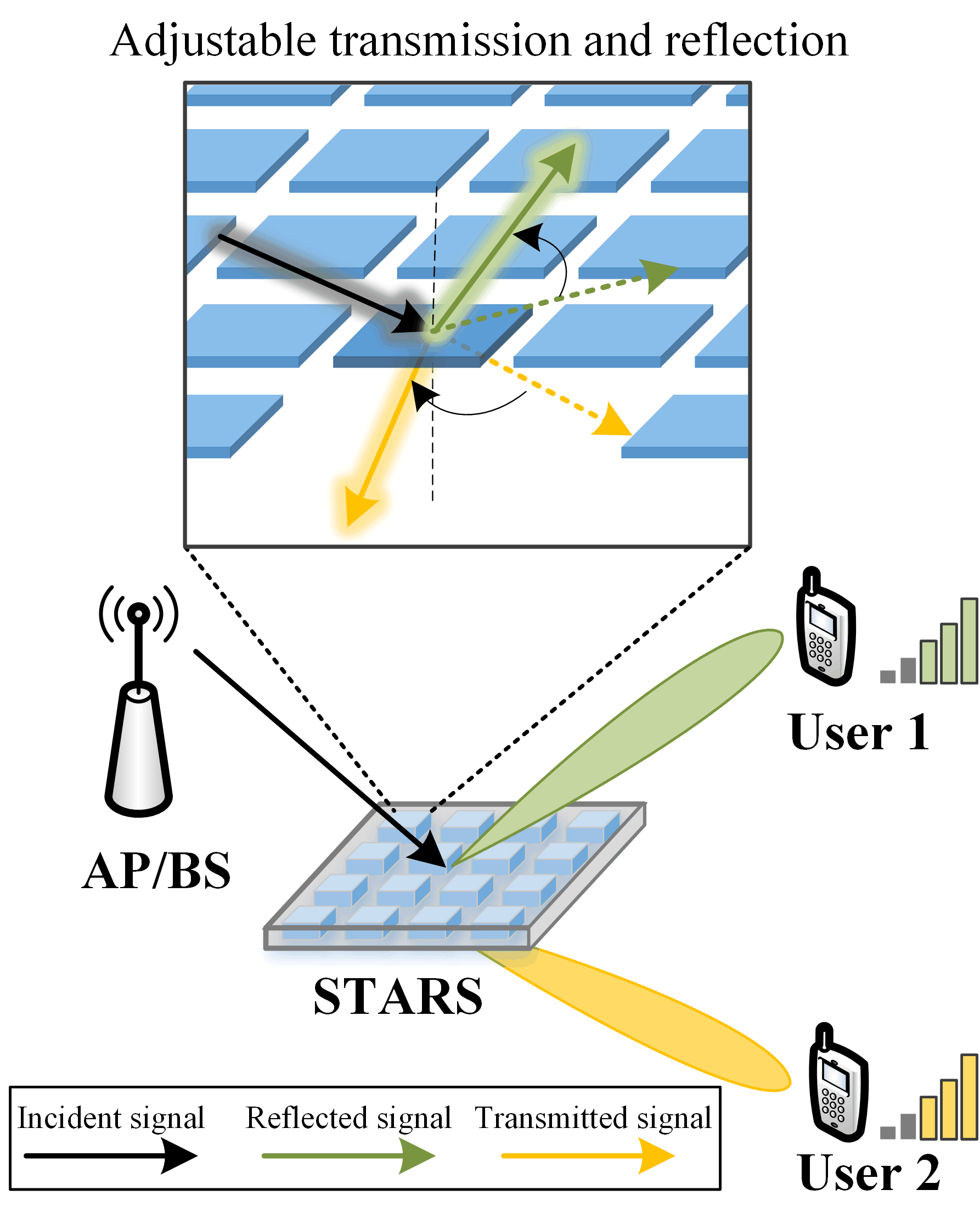}
    \caption{Illustration of the concept of STARS and the facilitated  360° smart radio environment.}
    \label{STAR_concept}
\end{center}
\end{figure}
\subsection{STARS: An Idea for 360° Smart Radio Environment}
Recently, reconfigurable intelligent surfaces (RISs) and their facilitated smart radio environment have attracted significant interest from both industries and academia~\cite{9140329,9424177,8910627,9475160}. Generally speaking, RISs have a planar surface structure embedded with a massive number of low-cost, low-energy consumption, and subwavelength electromagnetic (EM) elements (e.g., positive intrinsic negative (PIN) diodes and varactors). By changing the states of these elements, namely reconfigurable coefficients, the wireless signals incident upon RISs can be adjusted with the new propagation to achieve desired goals, such as useful signal enhancement and harmful interference mitigation, i.e., realizing a smart radio environment~\cite{9140329}. More importantly, RISs overall exhibit a nearly passive nature. The surface structure allows RISs to be easily deployed in wireless networks to beneficially 'reuse' those existing wireless signals to improve the communication performance instead of costly generating new additional wireless signals like expensive active relays. Given these advantages, RISs are regarded as the key technology for achieving sustainable, ubiquitous, and green NG wireless networks.

The cost- and energy-efficient smart radio environment provided by RISs has great potential for the realization of efficient NGMA~\cite{9693417,9779790}. Nevertheless, most of the existing RISs can merely reflect or transmit incident wireless signals, which inevitably leads to a 180° smart radio environment with degraded design flexibility~\cite{8910627,9475160}. To overcome this bottleneck, a novel concept of simultaneously transmitting and reflecting surfaces (STARS) (also known as STAR-RISs) was proposed~\cite{9437234,9690478}. As illustrated in Fig. \ref{STAR_concept}, in contrast to conventional reflecting/transmitting-only RISs, the incident wireless signals can be transmitted and reflected with adjusted propagation towards both sides of STARS~\cite{9437234,9690478}. As a result, with the aid of the 360° smart radio environment, STARS can improve the communication performance of users located on different sides compared to conventional reflecting/transmitting-only RISs. 

\subsection{Motivation and Contributions}
Based on the above observations, STARS is a promising technology for supporting NGMA among the family of RISs. Before reaping the benefits of STARS for NGMA in future wireless networks, a comprehensive review of STARS is essential to provide an in-depth understanding of the fundamentals, performance analysis, and beamforming design of STARS as well as a comprehensive discussion of possible applications of STARS and the interplay between STARS and other emerging wireless techniques for NGMA. This provides the main motivation for developing this survey paper. Our main contributions are summarized as follows.
\begin{itemize}	
	\item We introduce the fundamentals of STARS, encompassing their foundational principles, operating protocols, different categories, and state-of-the-art prototype implementations.
    \item We summarize existing research contributions on evacuating the performance of STARS-aided wireless communication networks. This research focuses on performance metrics, such as achievable rate, channel capacity, outage probability, and coverage probability.
	\item We review the existing literature on full-space STARS beamforming design with different optimization objectives and different mathematical optimization methods. Moreover, existing STARS beamforming approaches are summarized from the perspective of optimization dimensions and several open problems are highlighted.
	\item We put forward several advanced applications of STARS with unmanned aerial vehicle (UAV) communications, physical layer security (PLS), and simultaneous wireless information and power transfer (SWIPT) to address coverage, security, and sustainability requirements for NGMA.   
	\item We explore the interplay between STARS and emerging technologies, including integrated sensing and communications (ISAC), mobile edge computing (MEC), NOMA,  and near-field communications (NFC), against the backdrop of the anticipated paradigm shifts of NGMA towards accommodating multiple functionalities, achieving massive connectivity, and embracing new EM propagation techniques.
   
\end{itemize}
\begin{figure}[t!]
\begin{center}
    \includegraphics[width=3.5in]{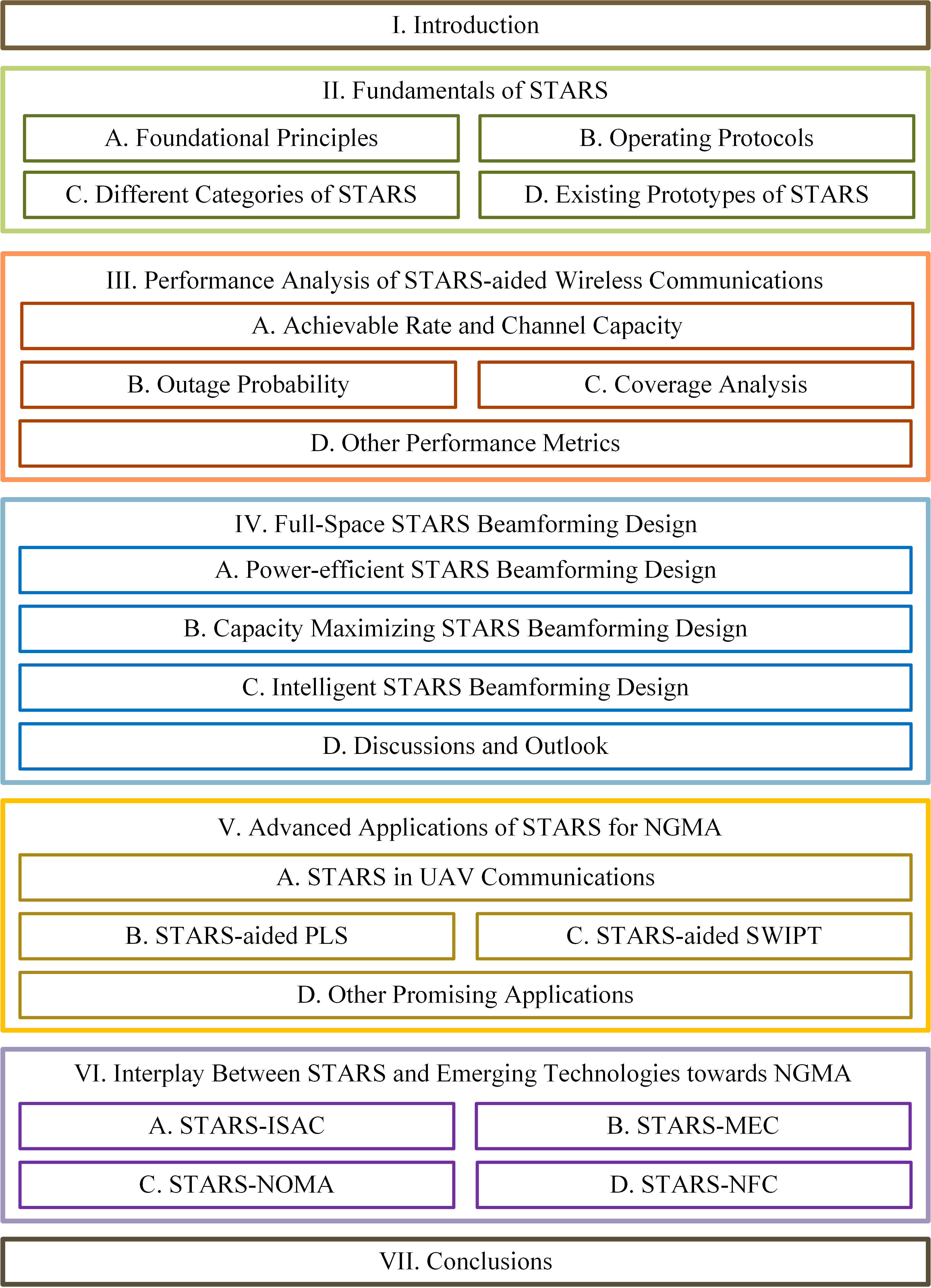}
    \caption{Organization of the survey.}
    \label{Organization}
\end{center}
\end{figure}
\subsection{Organization}
The remainder of this paper is organized as follows. In Section II, the fundamentals of STARS are elaborated with foundational principles, operating protocols, different categories, and prototypes. In Section III, performance evaluations of STARS-aided wireless communications are summarized in terms of different performance metrics. In Section IV, state-of-the-art research contributions on STARS beamforming designs are systematically discussed. To efficiently facilitate NGMA, promising applications of STARS for NGMA and the interplay between STARS and other emerging techniques towards NGMA are highlighted in Sections V and VI, respectively. Finally, Section VII concludes this survey. Fig. \ref{Organization} illustrates the organization of this paper and Table \ref{table:abbre} provides a list of acronyms used in this paper.

\begin{table}[!t]\footnotesize
\caption{List of Acronyms}
\begin{center}
\centering
\begin{tabular}{|l||l|}
\hline
AO& Alternating Optimization \\
AP&Access Point\\
BER& Bit Error Rate \\
BS & Base Station\\
CDMA & Code Division Multiple Access \\
CF&Cell-Free\\
CoMP&Coordinated Multi-Point \\
CRB & Cramér-Rao Bound \\
C\&S&Communication and Sensing \\
CSI & Channel State Information\\
DRL&Deep Reinforcement Learning \\
DoFs& Degrees-of-Freedom\\
EC & Effective Capacity \\
ES & Energy Splitting \\
EM & Electromagnetic \\
EE & Energy Efficiency\\
EDoFs& Effective DoFs \\
FBL&Finite Blocklength \\
FD& Full-Duplex\\
FDMA & Frequency Division Multiple Access\\
GAI&Generative Artificial Intelligence \\
ISAC & Integrated Sensing And Communications\\
IoT & Internet of Things\\
LoS & Line-of-Sight\\
MA & Multiple Access \\
MEC & Mobile Edge Computing\\
MIMO & Multiple-Input Multiple-Output\\
mMIMO&massive MIMO \\
MISO & Multiple-Input Single-Output\\
ML & Machine Learning\\
MS & Mode Switching \\
NFC& Near-Field Communications \\
NG & Next Generation \\
NGMA & Next Generation Multiple Access\\
NOMA & Non-Orthogonal Multiple Access\\
OFDMA & Orthogonal Frequency Division Multiple Access \\
OMA & Orthogonal Multiple Access\\
PIN & Positive Intrinsic Negative \\
PLS & Physical Layer Security \\
QoS& Quality of Service\\
RF& Radio Frequency \\
RIS& Reconfigurable Intelligent Surface\\
RL & Reinforcement Learning\\
RSMA & Rate Splitting Multiple Access\\
SCA & Successive Convex Approximation\\
SDMA& Space Division Multiple Access\\
SDR&Semi-Definite Relaxation \\
SIC& Successive Interference Cancellation\\
SINR& Signal-to-Interference-plus-Noise Ratio\\
SISO& Single-Input Single-Output\\
SNR& Signal-to-Noise Ratio \\
STARS&Simultaneously Transmitting And Reflecting Surfaces \\
SWIPT & Simultaneous Wireless Information and Power Transfer \\
TDMA & Time Division Multiple Access\\
THz & Terahertz\\
TS& Time Switching \\
TTD& True Time Delayer\\
UAV & Unmanned Aerial Vehicle\\
WSR&Weighted Sum Rate \\
\hline
\end{tabular}
\end{center}
\label{table:abbre}
\end{table}

\section{Fundamentals of STARS}

The unique feature of STARS is to produce bidirectional or multi-directional beams from a single incident beam in a controlled manner. This function is achieved through anomalous transmission and reflection of EM waves~\cite{9424177}. From the physics perspective, a surface is the boundary which separates two different media with distinct dielectric permeability. For a metasurface device, media with varying dielectric properties can be distributed in a periodic and multi-layer fashion. Wireless signals are transmitted and reflected within these devices. The overall phase delay and amplitude imposed on both the transmitted and reflected signals at each point are determined by the local dielectric permittivity and permeability of the metasurface~\cite{9140329}.
In this section, we first introduce the foundational principles of STARS, which leads to the independent and coupled phase-shift models. Based on this, we introduce three operating protocols for employing STARS in wireless communications. Then, we summarize different categories and existing prototypes of STARS in detail.

\begin{figure}[t!]
\begin{center}
    \includegraphics[width=3in]{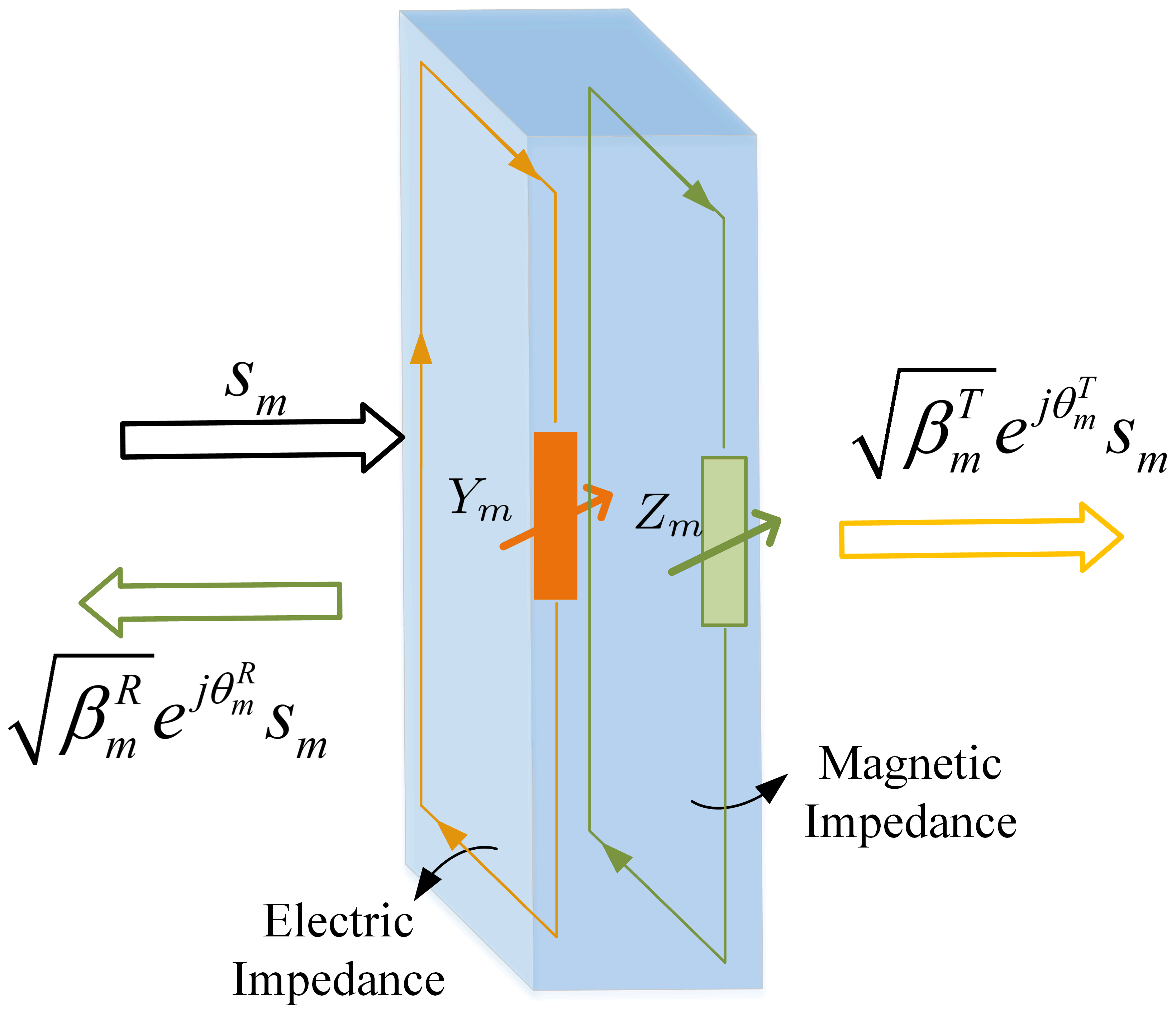}
    \caption{Conceptual illustration of the equivalent circuit of STARS element.}
    \label{STAR_0}
\end{center}
\end{figure}

\subsection{Foundational Principles}
STARS operate based on the foundational principles of RISs but introduce a groundbreaking capability that enables both signal transmission and reflection simultaneously.
The most common reflecting-only RISs are composed of a vast array of small and reconfigurable elements~\cite{9140329}. These elements can manipulate the incident EM waves by adjusting the phase shifts of their reflection coefficients.
In contrast, STARS expands upon this concept by enabling the simultaneously controlled transmission and reflection of signals via a single STARS element. To demonstrate the foundational principles of STARS, as illustrated in Fig.~\ref{STAR_0}, each element of a STARS can be treated as a lumped circuit with electric impedance, $Y_m$, and magnetic impedance, $Z_m$, where $ m \in \left\{ {1, \ldots ,M} \right\}$ and $M$ denotes the total number of STARS elements. By configuring the values of the two impedances, the corresponding transmission and reflection coefficients of the $m$th element (denoted by $T_m$ and $R_m$) can be adjusted through the following relation: 
\begin{subequations}\label{EM_STAR}
\begin{align}
&T_m = \frac{2-\eta_0Y_m}{2+\eta_0Y_m}-R_m, \\
&R_m = -\frac{2(\eta^2_0Y_m-Z_m)}{(2+\eta_0Y_m)(2\eta_0+Z_m)},
\end{align}
\end{subequations}
where $\eta_0$ is the impedance of free space~\cite{9437234}.

To further characterize the signal manipulation of each STARS element (i.e., the phase-shift and amplitude adjustment), the corresponding transmission and reflection coefficients can be re-expressed as follows:
\begin{subequations}\label{EM_STAR_2}
\begin{align}
&{T_m} = \sqrt {\beta _m^T} {e^{j\theta _m^T}}, \\
&{R_m} = \sqrt {\beta _m^R} {e^{j\theta _m^R}},
\end{align}
\end{subequations}
where $\sqrt {\beta _m^T} ,\sqrt {\beta _m^R}  \in \left[ {0,1} \right]$ and $\theta _m^T,\theta _m^R \in \left[ {0,2\pi } \right)$ characterize the amplitude and phase-shift adjustments\footnote{Note that in practice, the amplitude and phase-shift coefficients can only be taken from finite discrete values. Compared to continuous values, the finite resolution of amplitude and phase-shift adjustments will lead to performance loss, especially for 1 or 2 bits.} imposed on the incident signal by the $m$th element during transmission and reflection, respectively. As illustrated in Fig.~\ref{STAR_0}, let $s_m$ denote the signal incident upon the $m$th element. After being reconfigured by the corresponding transmission and reflection coefficients, the signals transmitted and reflected by the $m$th element are given by $\left( {\sqrt {\beta _m^T} {e^{j\theta _m^T}}} \right){s_m}$ and $\left( {\sqrt {\beta _m^R} {e^{j\theta _m^R}}} \right){s_m}$, respectively.

Moreover, according to the law of energy conservation, i.e., ${\left| {\sqrt {\beta _m^T} {e^{j\theta _m^T}}{s_m}} \right|^2} + {\left| {\sqrt {\beta _m^R} {e^{j\theta _m^R}}{s_m}} \right|^2} \le {\left| {{s_m}} \right|^2}$, the amplitude coefficients of the STARS element should follow $\beta_m^T + \beta_m^R \leq 1$.

Note that the adjustment capability of amplitudes and phase shifts in \eqref{EM_STAR_2} depends on the available values of $Y_m$ and $Z_m$ in \eqref{EM_STAR}, i.e., the practical implementation of STARS. Accordingly, there are two commonly used STARS models, namely the \emph{independent} and \emph{coupled} phase-shift models, which are discussed as follows.
\begin{itemize}	
	\item \textbf{Independent Phase-Shift Model}: As initially proposed in \cite{9437234}, the phase shifts for transmission and reflection of each STARS element can be adjusted independently of each other, i.e., $\theta _m^T \in \left[ {0,2\pi } \right)$ and $\theta _m^R \in \left[ {0,2\pi } \right)$ can be individually configured with the desired value.  The possible hardware designs for STARS with the independent phase-shift model were discussed in \cite{xu_vtmag}.
    \item \textbf{Coupled Phase-Shift Model}: As pointed out in ~\cite{zhu2014dynamic}, for STARS using passive lossless materials, the corresponding electric and magnetic impedances should be purely imaginary numbers. Here, 'lossless' means that there is no additional energy loss during transmission and reflection, i.e., $\beta _m^T + \beta _m^R = 1$. Under this constraint, it can be shown that the transmission phase shift, $\theta^T_m$, and reflection phase shift, $\theta^R_m$, are coupled subject to specific values of phase-shift differences as follows~\cite{xu_coupled}:
    \begin{equation}\label{pha}
    \left| {\theta _m^T - \theta _m^R} \right| = \frac{\pi }{2}{\text{ or }}\frac{{3\pi }}{2}.
    \end{equation}
\end{itemize}
\subsection{Operating Protocols}
\begin{figure*}[t!]
\centering
\subfigure[ES-STARS]{\label{ES} 
\includegraphics[width= 1.8in]{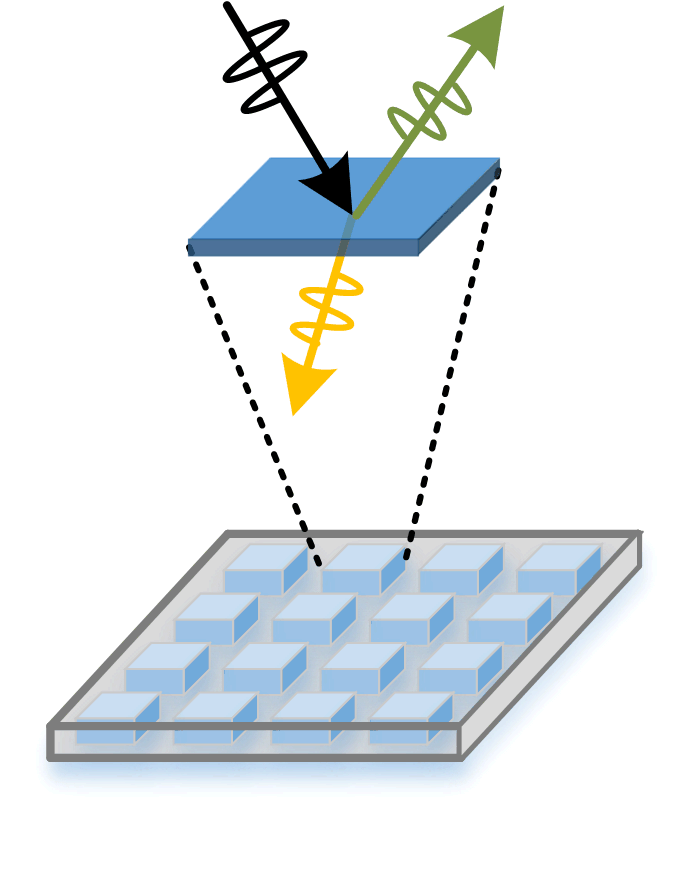}}
\subfigure[MS-STARS]{\label{MS}
\includegraphics[width= 2in]{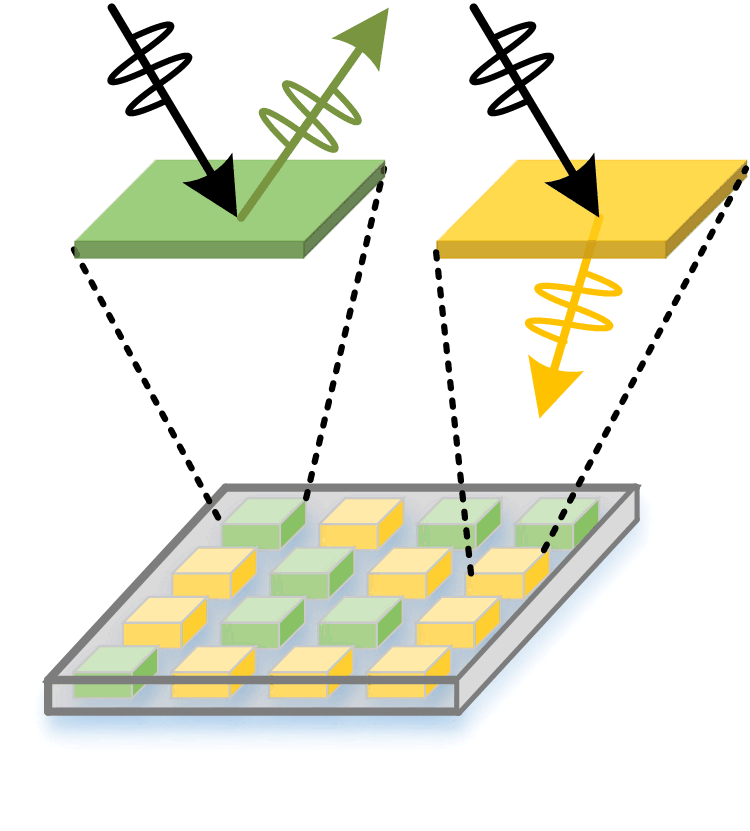}}
\subfigure[TS-STARS]{\label{TS}
\includegraphics[width= 2in]{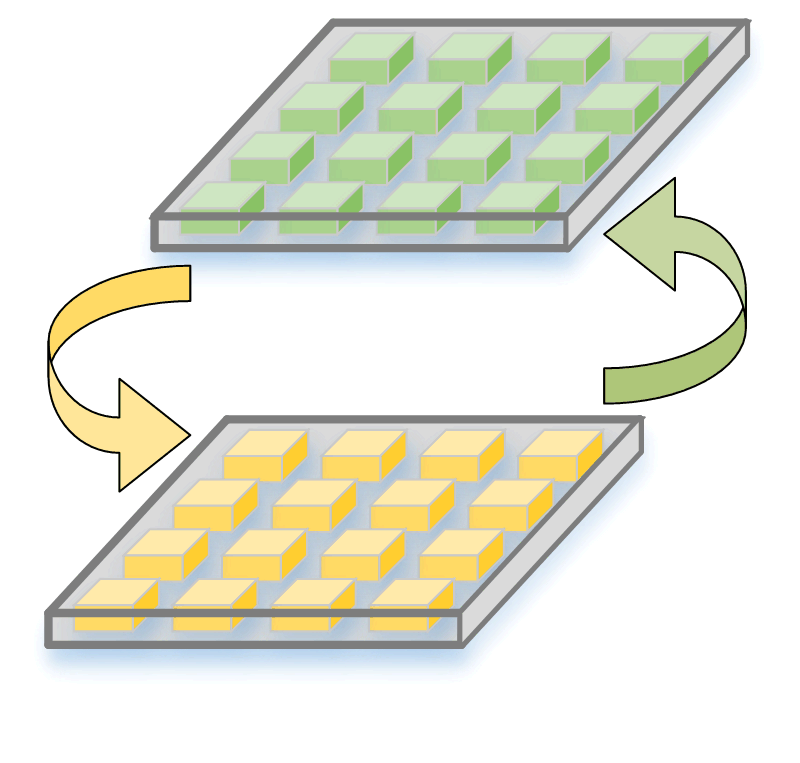}}
\caption{Three protocols for employing STARS.}\label{STARS_Protocols}
\end{figure*}

STARS provide enhanced degrees-of-freedom (DoFs) for wireless signal manipulation, which unifies conventional reflecting/transmitting-only RISs. To employ STARS in wireless communications, three operating protocols were proposed in \cite{9690478}, namely energy splitting (ES), mode switching (MS), and time switching (TS), which are described as follows.
\begin{itemize}	
	\item \textbf{ES}: As shown in Fig \ref{ES}, each element of ES-STARS is configured to simultaneously transmit and reflect the incident wireless signals. By doing so, the maximum DoFs for transmission and reflection can be exploited by ES-STARS, e.g., the maximum array gain can be achieved on both sides. However, ES-STARS generally leads to complicated beamforming design since a large number of variables for transmission and reflection have to be optimized. The case becomes even worse for ES-STARS under the coupled phase-shift model.
    \item \textbf{MS}: As shown in Fig \ref{MS}, MS-STARS can be regarded as special cases of ES-STARS, where each element is assumed to be only configured in the transmitting-only or reflecting-only mode, i.e., a binary configuration. Therefore, the configuration overhead of MS-STARS is significantly lower than that of ES-STARS, thus being more attractive for practical uses. Nevertheless, it is worth noting that MS-STARS has limited DoFs for communication design and thus degraded performance gain. 
    \item \textbf{TS}: As shown in Fig \ref{TS}, TS-STARS relies on the exploitation of the additional time domain, where the entire STARS successively work between the transmitting-only and reflecting-only modes. TS-STARS is appealing for use during the channel state information (CSI) estimation stage, where state-of-the-art CSI estimation approaches developed for conventional transmitting/reflecting-only RISs can be directly applied. Note that one main issue of TS-STARS is how to handle the high overhead and implementation complexity caused by frequent configurations.
\end{itemize}	
It is worth mentioning that the aforementioned coupled phase-shift model in \eqref{pha} is only valid and has to be considered in ES-STARS. This is because, for MS-STARS and TS-STARS, each element only works in either full transmission or full reflection mode, where the coupled phase-shift constraint becomes irrelevant. As a result, the transmission and reflection beamforming design associated with ES-STARS is more challenging than MS-STARS and TS-STARS. It is worth mentioning that compared to ES-STARS, although MS-STARS and TS-STARS do not introduce the additional coupled phase-shift constraint, i.e., less challenging STARS beamforming design, these two protocols have their own drawbacks. As discussed in \cite{9690478}, MS-STARS cannot achieve the full-dimensional transmission and reflecting beamforming gain on both sides and TS-STARS requires high hardware implementation complexity for frequently changing the working status of each element. The STARS beamforming design will be discussed in detail in Section IV.
\begin{table*}[!t]\small
\caption{Categories of STARS and Their Characteristics.}
\begin{center}
\centering
\resizebox{\textwidth}{!}{
\begin{tabular}{!{\vrule width0.8pt}l!{\vrule width0.8pt}l!{\vrule width0.8pt}l!{\vrule width0.8pt}}
\Xhline{0.8pt}
\centering
\makecell[c]{\textbf{Criteria}}  & \makecell[c]{\textbf{Categories}} & \makecell[c]{\textbf{Characteristics}} \\
\Xhline{0.8pt}
\centering
\multirow{2}{*}{\makecell[l]{Spatial Density}} & Patch-Array-Based & Low cost and easy to configure \\
\cline{2-3}
\centering
& {Metasurface-Based} & {Strong beamforming capability and enhanced DoFs}  \\
\Xhline{0.8pt}
\centering
\multirow{2}{*}{\makecell[l]{Power Consumption}} & Passive STARS & Low cost and high EE \\
\cline{2-3}
\centering
& {Active STARS} & {Able to mitigate double-fading effect}  \\
\Xhline{0.8pt}
\centering
\multirow{2}{*}{\makecell[l]{Reciprocity}} & Reciprocal STARS & Easy to implement but unavoidable energy loss in uplink \\
\cline{2-3}
\centering
& {Non-reciprocal STARS} & {High flexibility but high hardware complexity}  \\
\Xhline{0.8pt}
\end{tabular}
}
\end{center}
\label{Cat}
\end{table*}

\subsection{Different Categories of STARS}

After discussing the three major operating protocols of STARS, we focus on categorizing STARS using different criteria. 
Since its initial proposal, the novel technology of STARS has received significant research interest. As of today, there are various categories of STARS that have been proposed and studied~\cite{9570143, xu_vtmag, Zhang_2020, 10192541}. 
As summarized in Table~\ref{Cat}, in terms of their spatial density, STARS can be categorized as patch-array-based or metasurface-based. In terms of energy consumption, STARS can be passively or actively operated. Furthermore, considering reciprocity, we have reciprocal STARS, where the transmission coefficients on both sides are identical, and non-reciprocal STARS, where transmission and reflection coefficients from the two sides are different. 
In the following, we elaborate on the taxonomy of STARS.
\begin{figure}[t!]
\begin{center}
    \includegraphics[width=3in]{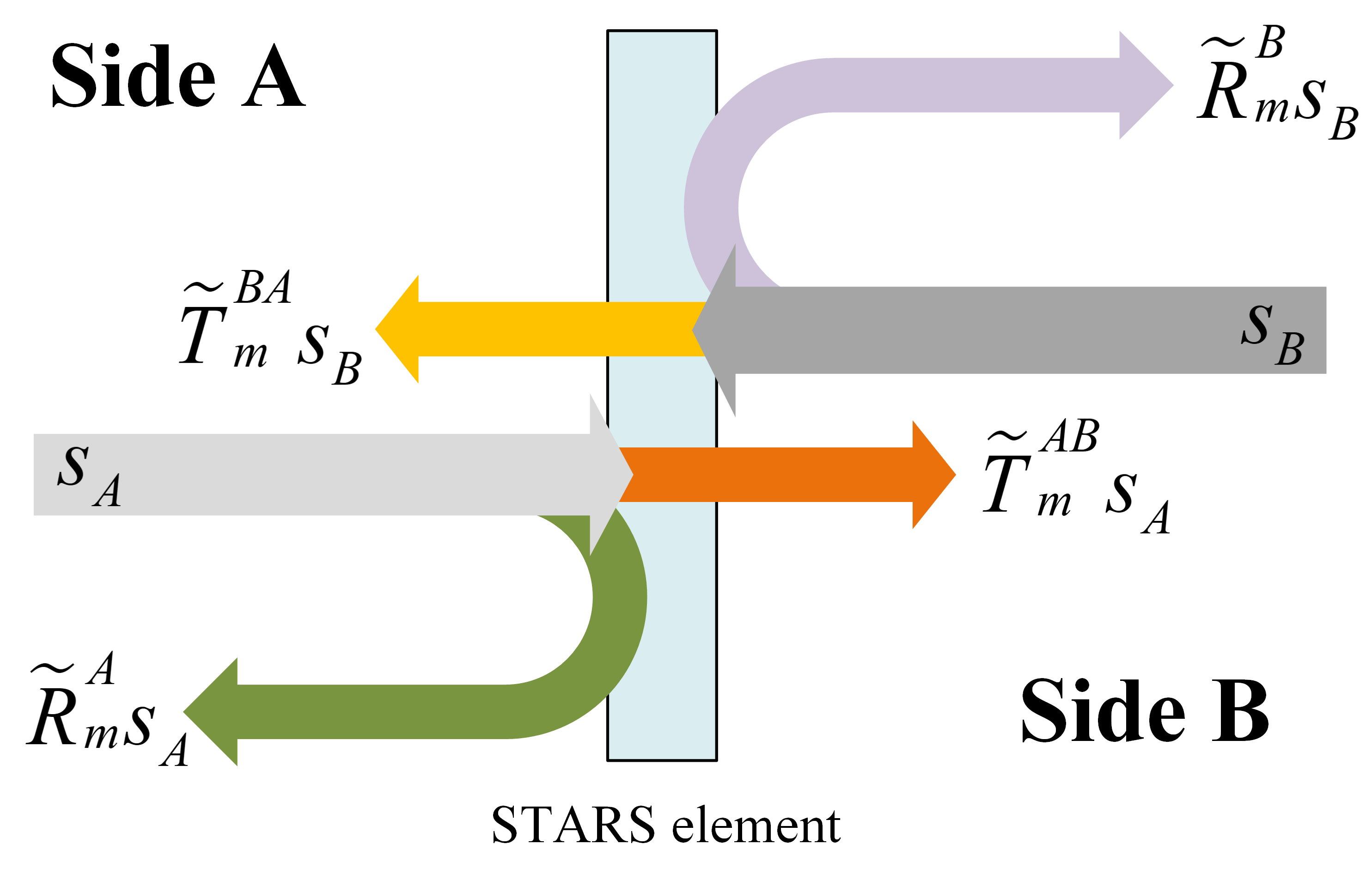}
    \caption{Signal models for STARS under dual-sided signal incidence.}
    \label{STAR_1}
\end{center}
\end{figure}

\subsubsection{Patch-Array-Based and Metasurface-Based STARS}
Patch-array-based STARS generally have a low spatial density with an element spacing in the order of several centimeters or more to accommodate multiple PIN diodes. In \cite{xu_vtmag}, existing prototypes, hardware models, and channel models for patch-array-based STARS were introduced. In contrast, metasurface-based STARS have a high spatial density with an element spacing in the order of several micrometers or less. In \cite{10192541}, a novel Green's function-based channel model was proposed for metasurface-based STARS. In this model, metasurface-based STARS are characterized by a continuous induced current distribution. The results in \cite{10192541} showed that metasurface-based STARS achieve higher numbers of spatial DoFs than patch-array-based STARS.

\subsubsection{Passive and Active STARS}
Depending on their power consumption, STARS can be nearly passive or active. Passive STARS require no dedicated power source to amplify incident signals, but an extremely small amount of energy to configure the attached EM elements. By contrast, active STARS usually contain amplifiers and even RF chains, which consume considerable power to amplify or process the incident signals. The key advantage of active STARS is that they can mitigate the unfavorable ``double-fading effect'' suffered by the passive STARS \cite{xu_active}, which motivates growing research efforts studying active STARS in different communication scenarios 
\cite{9961851,10264149,10153967,10227341}. In addition, unlike decode-and-forward relaying, no digital-to-analog or analog-to-digital converters are required at active STARS. Thus, the active STARS elements are still fairly low-cost compared to conventional antennas with RF chains.

\subsubsection{Reciprocal and Non-reciprocal STARS} STARS can be designed as either reciprocal or non-reciprocal. Under dual-sided incidence in Fig. \ref{STAR_1}, each STARS element has two pairs of transmission and reflection coefficients, i.e., $\left\{ {\widetilde T_m^{AB},\widetilde R_m^A} \right\}$ on side A, and $\left\{ {\widetilde T_m^{BA},\widetilde R_m^B} \right\}$ on side B. These four coefficients transform the wireless signals incident on sides A and B, $\left\{ {{S_A},{S_B}} \right\}$, into four output signals. For reciprocal STARS, the transmission and reflection coefficients on the two sides are symmetric, i.e., $\widetilde{T}^{AB}_m = \widetilde{T}^{BA}_m$ and $\widetilde{R}^{A}_m = \widetilde{R}^{B}_m$. For non-reciprocal STARS, the transmission and reflection coefficients on the two sides can be different. Note that for STARS-aided uplink communications, where users on both sides upload their signals to one base station (BS) via STARS. In this case, there is always a power leakage for reciprocal STARS. Otherwise, only the signals of users on the one side can be received by the BS at the same time. However, this issue can be mitigated by non-reciprocal STARS to have different transmission and reflection coefficients on each side. Although non-reciprocal STARS are more attractive, they usually have a higher hardware complexity than reciprocal STARS since multi-layer structures are needed to break the reciprocity~\cite{PhysRevLett}.
\subsection{Existing Prototypes of STARS}

In exploring the landscape of STARS prototypes, researchers have explored various materials, each offering unique capabilities. One of the early prototypes of STARS is the transparent dynamic metasurface, developed in collaboration with AGC Inc. and NTT DOCOMO, Japan~\cite{ntt}. This transparent dynamic metasurface operates at 28 GHz and allows dynamic control of transmission and reflection functions. As the transparent dynamic metasurface is based on glass, it is attractive to be integrated into windows for connecting outdoor and indoor users. In \cite{9895224}, another STARS prototype, namely intelligent omni-surfaces (IOS), was developed mainly relying on PIN diodes, where experimental results confirmed that desired transmitted and reflected beams can be achieved on both sides. Moreover, some researchers worked on developing STARS prototypes, which enable independent control for transmitted and reflected signals. For example, the authors of \cite{10288376} designed a STARS unit cell, which can independently achieve beamforming or beamsteering in both transmission and reflection modes. The designed unit cell utilizes PIN diodes to obtain two phase states 0° and 180° in the simultaneous transmission and reflection mode. In addition, the designed STARS unit cell was fabricated and characterized in a waveguide measurement, which showed excellent agreement between simulation and measurement at the operating frequency of 5.9 GHz. In \cite{10177915}, the authors designed and fabricated a STARS prototype with $16\times16$ elements. The designed prototype can realize bidirectional beams simultaneously, where the forward and backward beam directions can be controlled independently. The measured results showed a good beam-scanning capability of $\pm 60^\circ$ for both transmitting and reflecting beams.

\section{Performance Analysis of STARS-aided Wireless Communications}
In Section II, we have discussed the fundamentals of STARS. The enhanced DoFs provided by STARS have great potential to enhance the performance of wireless communications. In this section, we comprehensively review existing research contributions on the performance evaluation of STARS-aided wireless communications, encompassing the achievable rate, outage probability, coverage probability, and other key indicators across diverse scenarios. In Table.~\ref{Tab:Per}, we summarize these research contributions in terms of the considered performance metrics, channel models, and employed techniques.
\begin{table*}[!t]\tiny
\caption{Summary of Existing Research Contributions on Performance Analysis for STARS.}
\label{Tab:Per}
\begin{center}
    
\begin{tabular}{!{\vrule width1.5pt}l!{\vrule width1.5pt}l!{\vrule width1.5pt}l!{\vrule width1.5pt}l!{\vrule width1.5pt}p{8.5cm}!{\vrule width1.5pt}}
\Xhline{1.5pt}
\multicolumn{1}{!{\vrule width1.5pt}c!{\vrule width1.5pt}}{\textbf{Performance Metrics}} &
  \multicolumn{1}{c!{\vrule width1.5pt}}{\textbf{[Ref]}} &
  \multicolumn{1}{c!{\vrule width1.5pt}}{\textbf{System Setup}} &
  \multicolumn{1}{c!{\vrule width1.5pt}}{\textbf{Channel Model}} &
  \multicolumn{1}{c!{\vrule width1.5pt}}{\textbf{Characteristics/Techniques}} \\ \Xhline{1.5pt}
\multirow{10}{*}{Achievable Rate} &
  \cite{10156858} &
  STARS-NOMA &
  Rician fading &
  Linear minimum mean square error for channel estimation and HWIs were considered \\ \cline{2-5} 

&
  \cite{10466748} &
  STARS-NOMA &
  Rician fading &
 Both imperfect CSI and transceiver HWIs were considered.\\ \cline{2-5} 
  
 &
  \cite{Qingchao} &
  STARS-NOMA &
  Nakagami-m fading &
The asymptotically ergodic rate lower bound under HWIs was derived for infinite large and continuous aperture STARS.\\ \cline{2-5} 
  
 &
  \cite{9815097} &
  STARS-NOMA &
  Nakagami-m fading &
  EC was adopted as the metric to explore the delay requirements of NOMA users. \\ \cline{2-5} 
 &
  \cite{9869706} &
  STARS-NOMA &
  Rician fading &
  Increasing the size of STARS can effectively improve the ergodic rate. \\ \cline{2-5} 
 &
  \cite{10175074} &
  STARS-mMIMO &
  Correlated Rayleigh fading &
  Closed-form achievable rate expression was derived based on statistical CSI. \\ \cline{2-5} 
 &
  \cite{9843866} &
  STARS-NOMA &
  Nakagami-m fading &
  Closed-form expressions of ergodic rates and high SNR slopes for cell-edge NOMA users were derived. \\ \cline{2-5} 
 &
  \cite{10373089} &
  STARS-mMIMO &
  Correlated Rayleigh fading &
  The UatF bound of the sum rate was derived based on statistical CSI.\\ \cline{2-5} 
 &
  \cite{10297571} &
  STARS-CF-mMIMO &
  Correlated Rayleigh fading &
  Closed-form achievable rate expression was derived under correlated Rayleigh fading in CF mMIMO. \\ \cline{2-5} 
 &
  \cite{10264149} &
  Active STARS &
  Correlated Rayleigh fading &
  Active STARS can achieve higher spectral efficiency than conventional active RISs. \\ \Xhline{1.5pt}
\multirow{5}{*}{Outage Probability} &
  \cite{9437234} &
  STARS-NOMA/OMA &
  Rician fading &
  Outage probabilities and diversity order were derived for passive STARS in NOMA and OMA cases. \\ \cline{2-5} 
 &
  \cite{xu_active} &
  Active STARS &
  Rician fading &
  Scaling laws and diversity orders were derived for active STARS under the independent/coupled phase-shift model. \\ \cline{2-5} 
 &
  \cite{9864148} &
  STARS-D2D &
  Rayleigh fading &
  The impacts of the various parameters on the D2D system performance were examined. \\ \cline{2-5} 
 &
  \cite{10345673} &
  STARS-NOMA-HARQ &
  Nakagami-m fading &
  HARQ can improve the reliability-delay balance and security-reliability balance for STARS-NOMA. \\ \cline{2-5} 
 &
  \cite{9935303} &
  STARS-NOMA &
  Rician fading &
  The outage probability error floor can be reduced by adjusting the energy-splitting ratio of STARS. \\ \Xhline{1.5pt}
\multirow{4}{*}{Coverage Analysis} &
  \cite{9786058} &
  STARS-mMIMO &
  Correlated Rayleigh fading &
  Closed-form coverage probability expression was derived under phase-shift errors. \\ \cline{2-5} 
 &
  \cite{9462949} &
  STARS-NOMA/OMA &
  Richian fading &
  NOMA can further enhance the transmission and reflection coverage trade-off performance achieved by STARS. \\ \cline{2-5} 
 &
  \cite{ghadi2023analytical} &
  STARS-NOMA &
  Rayleigh fading &
  STARS-NOMA is more beneficial to extend the coverage region than STARS-OMA. \\ \cline{2-5} 
 &
  \cite{9808307} &
  STARS-NOMA &
  Any fadings &
  A stochastic geometry-based analytical framework was provided for STARS-aided multi-cell communication networks. \\ \Xhline{1.5pt}
Effective DoFs &
  \cite{10192541} &
  STARS-NFC &
  Near-field channel &
  Metasurface-based STARS can achieve higher EDoFs than patch-array-based STARS in the near-field region.  \\ \Xhline{1.5pt}
  
Bit Error Rate &
  \cite{9786807} &
  STARS-NOMA &
  Rayleigh fading &
  Closed-form BER expressions were derived under perfect and imperfect SIC cases. \\ \Xhline{1.5pt}
  
Block Error Rate &
  \cite{10049460} &
  STARS-NOMA-FBL &
  Nakagami-m fading &
  PDF and CDF were derived for analyzing the block error rate under the FBL regime. \\ \Xhline{1.5pt}
\end{tabular}
\end{center}
\end{table*}

\subsection{Achievable Rate and Channel Capacity}

The achievable rate or channel capacity is one of the most important performance metrics when evaluating the performance gain of new wireless technologies. In the context of STARS-aided wireless communications, many studies have contributed valuable insights. For example, the authors of \cite{10156858} analyzed the ergodic sum rate in a STARS-assisted NOMA uplink communication system considering practical factors, namely channel estimation error and hardware impairments (HWIs). They showed that the transceiver HWIs result in significant performance limitations for both channel estimation and ergodic sum rate at the high transmit power regime. The ergodic spectral efficiency was further analyzed in~\cite{10466748} considering HWIs. It was shown that the HWIs at the AP significantly limit the ergodic spectral efficiency and the negative effect of HWIs can be effectively compensated by having more antennas at the AP. Moreover, the authors of \cite{Qingchao} proposed to employ the STARS at the BS to serve two users using the NOMA technique. Considering HWIs, the asymptotically ergodic rate lower bound was derived for the regime of an infinite number of STARS elements and the continuous aperture STARS.
Considering ultra-reliable low-latency communications, the authors of \cite{9815097} analyzed the effective capacity (EC) of STARS-aided NOMA communication systems. Simulation results revealed a linear increase of EC for the near NOMA user with the increase of transmit signal-to-noise ratio (SNR). For the downlink case, \cite{9869706} provided an ergodic rate approximation for a STARS-aided NOMA communication system. The statistics of signal-to-interference-plus-noise ratio (SINR) were derived using the statistical characteristics of channels, which were maximized via the optimization of transmission and reflection coefficients. Noteworthy findings from the analytical expressions include the positive impact of increasing the size of the uniform linear array (ULA) and STARS size on the system ergodic rate. To analyze the impact of correlated fading, the authors of \cite{10175074} studied the achievable rate of a STARS-assisted downlink massive multiple-input multiple-output (mMIMO) system under spatially correlated fading. A closed-form achievable rate expression was derived based on statistical CSI, which was further maximized using the projected gradient ascent method. 
Furthermore, the authors of \cite{9843866} analyzed the ergodic rates of a STARS-aided NOMA communication system, where closed-form expressions for ergodic rates were derived and high SNR slopes were obtained. This study revealed that a significant performance enhancement can be achieved by increasing the number of STARS elements. Investigating the impact of phase noise in STARS-aided mMIMO downlink systems, the authors of \cite{10373089} derived the use-and-then-forget (UatF) bound of the sum rate under statistical CSI. They showed that an improved sum rate can be achieved with an increased noise concentration parameter. Moreover, the authors of \cite{10297571} investigated the performance of STARS-assisted cell-free (CF) mMIMO systems under correlated Rayleigh fading. Through the design of passive beamforming based on statistical CSI, the authors derived a closed-form expression for the achievable rate, which offered insights into the impact of channel correlation, the number of surface elements, and pilot contamination. Considering a similar system setup to \cite{10297571}, the spectral efficiency was further analyzed by the authors of \cite{10264149} with active STARS. Simulation results showed that active STARS significantly improve the spectral efficiency of the CF-mMIMO system than conventional active RISs.

\begin{figure}[t!]
\begin{center}
    \includegraphics[width=3.5in]{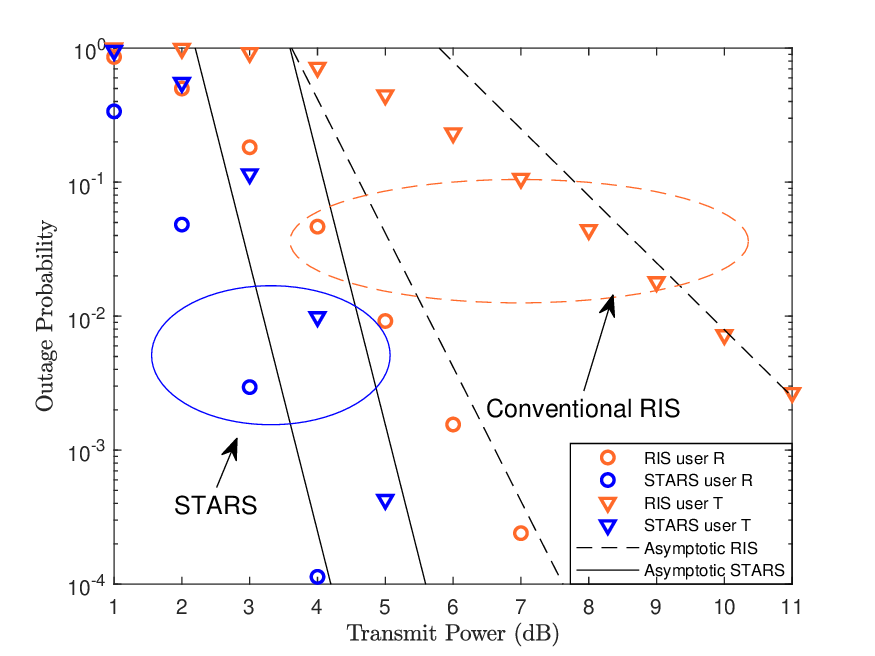}
    \caption{Outage probabilities of users with the aid of STARS and conventional RIS, where all links are assumed to follow Rician fading. The number of STARS elements is 8 and the conventional RIS has 3 transmitting-only elements and 5 reflecting-only elements. The other parameters adopted can be found in~\cite{9437234}.}
    \label{CL_OUT}
\end{center}
\end{figure}

\subsection{Outage Probability}
Outage probability in wireless communications quantifies the likelihood of the instantaneous channel performance falling below a specified threshold, serving as a critical metric for assessing system reliability and robustness. Several studies contributed insights into the outage probability performance analysis of STARS-aided wireless communications.
Beginning with a STARS-aided two-user downlink communication system, the authors of~\cite{9437234} derived the outage probabilities and diversity orders achieved by STARS in both NOMA and orthogonal multiple access (OMA) cases. As shown in Fig.\ref{CL_OUT}, for STARS, both users can achieve full diversity order. However, for the conventional RIS, the diversity orders achieved on both sides are significantly reduced. In \cite{xu_active}, the authors presented closed-form expressions for outage probabilities achieved by active STARS under both independent and coupled phase-shift models. Moreover, scaling laws and diversity orders for users on both sides were derived, which demonstrated that active STARS achieve the same diversity orders as passive STARS but with different scaling laws. Moreover, the authors of \cite{9864148} studied a STARS-aided two-way full-duplex (FD) device-to-device (D2D) communication system, where the closed-form expressions of the outage probability, sum throughput, ergodic capacity, and energy efficiency (EE) were derived under optimal and uncertain phase-shift alignments. Insights into the diversity order and ergodic slope were provided, examining the influence of transmit power configurations, STARS deployments, target data rate transmission allocation, and the number of user deployments on system performance. The authors of \cite{10345673} conducted performance analysis for STARS-aided hybrid automatic repeat request (HARQ) assisted cognitive NOMA systems. The study focused on a security-required user (SRU) with secure transmission requirements paired with a quality of service (QoS)-sensitive user with low delay requirements. Analytical expressions for connection outage probability (COP), average number of transmissions, and secrecy outage probability were derived for the SRU in the randomized retransmission NOMA scheme. Similarly, the authors of \cite{9935303} provided insights into the asymptotic outage probability for STARS-aided uplink communication systems with NOMA and OMA. It demonstrated that adjusting the energy splitting ratio between transmission and reflection signals can help to reduce the outage probability error floor occurring in uplink NOMA.

\subsection{Coverage Analysis}
\begin{figure}[t!]
\begin{center}
    \includegraphics[width=3.5in]{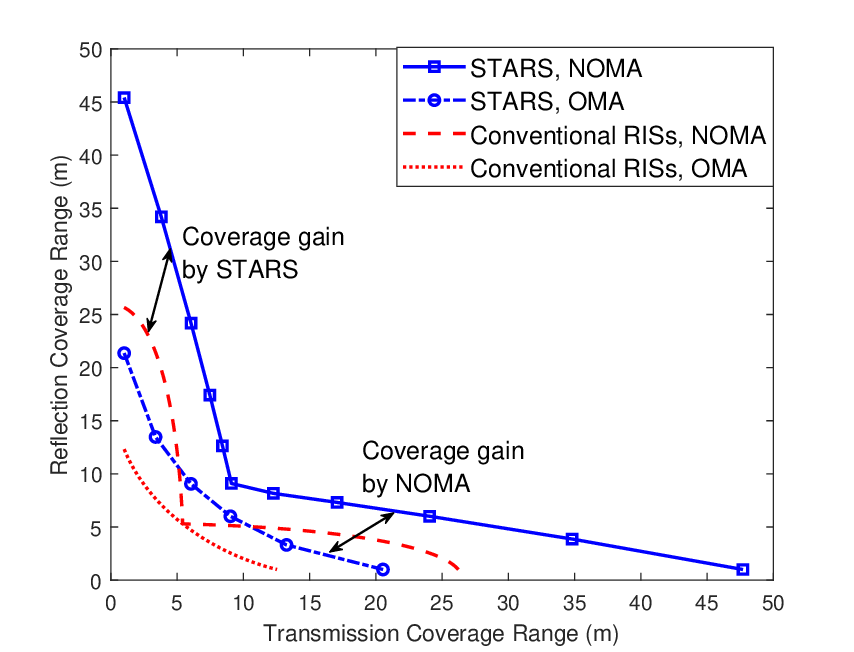}
    \caption{Illustration of transmission and reflection coverage trade-off achieved by STARS and conventional RISs, where the number of RIS elements is 100 and the users' minimum rate requirement is 5 bps/Hz. The direct BS-user link is assumed to be blocked. The other parameters adopted can be found in~\cite[Section IV]{9462949}.}
    \label{coverage}
\end{center}
\end{figure}

As one of the key advantages of STARS over conventional reflecting/transmitting-only RISs, the extended coverage performance of STARS was characterized by many research contributions. For example, the authors of \cite{9786058} investigated the coverage performance of a STARS-assisted mMIMO system. In particular, the authors derived a closed-form expression for the coverage probability under correlated fading and STARS phase-shift errors. Focusing on a basic STARS-aided single-input single-output (SISO) two-user communication system, the authors of \cite{9462949} characterized the sum of maximum transmission and reflection coverage ranges subject to users' minimum rate requirements. In particular, both the capacity-achieving NOMA and the suboptimal OMA were considered. As depicted in Fig. \ref{coverage}, the transmission and reflection coverage trade-off performance achieved by STARS and conventional RISs were compared in NOMA and OMA. Here, conventional RISs consist of one reflecting-only RIS and one transmitting-only RIS, each of which has half the size of STARS. As illustrated in Fig. \ref{coverage}, the maximum transmission/reflection coverage range achieved by STARS is significantly enlarged compared to conventional RISs. Moreover, employing NOMA can further enhance the coverage performance of STARS. Furthermore, the authors of \cite{ghadi2023analytical} provided an analytical characterization of the coverage region of a STARS-aided two-user downlink communication system. Both OMA and NOMA were considered under the ES protocol. Results confirmed that the use of STARS is beneficial to extend the coverage region and the use of NOMA provides better performance compared to OMA. In the context of STARS-aided multi-cell communication networks, the authors of \cite{9808307} proposed a fitting method for characterizing the small-scale fading power distribution and developed an analytical framework based on stochastic geometry. This framework captures the randomness of STARS, BSs, and user equipment, which provides insights into coverage probability and ergodic rates achieved by STARS. 
\subsection{Other Performance Metrics}
A few studies investigated valuable insights using other performance indicators, which are discussed as follows.

\begin{itemize}
    \item \textbf{Effective DoFs (EDoFs):}
    The number of EDoFs of a STARS-aided MIMO channel is of vital importance to determine the capacity. Especially when near-field propagation is considered, the EDoFs achieved by the STARS-aided MIMO channel are distance-dependent. To address this challenge, the authors of \cite{10192541} derived the near/far-field boundary, end-to-end channel gain, and the numbers of EDoFs for metasurface-based STARS. Exploiting the Green's function method, they showed that metasurface-based STARS can achieve higher EDoFs than patch-array-based STARS in the near-field regime.
    \item \textbf{Bit Error Rate (BER):}
    The authors of \cite{9786807} examined the BER performance of a STARS-aided NOMA communication system, where closed-form BER expressions were derived for both perfect and imperfect successive interference cancellation (SIC) cases. Asymptotic analyses are conducted to gain deeper insights into BER behavior in the high SNR region. The results showed that in the case of perfect SIC, the performance of the far NOMA user is more sensitive to the power allocation than the change of STARS size. 
    \item \textbf{Block Error Rate:} Shifting the focus to the finite blocklength (FBL) regime, the authors of \cite{10049460} investigated a STARS-aided downlink NOMA communication system. Probability density functions (PDF) and cumulative distribution functions (CDF) of the SINR for each user were derived, providing a foundation for analyzing system performance through block error rate.
\end{itemize}

\section{Full-Space STARS Beamforming Design}

The diverse performance analysis work presented in the previous section has revealed the great potential of deploying STARS in NG wireless networks. Nevertheless, to fully unlock these promising benefits of STARS, efficient STARS beamforming approaches have to be conceived, which have attracted extensive efforts from researchers. Compared to conventional reflecting/transmitting-only RISs, the STARS beamforming design imposes unique challenges. Firstly, there are significantly increased numbers of optimization variables, i.e., the optimization of amplitudes and phase shifts for both transmission and reflection, which further increases the design complexity. Secondly, additional coupling constraints exist in the STARS beamforming configuration (e.g., the energy conservation constraint, the coupled phase-shift model, etc), which makes the design quite challenging. To overcome these obstacles, extensive research efforts have been devoted. In this section, we provide an overview of state-of-the-art full-space STARS beamforming design in terms of different objectives and mathematical methods. Table \ref{STARS beamforming} summarizes existing research contributions on STARS beamforming.
\begin{table*}[!t]\large
\caption{Summary of Existing Research Contributions on STARS Beamforming Design.}
\begin{center}
\centering
\resizebox{\textwidth}{!}{
\begin{tabular}{!{\vrule width1.5pt}l!{\vrule width1.5pt}l!{\vrule width1.5pt}l!{\vrule width1.5pt}l!{\vrule width1.5pt}l!{\vrule width1.5pt}}
\Xhline{1.5pt}
\centering
\makecell[c]{\textbf{Category}}  & \makecell[c]{\textbf{Ref.}} & \makecell[c]{\textbf{Design Objectives}} &\makecell[c]{\textbf{Phase-shift Model}} & \makecell[c]{\textbf{Characteristics/Techniques}} \\
\Xhline{1.5pt}
\centering
\multirow{6}{*}{\makecell[l]{Power-efficient \\STARS Beamforming \\(convex optimization-based)}} & \makecell[c]{\cite{9570143}} & {Power} & {Independent}  & {Three operating protocols and STARS beamforming for unicast and multicast scenarios }\\
\cline{2-5}
\centering
& \makecell[c]{\cite{9920228}}   & {Power}    & {Independent} &{AO-based iterative amplitude and phase-shift coefficients design for STARS beamforming}\\
\cline{2-5}
\centering
& \makecell[c]{\cite{10130543}}   & {EE}    & {Independent} & {AO-based STARS beamforming with SDR and SCA for EE fairness optimization}\\
\cline{2-5}
\centering
& \makecell[c]{\cite{9785636}}   & {Power}    & {Independent} & {STARS beamforming for uplink-to-downlink interference mitigation in FD systems}\\
\cline{2-5}
\centering
& \makecell[c]{\cite{9838767}}   & {Power}    & {Coupled} & {Low-complexity element-wise algorithm for STARS beamforming}\\
\cline{2-5}
\centering
& \makecell[c]{\cite{10224271}}   & {Power}    & {Coupled} & {Dual-functional STARS for enhancing primary transmission and enabling secondary transmission}\\
\Xhline{1.5pt}
\centering
\multirow{7}{*}{\makecell[l]{Capacity Maximizing \\STARS Beamforming \\(convex optimization-based)}}& \makecell[c]{\cite{9629335}}   & {WSR}    & {Independent} & {CCCP-based algorithm for STARS beamforming in MIMO communications}\\
\cline{2-5}
\centering
& \makecell[c]{\cite{9751144}}   & {WSR}    & {Coupled} & {Exhaustive search-based element-wise algorithm for STARS beamforming with discrete phase shifts}\\
\cline{2-5}
\centering
& \makecell[c]{\cite{10325546}}   & {Sum rate}    & {Coupled} & {Element-wise alternating amplitude and phase-shift coefficient optimization}\\
\cline{2-5}
\centering
& \makecell[c]{\cite{9935266}}   & {WSR}    & {Coupled} &{A general penalty-based optimization framework with provable optimality for STARS beamforming}\\
\cline{2-5}
\centering
& \makecell[c]{\cite{10093070}}   & {Sum rate}    & {Independent} & {Exploiting statistical CSI for achieving low-overhead STARS beamforming}\\
\cline{2-5}
\centering
& \makecell[c]{\cite{10050140}}   & {Sum rate}    & {Independent} & {Joint STARS beamforming and deployment location optimization for the single-STARS case}\\
\cline{2-5}
\centering
& \makecell[c]{\cite{10254537}}   & {Sum rate}    & {Independent} & {Joint STARS beamforming design and candidate deployment selection for the multiple-STARS case}\\
\Xhline{1.5pt}
\centering
\multirow{5}{*}{\makecell[l]{Intelligent \\STARS Beamforming \\(ML-based)}}& \makecell[c]{\cite{9837935}}   & {Power}    & {Coupled} & {A pair of hybrid RL algorithms with continuous and discrete actions}\\
\cline{2-5}
\centering
& \makecell[c]{\cite{9964251}}   & {EE}    & {Independent} & {DRL for EE maximization in STARS-aided NOMA communications}\\
\cline{2-5}
\centering
& \makecell[c]{\cite{10306287}}   & {Date rate}    & {Independent} & {DRL for joint spectrum allocation and STARS beamforming in V2X communications}\\
\cline{2-5}
\centering
& \makecell[c]{\cite{10187159}}   & {Capacity/Coverage}    & {Independent} & {MO-PPO algorithm for characterizing the capacity-versus-coverage peformance}\\
\cline{2-5}
\centering
& \makecell[c]{\cite{10049110}}   & {Throughput}    & {Independent} & {A tile-based STARS beamforming approach and a distributed learning approach for multi-cell cases}\\
\Xhline{1.5pt}
\end{tabular}
}
\end{center}
\label{STARS beamforming}
\end{table*}

\subsection{Power-efficient STARS Beamforming Design}
With the increasing carrier frequencies and the employment of power-hungry hardware components at access points (APs)/BSs, the minimization of transmit power and/or the maximization of EE becomes a main design objective for wireless networks. To this end, growing research efforts have been devoted to the topic of ``power-efficient STARS beamforming design''. The authors of \cite{9570143} first proposed three protocols and studied the corresponding joint active and passive beamforming design in a STARS-aided downlink multiple-input single-output (MISO) multi-user communication system. Considering both unicast and multicast communication scenarios, penalty-based iterative joint optimization algorithms were developed to minimize the transmit power of the AP required to satisfy the minimum communication rate requirements of users. The results obtained in \cite{9570143} characterized significant performance gains that can be achieved by STARS over conventional reflecting/transmitting-only RISs and also showed that TS and ES protocols are generally preferred to be employed for unicast and multicast scenarios, respectively. The same problem was further studied in \cite{9920228} for ES-STARS. In particular, an alternating optimization (AO)-based iterative algorithm was proposed, where the amplitude coefficients and phase-shift coefficients of STARS are alternately optimized. Moreover, the authors of \cite{10130543} studied EE fairness optimization for a STARS-aided MISO multi-user communication system, where the semi-definite relaxation (SDR) and successive convex approximation (SCA) methods were used in the developed AO-based algorithm. To further utilize the interference management capability of STARS, the transmit power minimization problem was studied for a STARS-aided FD communication system in \cite{9785636}, where the STARS beamforming was optimized to mitigate the interference from the uplink user to the downlink user. In contrast to the aforementioned works assuming the independent phase-shift model, the authors of \cite{9838767} first studied the STARS beamforming design under the coupled phase-shift model. Considering a STARS-aided downlink SISO two-user communication system with OMA and NOMA schemes, an element-wise iterative algorithm was developed for minimizing the transmit power, whose computational complexity only increases linearly with the number of STARS elements, thus being promising for practical implementation. Focusing on the coupled phase-shift model, the authors of \cite{9785636} continued to study the transmit power minimization problem in a STARS-aided symbiotic radio communication system. Specifically, the primary and secondary users are located on different sides of the STARS, where the STARS not only enhances the performance of the primary network but also acts as a transmitter for the secondary network.

\subsection{Capacity Maximizing STARS Beamforming Design}
Capacity is another critical metric for evaluating the performance gain that can be achieved by STARS, which leads to research on ``capacity maximizing STARS beamforming design''. As a first step, the authors of \cite{9629335} studied the weighted sum rate (WSR) maximization problem for a STARS-aided two-user MIMO communication system under three operating protocols and the independent phase-shift model. Following a block coordinate descent optimization framework, the constrained concave-convex procedure (CCCP) was invoked for the STARS beamforming optimization, which showed a better performance than the commonly used SDR method. The similar WSR maximization problem was further studied in \cite{9751144,10325546,9935266} by taking the coupled phase-shift model into consideration. To address this challenging constraint in STARS beamforming design, the authors of \cite{9751144} proposed an exhaustive search-based element-wise iterative algorithm to obtain the optimal discrete amplitude and phase-shift coefficients. The element-wise alternating amplitude and phase-shift coefficient optimization algorithm was further extended for STARS beamforming design in \cite{10325546} to maximize the average sum rate of the STARS-aided multi-user MISO-OFDM system. Note that for the coupled phase-shift model, STARS beamforming optimization algorithms developed in the aforementioned works \cite{9838767,10325546,9751144} are only applicable to specific communication scenarios (i.e., the two-user case in \cite{9838767}) and/or do not have provable optimality. To address this issue, the authors of \cite{9935266} proposed a general penalty-based optimization framework for STARS beamforming design with the coupled phase-shift model, which can obtain the Karush-Kuhn-Tucker (KKT) optimal solution under some mild conditions. Note that the beauty of the general optimization framework is that it can be directly extended to other STARS-aided communication designs. Furthermore, considering the channel estimation overhead issue, the authors of \cite{10093070} maximized the sum spectral efficiency of a cooperative RIS and STARS-aided mMIMO communication system, where the RIS and STARS beamforming were designed based on statistical CSI. As a further advance, the STARS deployment location design was investigated for single-STARS \cite{10050140} and multiple-STARS \cite{10254537} cases. In \cite{10050140}, the STARS beamforming, deployment location, and active BS beamforming were jointly optimized with the developed AO-based algorithm in a STARS-aided NOMA communication system. It shows that the WSR can be further enhanced via the deployment design and asymmetric STARS deployment strategy is preferred when employing NOMA communication. Focusing on a multiple-STARS-aided downlink communication system in \cite{10254537}, the sum rate of blocked users was maximized via optimizing the STARS beamforming and STARS candidate deployment locations as well as the BS active beamforming.

\subsection{Intelligent STARS Beamforming Design}
The above research contributions utilized conventional convex optimization-based methods for STARS beamforming design. However, these developed algorithms may rely on some ideal assumptions (e.g., perfect CSI, ignored STARS configuration overhead, etc) and sometimes suffer from high computational complexity, thus becoming inefficient in practical implementation. As a remedy, with the rapid development of artificial intelligence, machine learning (ML) techniques have become an efficient enabler for 6G and beyond wireless communication designs. Therefore, ML-based intelligent STARS beamforming design has attracted significant research interest in recent years. The authors of \cite{9837935} proposed a pair of hybrid reinforcement learning (RL) algorithms for power-efficient STARS beamforming design under the coupled phase-shift model, where the transmission and reflection phase-shift optimization is decomposed into one continuous control from $\left[ {0,2\pi } \right)$ and one discrete control, i.e., $ \pm \frac{\pi }{2}$. The EE maximization problem of a STARS-aided NOMA communication system was studied in \cite{9964251}, where deep reinforcement learning (DRL) was employed for the joint STARS and BS beamforming design. In \cite{10306287}, the DRL method was invoked for the joint spectrum allocation, STARS beamforming, and BS beamforming optimization in a STARS-aided vehicle-to-everything (V2X) communication system. It showed that DRL can enhance the data rate achieved by each vehicle user while satisfying the latency and reliability requirements between vehicle users. Furthermore, the capacity-versus-coverage performance of STARS-aided wireless communications was studied in \cite{10187159}, where a multi-objective proximal policy optimization (MO-PPO) algorithm was developed. The STARS beamforming design was also studied in multiple-STARS-aided multi-cell wireless networks in \cite{10049110}. To address the high computational complexity and configuration overhead caused by the massive STARS elements, the authors of \cite{10049110} proposed a tile-based STARS beamforming architecture, where the entire STARS is partitioned into several tiles of different sizes sharing the same coefficients. Based on this scheme, a novel distributed learning approach with DRL and access-free federated learning (AFFL) was developed to optimize the tile partition and beamforming of the STARS deployed in each cell.

\subsection{Discussions and Outlook}
\begin{figure*}[t!]
\begin{center}
    \includegraphics[width=7in]{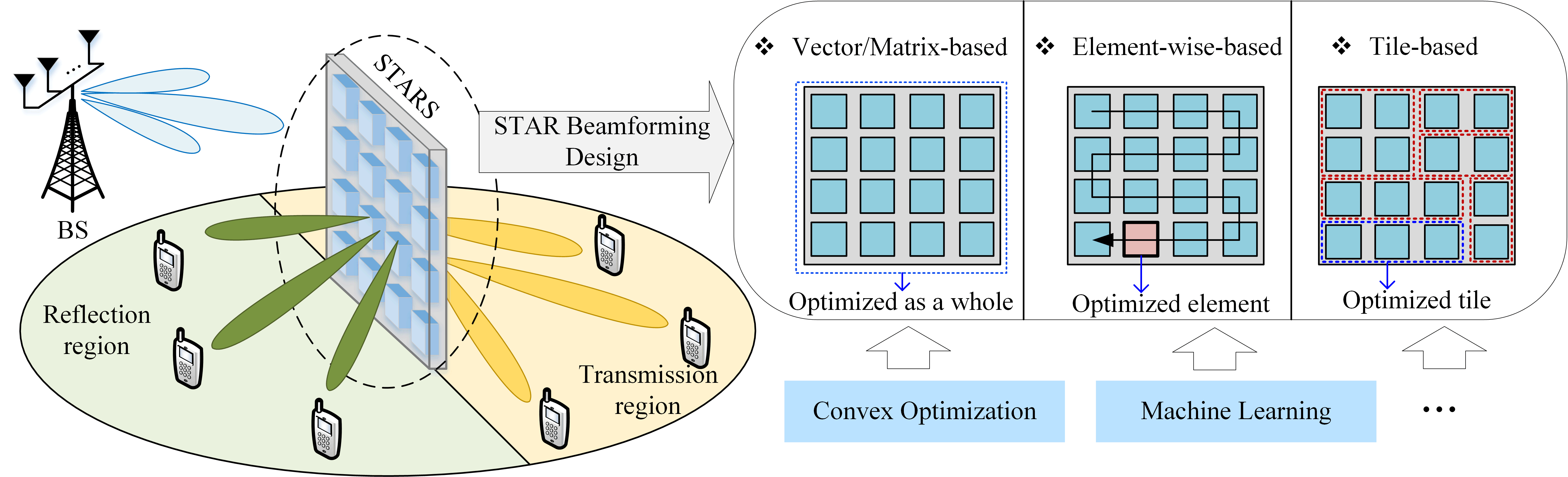}
    \caption{Illustration of three STARS beamforming optimization schemes in terms of the dimensions of optimized STARS elements.}
    \label{STARS schemes}
\end{center}
\end{figure*}
\begin{table*}[!t]\tiny
\caption{Summary of STARS Beamforming Optimization Schemes.}
\begin{center}
\centering
\resizebox{\textwidth}{!}{
\begin{tabular}{!{\vrule width0.8pt}l!{\vrule width0.8pt}l!{\vrule width0.8pt}l!{\vrule width0.8pt}l!{\vrule width0.8pt}}
\Xhline{0.8pt}
\centering
\makecell[c]{\textbf{Schemes}}  & \makecell[c]{\textbf{Advantages}} & \makecell[c]{\textbf{Disadvantages}} &\makecell[c]{\textbf{Ref.}}  \\
\Xhline{0.8pt}
\centering
\makecell[l]{Vector/Matrix-based Optimization} & {Easy to use with provable optimality} & {Potentially high computational complexity} & \makecell[c]{\cite{9570143,9920228,10130543,9785636}, etc}  \\
\cline{1-4}
\centering
\makecell[l]{Element-wise-based Optimization}& {Low computational complexity}   & {Unstable performance}    & \makecell[c]{\cite{9838767,9751144,10325546}, etc} \\
\cline{1-4}
\centering
\makecell[l]{Tile-based Optimization}& {Low CSI estimation and configuration overhead}   & {Additional tile partition design and reduced DoFs}    & \makecell[c]{\cite{10049110}} \\
\Xhline{0.8pt}
\end{tabular}
}
\end{center}
\label{STARS optimization schemes}
\end{table*}

It can be observed that the STARS beamforming design critically affects the performance of STARS-aided wireless communications. As shown in Fig. \ref{STARS schemes}, from the perspective of optimization dimensions of STARS beamforming, existing works can be classified into three main schemes, namely vector/matrix-based optimization, element-wise-based optimization, and tile-based optimization, which are discussed as follows. 
\begin{itemize}	
	\item \textbf{Vector/Matrix-based Optimization:} The most common method to directly iteratively optimize the entire STARS beamforming vector or matrix is with the aid of convex optimization methods (e.g., penalty-based optimization in \cite{9570143} and SDR in \cite{10130543}), and/or ML tools (e.g., RL in \cite{9964251}), as depicted in Fig. \ref{STARS schemes}. The advantages of vector/matrix-based STARS beamforming optimization are that it is easy to employ state-of-the-art optimization methods and normally a locally optimal solution can be obtained through iteration. However, it is worth noting that although the computational complexity of these optimization methods is generally of polynomial complexity, it would be unacceptable for practical implementation given a large number of STARS elements, especially when a high number of iterations is needed for convergence. 		
	\item \textbf{Element-wise-based Optimization:} To address the potentially high computational complexity of STARS beamforming, one efficient solution is to carry out the element-wise-based optimization. As shown in Fig. \ref{STARS schemes}, the key idea is to iteratively optimize the amplitude/phase-shift coefficient of one of the STARS elements with others' coefficients fixed until convergence (e.g., \cite{9838767,9751144,10325546}). By doing so, the overall computational complexity can be greatly reduced and only linearly scales with the number of STARS elements. Nevertheless, the drawback of the element-wise-based optimization is unstable performance since it is difficult to prove the optimality of the obtained solution.
	\item \textbf{Tile-based Optimization:} Another promising solution to reduce the computational complexity is the tile-based optimization employed in \cite{10049110}, where the optimization dimensions are reduced according to the used tile partition scheme. The tile-based scheme is very attractive to be adopted to reduce the overheads caused by CSI estimation and STARS configuration. However, it also introduces additional tile partition design, which is a challenging mixed-integer optimization problem. To address this problem, convex optimization generally becomes inefficient and therefore advanced ML tools have to be employed. Moreover, as several STARS elements are partitioned into one tile sharing the same coefficients, the DoFs for communication designs are reduced, which leads to performance loss.
\end{itemize}
In Table \ref{STARS optimization schemes}, we summarize the representative advantages and disadvantages as well as the corresponding research contributions of the three STARS beamforming optimization schemes.

With the stringent requirements and diverse applications in NG wireless networks, there are still many open problems and future research directions for STARS beamforming. Some of them are highlighted as follows. 

\begin{itemize}	
	\item \textbf{Multiple-Objective STARS Beamforming:} Most of the existing research contributions on STARS beamforming design only considered single metric (e.g., either transmit power or sum rate). However, for future STARS-aided wireless networks, multiple objectives have to be jointly optimized to strike a good trade-off between different and even conflicting metrics, such as capacity, EE, coverage, and latency. How to obtain the Pareto-optimal solution for such a set of multiple-objective STARS beamforming design problems constitutes an interesting and challenging research topic. Considering the high-dimensional search spaces caused by multiple objectives, ML tools can be potential solutions compared to conventional convex optimization-based methods. 
	\item \textbf{Multiple-Functional STARS Beamforming:} Supporting multiple functionalities (e.g., communication, sensing, computation, etc) is the developing trend of NG wireless networks. This imposes additional challenges for STARS beamforming design. For example, a narrow beam with minimum interference leakage is preferred for communication, while a wide beam is efficient for sending echo signals for target sensing. How to achieve optimal spatial multiplexing and interference management among different wireless functionalities, namely multiple-functional STARS beamforming design, is another important research direction. 
	\item \textbf{Generative Artificial Intelligence (GAI) empowered STARS Beamforming:} GAI has received significant attention with many successful applications in computer vision and natural language processing. Recently, the employment of GAI in wireless communication designs has been regarded as a promising solution in 6G~\cite{10172151,GAI}. It is unveiled that GAI can achieve better performance for channel estimation and beam training problems than conventional discriminative AI. This is of vital importance for STARS beamforming due to the nearly passive nature of STARS. Exploiting GAI for STARS beamforming is a promising solution but requires further research.
\end{itemize}

\section{Advanced Applications of STARS for NGMA}
The fundamental performance analysis and beamforming design of STARS reviewed above underscores its significant potential for NG wireless networks. In this section, we focus our attention on advanced applications of STARS that contribute to the efficient realization of NGMA. These applications address requirements in aerial-ground coverage, security, and sustainability, utilizing associated techniques, namely UAV communications, PLS, and SWIPT.
\begin{figure*}[t!]
\centering
\subfigure[UAV-mounted STARS-aided communications.]{\label{UAV_mounted_STARS} 
\includegraphics[width= 2.8in]{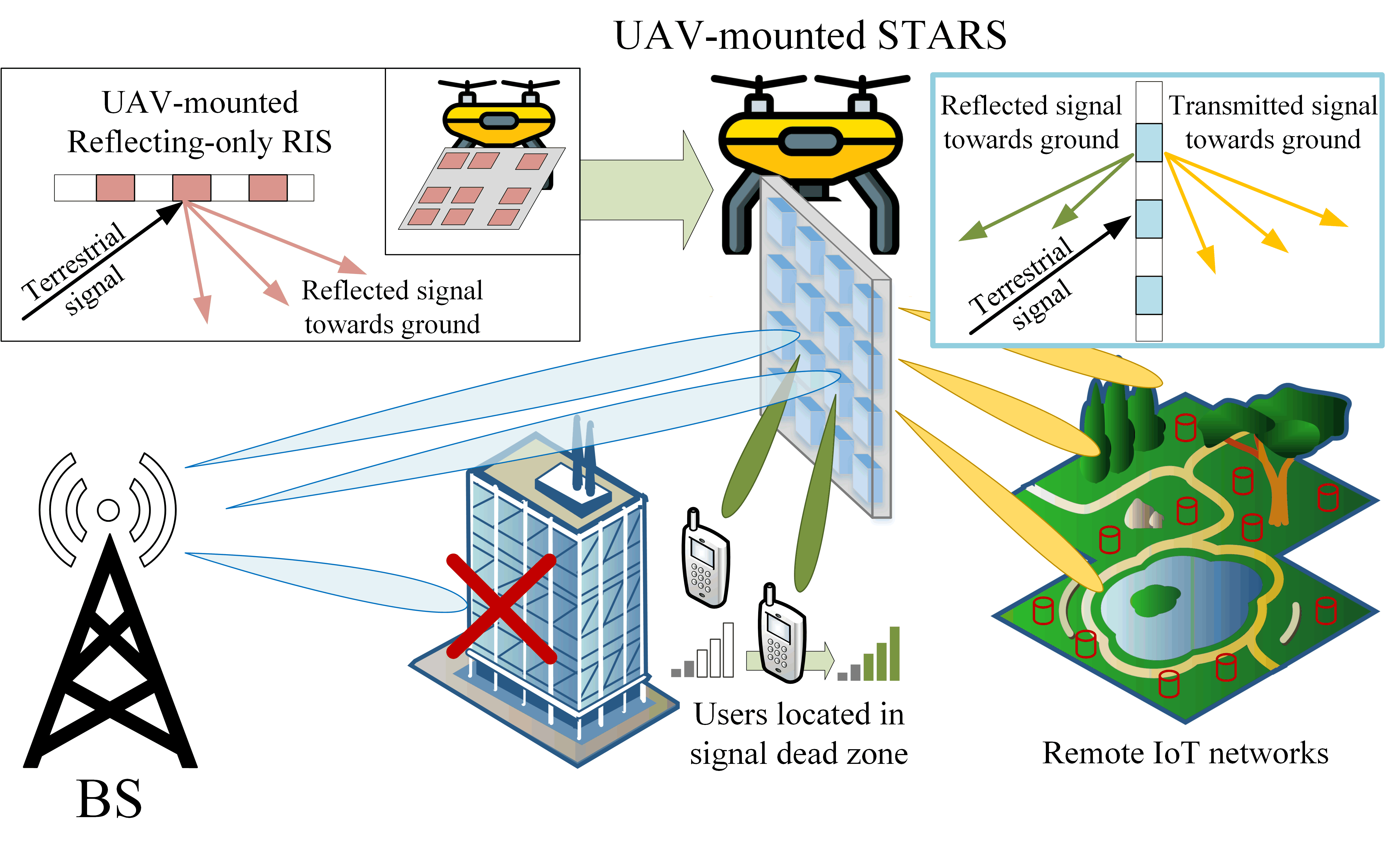}}
\subfigure[STARS-aided UAV communications.]{\label{STARS_aided_UAV}
\includegraphics[width= 3.9in]{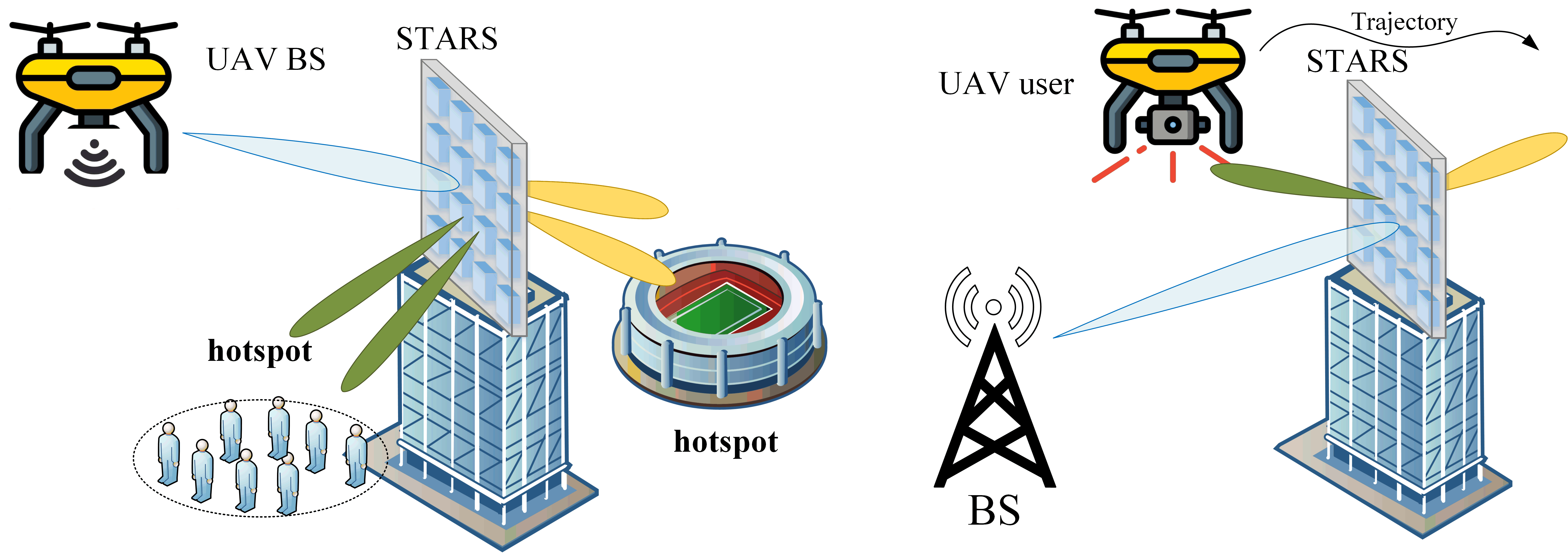}}
\caption{Two STARS applications in UAV communications.}\label{STARS_UAV}
\end{figure*}
\subsection{STARS in UAV Communications}
Given the attractive features, such as high mobility and high manoeuvrability, UAVs bring great convenience and support to our daily lives with numerous applications, such as cargo delivery, remote monitoring, and search and rescue. Among these applications, UAV communications have become one of the most critical technologies, where a UAV carrying on wireless transceivers can either provide wireless services to ground users or act as cellular users to accomplish specific tasks via the line-of-sight (LoS) dominated air-to-ground link~\cite{8918497}. Compared to conventional costly active communication devices, the cost-efficient STARS constitutes a promising candidate technology in UAV communications. According to whether the STARS is mounted on the UAV or not, existing research contributions can be classified into two categories, namely UAV-mounted STARS-aided communications and STARS-aided UAV communications, which are illustrated in Fig. \ref{STARS_UAV}. On the one hand, UAV-mounted STARS can leverage the mobility of UAVs to enhance the deployment flexibility of STARS when assisting wireless communications. On the other hand, STARS-aided UAV communications can utilize STARS to address the blockage issue and enhance the wireless coverage of UAV communications. A summary of the existing work on applying STARS in UAV communications is provided in Table \ref{STAR_UAV_TABLE}.

\begin{table*}[!t]\large
\caption{Summary of existing works on the applications of STARS in UAV communications.}
\begin{center}
\centering
\resizebox{\textwidth}{!}{
\begin{tabular}{!{\vrule width1.5pt}l!{\vrule width1.5pt}l!{\vrule width1.5pt}l!{\vrule width1.5pt}l!{\vrule width1.5pt}}
\Xhline{1.5pt}
\centering
\makecell[c]{\textbf{Category}}  & \makecell[c]{\textbf{Ref.}} &\makecell[c]{\textbf{UAV Mobility}} & \makecell[c]{\textbf{Characteristics/Techniques}} \\
\Xhline{1.5pt}
\centering
\multirow{6}{*}{\makecell[c]{UAV-mounted STARS-aided Communications}} & \makecell[c]{\cite{10354335}} & {Static} & {A novel foldable STARA architecture for combating the air resistance}  \\
\cline{2-4}
\centering
& \makecell[c]{\cite{10376206}}  &{Static/Mobile}  &{Convex optimization and DDQN algorithms are proposed for SEE maximization in two scenairos} \\
\cline{2-4}
\centering
& \makecell[c]{\cite{STAR_UAV_3D}}   & {Static}    & {Quantum sensing-based beam training for hybrid active and passive STARS} \\
\cline{2-4}
\centering
& \makecell[c]{\cite{10453937}}   & {Static}    & {UAV-mouted STARS-aided multicast satellite communications} \\
\cline{2-4}
\centering
& \makecell[c]{\cite{STAR_UAV_MEC1}}   & {Mobile}    & {DRL-based joint UAV trajectory, transmit power, and STARS beamforming design for MEC} \\
\cline{2-4}
\centering
& \makecell[c]{\cite{STAR_UAV_MEC2}}   & {Mobile}    & {A novel bi-directional BS and UAV task offloading MEC framework via STARS} \\
\Xhline{1.5pt}
\centering
\multirow{5}{*}{\makecell[c]{STARS-aided UAV Communications}}& \makecell[c]{\cite{9849460}}   & {Mobile}    & {DRRL-based joint trajectory and STARS beamforming design for uncertain environment} \\
\cline{2-4}
\centering
& \makecell[c]{\cite{9878137}}   & {Mobile}    & {Convex optimization-based sum rate maximization with three STARS operating protocols} \\
\cline{2-4}
\centering
& \makecell[c]{\cite{10286085}}   & {Mobile}    & {A STARS-aided UAV wireless powered communication framework for self-sustainable IoT networks} \\
\cline{2-4}
\centering
& \makecell[c]{\cite{10083240}}   & {Mobile}    & {A STARS-aided UAV outdoor-indoor NOMA communication framework} \\
\cline{2-4}
\centering
& \makecell[c]{\cite{10320337}}   & {Mobile}    & {A multiple-STARS-aided UAV emergency communication network with NOMA} \\
\Xhline{1.5pt}
\end{tabular}
}
\end{center}
\label{STAR_UAV_TABLE}
\end{table*}

\subsubsection{UAV-mounted STARS-aided Communications} As depicted in Fig. \ref{STARS_aided_UAV}, UAV-mounted STARS can redirect the wireless signals from terrestrial BSs via the full-space transmission and reflection for enhancing the signal strength (e.g., for users who are blocked) and extending the wireless coverage (e.g., remotely located Internet of Things (IoT) networks). It is worth mentioning that conventional reflecting-only RISs can be affixed on the UAV in parallel to the ground, reflecting all incident signals towards the ground users underneath and achieving a 360$^\circ$-like coverage, as shown in the left top of Fig. \ref{STARS_aided_UAV}. However, the maximum DoFs for communication design are limited compared to UAV-mounted STARS having the transmission and reflection beamforming. The aforementioned benefits have drawn significant research interests to study the UAV-mounted STARS-aided communications. Considering STARS in general has to be vertically attached to the UAV (see Fig. \ref{STARS_aided_UAV}), one practical challenge is how to tackle the air resistance caused by the STARS panel. To address this issue, the authors of \cite{10354335} developed a novel foldable UAV-mounted STARS architecture, where the position of STARS can be changed either parallel to or vertical to the UAV depending on the corresponding circumstances. Based on this architecture, a UAV-mounted STARS-aided NOMA communication system was studied in \cite{10354335}, where the system sum rate was maximized by jointly optimizing the user power allocation, NOMA clustering, STARS beamforming, and UAV hovering location. The authors of \cite{10376206} continued to explore the employment of UAV-mounted STARS in secure communications, where the secrecy energy efficiency (SEE) was maximized for both static and mobile UAV scenarios with the aid of convex optimization and double deep Q-network (DDQN) algorithms. To address the channel estimation issue, the authors of \cite{STAR_UAV_3D} employed quantum sensing for achieving accurate beam training in UAV-mounted STARS-aided Terahertz (THz) MIMO communications, where the STARS uses a hybrid active and passive structure for enhancing the STARS beamforming gain. Moreover, the authors of \cite{10453937} studied the UAV-mounted STARS-aided multicast satellite communications, where the system average sum rate was maximized via jointly optimizing the user scheduling and STARS beamforming. As a further advance, the novel idea of UAV-mounted STARS is exploited in MEC networks in \cite{STAR_UAV_MEC1,STAR_UAV_MEC2}. The authors of \cite{STAR_UAV_MEC1} investigated a UAV-mounted STARS-aided uplink MEC network, where the aerial STARS forwards the computation tasks from ground IoT devices to the remote MEC server. A DRL approach was proposed for jointly optimizing the task offloading schemes, transmit power, STARS beamforming, and aerial STARS trajectory to minimize the total energy consumption of IoT devices and the UAV. The authors of \cite{STAR_UAV_MEC2} further developed a novel UAV-mounted STARS-aided bi-directional MEC architecture, where the UAV acts as one mobile MEC server and the STARS is still attached in parallel with the UAV. By doing so, the computation task of the ground users can be offloaded not only to the BS via reflection but also to the UAV via transmission, which is impossible to realize for reflecting-only RISs. The results showed that the system EE can be greatly improved by the proposed scheme. 
\subsubsection{STARS-aided UAV Communications} As depicted in Fig. \ref{STARS_aided_UAV}, STARS can also be deployed in the environment to assist the communication between the UAV-BS and ground users (e.g., providing service for temporary hot spots) as well as ground BSs and UAV users (e.g., controlling UAV for accomplishing tasks). For conventional reflecting-only RIS-aided UAV communications, the UAV has to be located on the same side of the RIS as the ground node, which leads to an unwanted no-fly zone and reduces the flexibility of the UAV. By contrast, the full-space coverage facilitated by STARS will not impose additional constraints for UAV trajectory planning and deployment location design, as shown in Fig. \ref{STARS_aided_UAV}. To reap the benefits of STARS-aided UAV communication, the authors of \cite{9849460} first studied the joint UAV-BS trajectory and beamforming design with the aim of maximizing the long-term sum rate, where a distributionally-robust RL (DRRL) algorithm was proposed against uncertain flying environment. A similar problem was further studied by the authors of \cite{9878137} employing convex optimization methods and considering three STARS operating protocols. The results obtained in \cite{9849460,9878137} both revealed that STARS can further enhance the UAV communication performance over conventional RISs. Furthermore, the authors of \cite{10286085} proposed a STARS-aided UAV wireless-powered communication framework, where the indoor and outdoor IoT devices are first charged by the UAV wireless signals and then they upload their information with the aid of STARS. In particular, ES and TS protocols are employed during the wireless charging and information uploading stages, respectively. They showed that the proposed framework can greatly improve the sum rate of energy-limited IoT networks. As a further advance, STARS-aided UAV NOMA communications are further studied in \cite{10083240,10320337}. By installing the STARS on the building facade, a STARS-aided UAV indoor-outdoor communication system was studied in \cite{10083240}, where the UAV sends information to an indoor user and an outdoor user via STARS using the NOMA technique. A multiple-STARS-aided UAV emergency communication network was investigated by the authors of \cite{10320337}, where the long-term uplink throughput was maximized with the developed Lagrange-based reward-constrained proximal policy optimization (LRCPPO) algorithm. 

\subsection{STARS-aided PLS}
\begin{figure}[t!]
\begin{center}
    \includegraphics[width=3in]{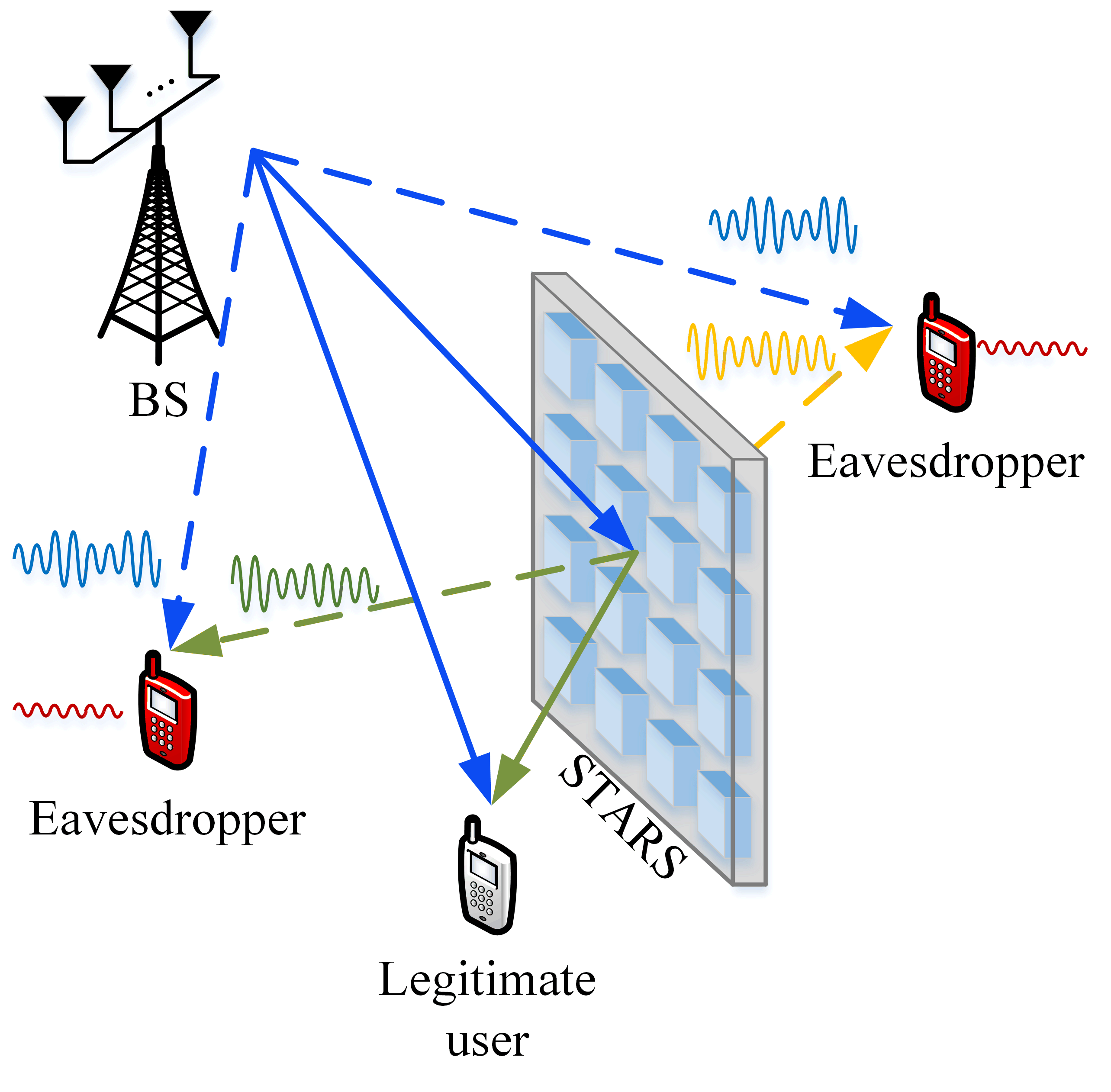}
    \caption{Illustration of STARS-aided full-space PLS, where the received signal strength of eavesdroppers distributed on both sides can be mitigated via transmission and reflection.}
    \label{STAR_PLS}
\end{center}
\end{figure}
Due to the broadcast nature of wireless channels, it is critical to prevent private data from being eavesdropped through wireless communications. To this end, PLS has been proposed as an efficient technology for achieving secrecy transmission~\cite{8918497}. Compared to conventional encryption methods and secret key management, the key idea of PLS is to exploit the characteristics of wireless channels (e.g., fading, interference, and noise) to combat malicious information eavesdropping. Therefore, with the capability of adjusting the signal propagation, RISs are beneficial for improving the performance of PLS. However, given the uncertain locations of malicious eavesdroppers, conventional reflecting-only RISs can only provide a half-space safeguarding wireless environment. As a remedy, a full-space safeguarding wireless environment can be facilitated by STARS and secrecy transmission can be achieved regardless of eavesdroppers' locations, as shown in Fig. \ref{STAR_PLS}. Note that the extended full-space PLS also introduces complicated secrecy communication design problems, which has led to many research contributions on STARS-aided PLS in recent years. For example, the authors of \cite{9525400} first studied the joint beamforming design for a STARS-aided secure MISO communication network, where the weighted sum secrecy rate was maximized under the three STARS operating protocols of STARS.  Considering the imperfect CSI of both legitimate and eavesdropping channels, the authors of \cite{9774882} further investigated robust joint beamforming design to maximize the SEE of a STARS-aided NOMA network. They showed that TS and ES/MS protocols are preferable for limited and high power regimes, respectively. A similar STARS-aided secure NOMA communication design was also studied in \cite{10005206}, where the worst secrecy rate among NOMA users was maximized given the imperfect CSI of eavesdroppers. The STARS-aided PLS was studied by the authors of \cite{10040906} with the coupled phase-shift model, where the fair secrecy rate among users was maximized with the developed penalty-based optimization algorithm. Moreover, the study of STARS-aided PLS was extended to many advanced scenarios. The authors of \cite{10304291} studied the employment of STARS for simultaneously anti-jamming and anti-eavesdropping, where ES was shown to be the optimal protocol to achieve the minimum power consumption for satisfying the secrecy communication requirement. Instead of using STARS for combating eavesdroppers, the authors of \cite{10002889} exploited the STARS for proactive eavesdropping of malicious transmission. In particular, STARS is deployed surrounding a legitimate monitor to not only assist in the jamming of malicious transmission but also eavesdrop the malicious information. The authors of \cite{10315044} further studied the STARS-aided PLS in a multi-cell NOMA network using the tool of stochastic geometry. In \cite{10315044}, the secrecy outage probability and average secrecy capacity were derived, which showed that the ES protocol can achieve a better secrecy performance than the TS protocol.

\subsection{STARS-aided SWIPT}
\begin{figure}[t!]
\begin{center}
    \includegraphics[width=3in]{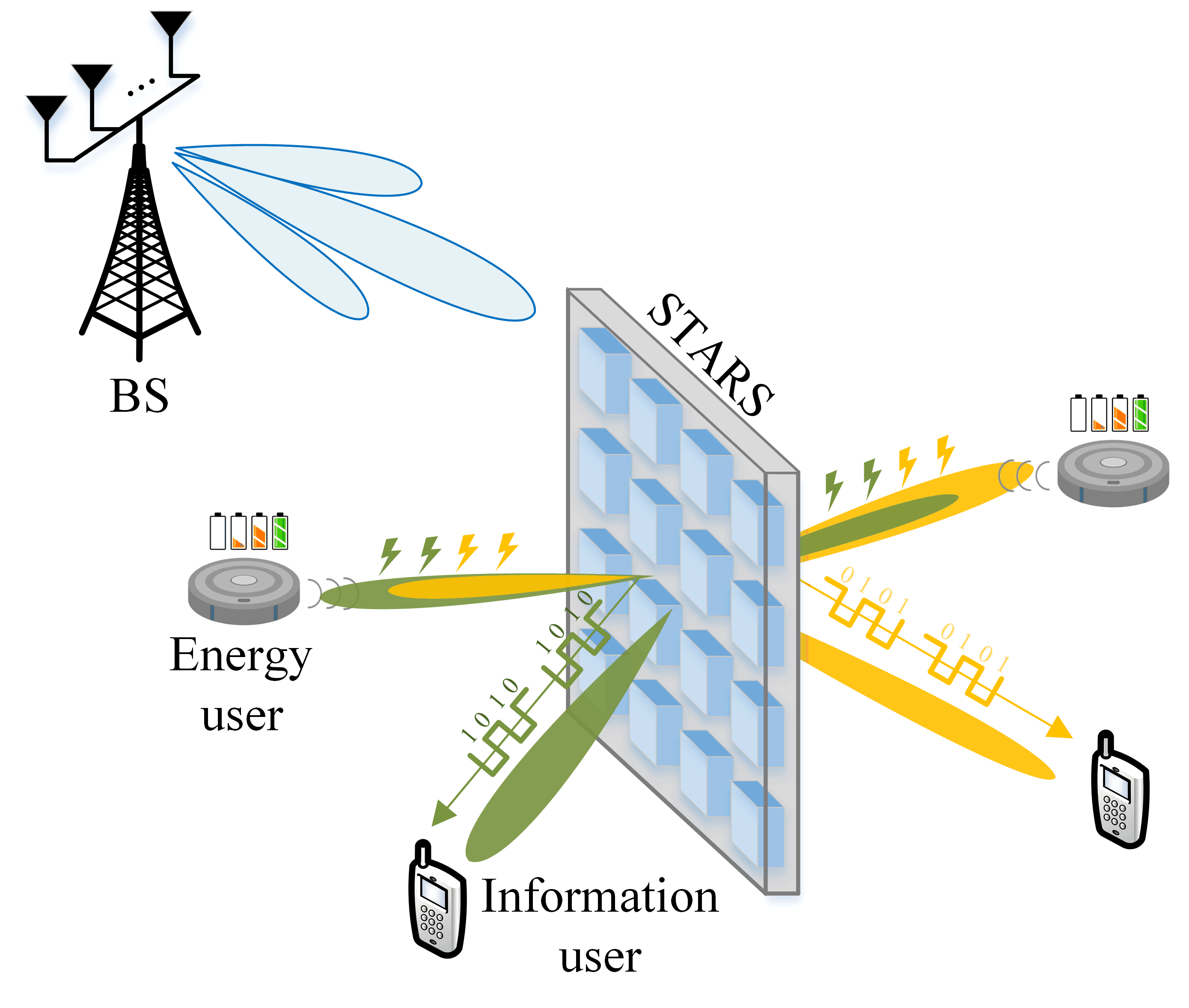}
    \caption{Illustration of STARS-aided SWIPT.}
    \label{STAR_SWIPT}
\end{center}
\end{figure}

\begin{figure}[t!]
\begin{center}
    \includegraphics[width=3.5in]{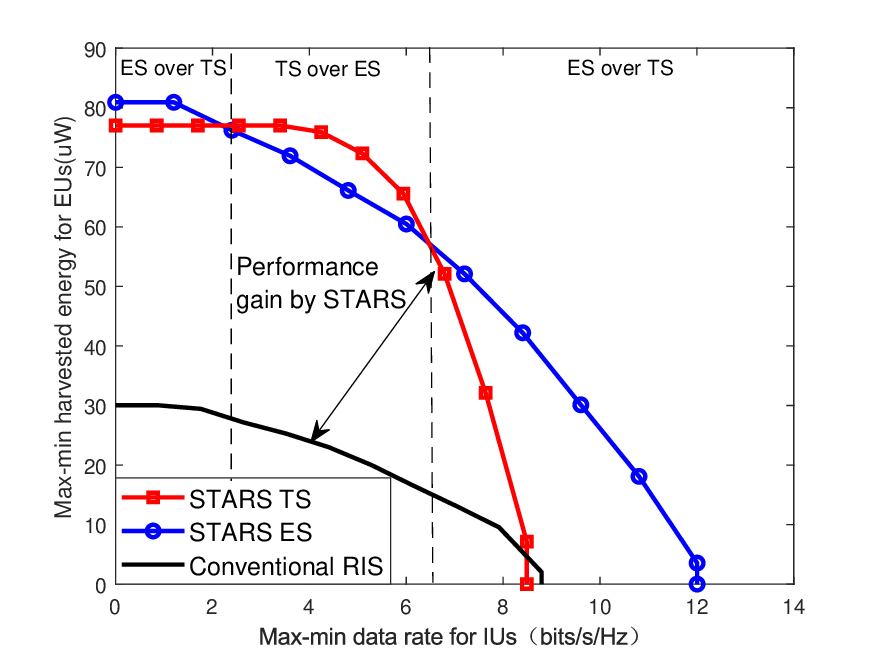}
    \caption{Rate-energy region achieved by ES-STARS, TS-STARS, and conventional RIS-aided SWIPT, where the direct links between the BS and the 2 energy/information users are assumed to be blocked. The number of RIS elements and BS antennas are 32 and 4. Here, the conventional RIS consists of 16 transmitting-only elements and 16 reflecting-only elements. The other parameters adopted can be found in~\cite[Section IV]{10304608}.}
    \label{SWIPT}
\end{center}
\end{figure}
Sustainability plays a crucial role in the advancement of IoT and wireless sensor networks, laying the groundwork for numerous promising applications in the future, including smart homes, cities, factories, and environmental monitoring. SWIPT is widely recognized as a key enabler in extending the lifespan and mitigating the energy challenges faced by IoT and wireless sensor networks, especially in remote and inaccessible areas~\cite{8214104}. Note that employing the STARS in SWIPT can compensate for the disadvantage of energy leaking in conventional STARS-aided communication systems. For demonstration, let us take a STARS-aided unicast communication system as an example. Due to the STARS feature, the energy of signals intended for users on one side of the STARS will generally leak to the other side as harmful interference. This not only calls for interference management but also reduces EE. Fortunately, in STARS-aided SWIPT, the aforementioned leaked energy can be redesigned to improve the energy harvesting performance of energy-harvesting users, as shown in Fig. \ref{STAR_SWIPT}. This, however, leads to a complicated STARS beamforming design for striking a good trade-off between information and energy-harvesting performance. To address this issue, the authors of \cite{10304608} studied robust resource allocation for STARS-aided SWIPT under the imperfect CSI condition. Dual-objective resource allocation optimization problems for ES-STARS and TS-STARS were investigated in \cite{10304608} to simultaneously maximize the energy-harvesting fairness and communication rate fairness, which characterized the fundamental rate-energy trade-off. It unveiled that ES-STARS is the optimal protocol for information-priority and energy-harvesting-priority cases, while TS-STARS is preferable to be employed for achieving a better rate-energy trade-off. In Fig. \ref{SWIPT}, we depict the rate-energy trade-off achieved by ES-STARS, TS-STARS, and the conventional RIS in \cite{10304608}. It can be observed that STARS can significantly improve the SWIPT performance compared to conventional RISs and ES/TS protocols are suitable to be employed for different cases. The above results confirm the benefits of STARS-aided SWIPT. Moreover, the authors of \cite{10283600} studied the joint beamforming design for STARS-aided SWIPT for minimizing the transmit power subject to the minimum information and energy-harvesting requirements. Focusing on a basic STARS-aided SWIPT system with a single information/energy-harvesting user, the authors of \cite{10278936} conducted the performance analysis of the rate outage probability and energy outage probability at the high SNR regime, which also demonstrated the significant performance gain of STARS for SWIPT. Furthermore, STARS were deployed in wireless-powered communication systems in \cite{10086660,10333826}. In \cite{10086660}, a TS-STARS was deployed in a wireless-powered IoT network to successively assist the wireless power charging phase from a power station to IoT devices and the wireless information uploading phase from IoT devices to an AP. The authors of \cite{10333826} further exploit active STARS in a wireless-powered communication system, which consists of one downlink SWIPT stage and one uplink wireless information uploading stage. The results showed that active STARS can greatly improve the uplink sum rate compared to passive STARS.

\subsection{Other Promising Applications}
Besides the aforementioned STARS applications for NGMA, there are two other promising scenarios for employing STARS, which are discussed as follows.
\subsubsection{STARS-aided THz Communications} Along with the evolution of each generation of wireless systems, spectra have been regarded as the most valuable resource. Nowadays, with the emergence of many bandwidth-demanding applications, the THz band, generally ranging between 0.1-10 THz, is regarded as a promising candidate for NG wireless networks. Although extremely wide THz bandwidths are available, one main challenge for practical THz communications is how to address the blockage problem. Compared to conventional reflecting-only RIS, the full-space coverage makes STARS more efficient for assisting THz communications. For example, STARS installed in the wall/window can redirect the THz signal from BSs/APs to surrounding users via only one hop. However, STARS-aided wideband THz communications also introduce the beam split issue due to the lack of the frequency-selective capability of STARS elements, which attracted some initial research efforts in \cite{10133914,STAR_THZ}. The authors of \cite{10133914} studied the STARS-aided THz communication design in both narrowband and wideband systems. To evaluate the EE performance, two power consumption models were proposed for STARS elements under the independent and coupled phase-shift models. Two joint BS and STARS beamforming design algorithms were developed for narrowband and wideband THz communication systems, where true time delayers (TTDs) are used at the BS to combat the beam split issue in the wideband system. Simulation results showed that there is only a slight performance gap between the coupled and independent phase-shift models in both STARS-aided narrowband and wideband THz systems. To further combat the beam split issue associated with STARS, the authors of \cite{STAR_THZ} proposed a TTD-based STARS structure, where each or several STARS elements are connected with one TTD. Based on this, the TTD-based hybrid beamforming at the BS and the TTD-based STARS beamforming were jointly optimized for maximizing the sum rate. This study showed that the proposed sub-connected TTD-based STARS has a better communication performance and hardware cost trade-off than fully-connected TTD-based STARS.
\subsubsection{STARS-aided Coordinated Multi-Point (CoMP)} CoMP technique is proposed for inter-cell interference mitigation and thus enhancing the communication performance of cell-edge users~\cite{6146494}. STARS-aided CoMP provides new design opportunities. With the STARS capability, STARS can be deployed at each cell and allow the cell-edge and cell-center users to be located on the two sides. For example, the transmission function of multiple STARS can be exploited to serve cell-edge users while the remaining reflection function of STARS can be used to serve cell-center users located in each cell. As an initial step, the performance of a STARS-aided NOMA-CoMP network was studied in \cite{9622133}, where a novel simultaneous-signal-enhancement-and-cancellation-based (SSECB) design was proposed for STARS beamforming to eliminate the multi-user interference and enhance the desired signal strength. Simulation results presented in \cite{9622133} showed that the employment of STARS can achieve significant performance improvement for CoMP.

\section{Interplay Between STARS and Emerging Technologies towards NGMA}
To meet the evolving demands of NG wireless networks, several key paradigm shifts are required for NGMA. First, beyond the core communication function, NG wireless networks are expected to incorporate additional capabilities such as sensing and computing \cite{9687468}. Consequently, NGMA has to coordinate diverse user types with different performance metrics and QoS requirements. Second, the massive connectivity requirements of future machine-type communications demand that NGMA needs to have strong resource and interference management capabilities. NOMA and its derivatives are recognized as promising approaches to fulfil this need \cite{9693417}. Finally, the emerging trend of utilizing supermassive MIMO and ultra-high frequency bands (e.g., THz bands) makes it imperative for NGMA designs to take into account near-field effects \cite{10220205}. Against the above background, in this section, we explore the interplay between STARS and emerging technologies in these NGMA paradigm shifts, including ISAC, MEC, NOMA, and NFC, and discuss what fundamental changes STARS can bring to these technologies towards NGMA.

\begin{figure*}[!t]
    \centering
    \subfigure[Integrated full-space configuration.]{
        \includegraphics[width=0.4\textwidth]{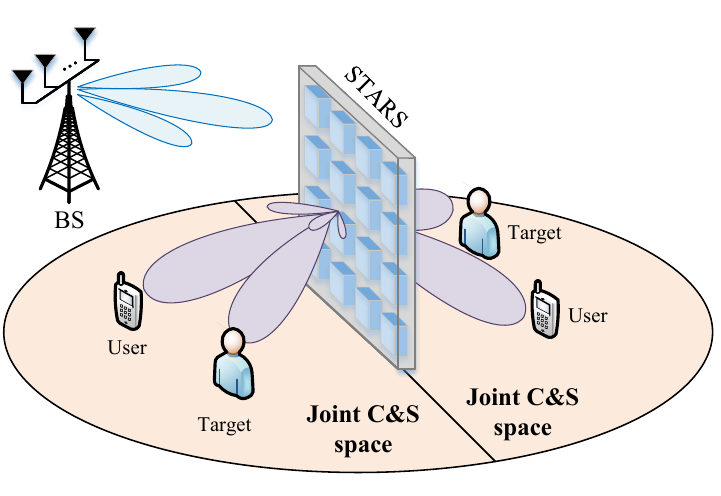}
    }
    \subfigure[Separated half-space configuration.]{
        \includegraphics[width=0.4\textwidth]{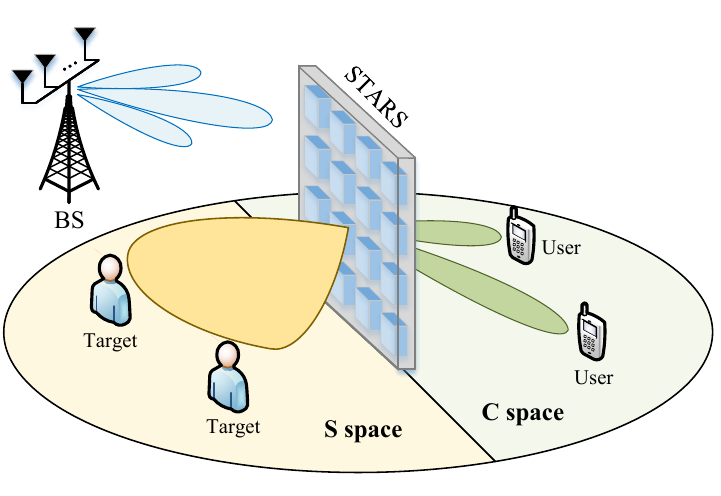}
    }
    \caption{Illustration of two systems configurations for STARS-ISAC systems.}
    \label{fig:STAR_ISAC}
\end{figure*}

\subsection{STARS-ISAC}
NG wireless networks are envisioned to support not only communication but also sensing functions, which has made the concept of ISAC a hot research topic recently \cite{9737357}. The key objective of ISAC is to carry out communication and sensing (C\&S) functions using a shared hardware platform and signals. However, these two functions have different requirements for system design. For example, the communication function can effectively exploit the non-LoS path for data delivery, while the sensing function generally relies on the LoS path for target detection and parameter estimation. In practice, the LoS path does not always exist due to the blockage, which makes target sensing quite challenging. RISs are a promising solution to address this issue by establishing a virtual LoS path from the BS to the target \cite{9732186}. However, employing conventional reflecting/transmitting-only RISs has the following two disadvantages. On the one hand, as discussed in Section \ref{sec:intro}, conventional RISs can only achieve half-space coverage for both C\&S. On the other hand, in RIS-ISAC systems, the communication users and sensing targets have to be located on the same side of the RIS and share the same passive beamforming matrix, which restricts design flexibility. Compared to conventional RISs, STARSs provide a paradigm shift for ISAC design. In the following, we first introduce some basic system configurations of STARS-ISAC, which is followed by an introduction of a promising \emph{sensing-at-STARS} architecture. 

\subsubsection{System Configurations}

As shown in Fig. \ref{fig:STAR_ISAC}, based on the full-space coverage and signal splitting nature, there are two possible configurations of STARS-ISAC, namely integrated full-space (IFS) configuration \cite{10155669, wang2023dual} and separated half-space (SHS) configuration \cite{10050406, 10178069, 10311519, 10188900, 10238433, 10226306}. In the following, we elaborate on these configurations.

\begin{itemize}
    \item \textbf{Integrated Full-Space Configuration:} As illustrated in Fig. \ref{fig:STAR_ISAC}(a), in the IFS configuration, communication users and sensing targets exist on both sides of STARS. This configuration enables $360^\circ$ coverage and full DoFs for both C\&S functions. Drawing on this configuration, \cite{10155669} crafted a STARS beamforming design based on the TS protocol, which minimized the Cramér-Rao bound (CRB) for target sensing while ensuring the QoS requirements of uplink communications. In particular, CRB provides a closed-form theoretical tight lower bound on the parameter estimation error variance achieved by unbiased estimators. In another study, \cite{wang2023dual} proposed an ES-based STARS beamforming design, aiming to maximize the signal-to-clutter-plus-noise ratio (SCNR) for sensing under the constraint of minimum QoS requirements for downlink communications. While the IFS configuration can achieve full DoFs for ISAC, it is important to note that the joint C\&S beamforming design for STARS can lead to substantial computational complexity that may hinder practical implementation, especially considering the inherently passive nature of STARS.   

    \item \textbf{Separated Half-Space Configuration:} In contrast to the IFS configuration, the SHS configuration separates the entire space into two separated spaces (termed as S and C spaces) for C\&S, respectively, as depicted in Fig. \ref{fig:STAR_ISAC}(b). The SHS configuration is a new possibility introduced by STARS compared to conventional RISs, where two generally independent passive STARS beamforming matrices can be tailored for C\&S, respectively. This is important for a flexible and low-complexity ISAC design since the C\&S functions generally have distinct beamforming requirements, e.g., wide beams for sensing and narrow beams for communication in \ref{fig:STAR_ISAC}(b). The authors of \cite{10050406} first studied the SHS configuration with one sensing target in the S space and multiple communication users in the C space. Two tailored STARS beamforming designs were proposed to minimize the sensing CRB subject to the communication QoS requirements for independent and coupled phase-shift models. Building on this model, \cite{10178069} further studied SNR maximization for target sensing. As a further advance, NOMA was exploited in \cite{10311519} to mitigate both sensing-to-communication interference and inter-user interference in SHS-based STARS-ISAC systems. The PLS issue in SHS-based STARS-ISAC systems was explored in \cite{10188900} and \cite{10238433}, with a focus on independent and coupled phase-shift models, respectively. In a separate study, \cite{10226306} conceived a novel SHS-based sensing-assisted communication framework for vehicular networks. In this framework, STARS are mounted on vehicles to enhance the sensing echoes to the BS through the reflection beamforming and improve the communication performance inside the vehicle through the transmission beamforming.
\end{itemize}

\begin{figure}[!t]
    \centering
    \includegraphics[width=0.4\textwidth]{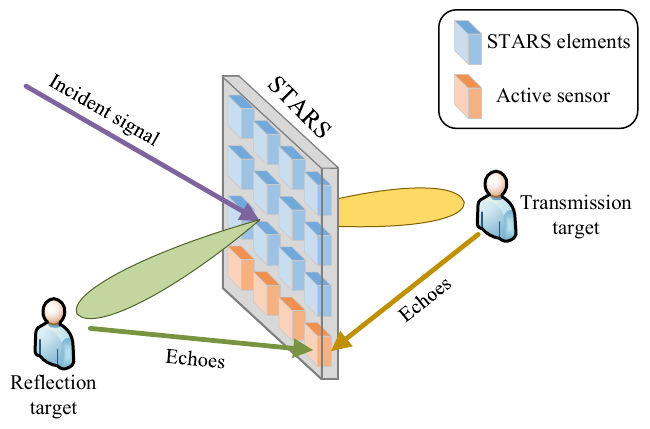}
    \caption{Illustration of the sensing-at-STARS architecture.}
    \label{fig:sensing_at_STARS}
\end{figure}

\subsubsection{Sensing-at-STARS Architecture}

While STARS provide notable advantages for ISAC designs, it also introduces specific challenges, particularly for sensing. Firstly, akin to conventional RIS-ISAC \cite{9724202}, STARS-ISAC exhibits the issue of multi-hop path loss. When carrying out the sensing function at the BS, the sensing echoes experience substantial path loss through a cascaded BS$\rightarrow$STARS$\rightarrow$target$\rightarrow$STARS$\rightarrow$BS path. Secondly, due to the inherent energy-splitting effect, there is a power leakage issue at STARS that further diminishes the echo power. Lastly, the mixing of sensing echoes from both transmission and reflection sides in the STARS$\rightarrow$BS link complicates the accurate estimation of target locations.

\begin{figure}[!t]
    \centering
    \includegraphics[width=0.45\textwidth]{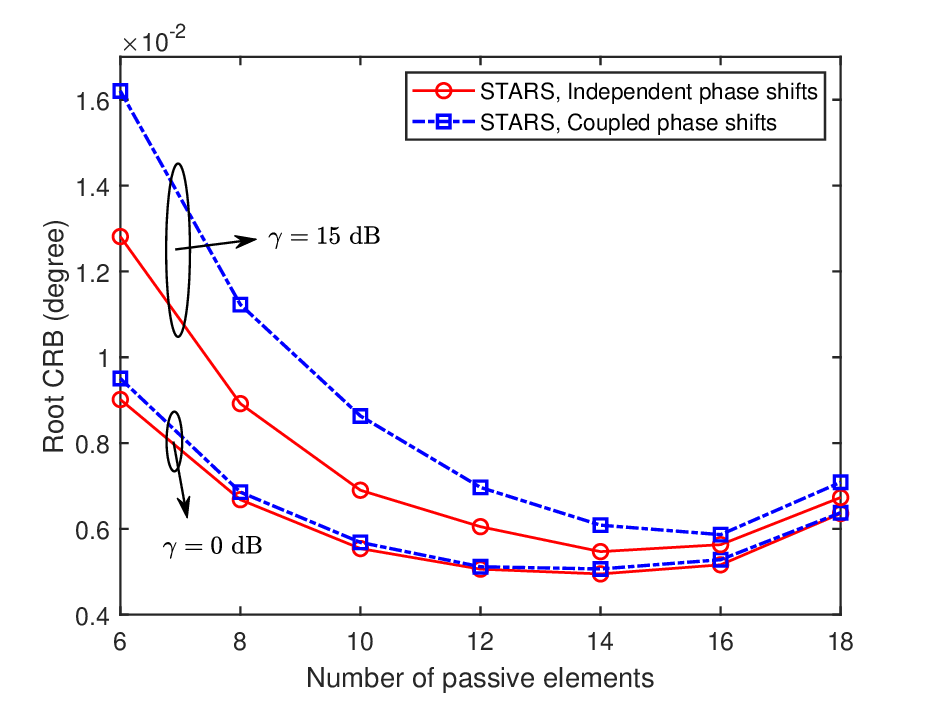}
    \caption{Trade-off between the number of passive elements and active sensors in STARS-ISAC using the separated half-space configuration and the sensing-at-STARS architecture. Here, $\gamma$ denotes the minimum SINR of communication users and the total number of passive elements and active sensors is 20. The other parameters adopted can be found in \cite[Table I]{10050406}.}
    \label{fig:element_tradeoff}
\end{figure}

To overcome these challenges, the authors of \cite{10050406} proposed a \emph{sensing-at-STARS} architecture, as shown in Fig. \ref{fig:sensing_at_STARS}. This architecture, where dedicated active sensors are installed on STARS, enjoys the following benefits:
\begin{itemize}
    \item \textbf{Enhanced Echo Power:} By processing echoes directly at STARS rather than at BSs, this architecture significantly reduces multi-hop path loss and mitigates power leakage issues at STARS, thereby substantially boosting echo power and improving sensing performance.
    \item \textbf{Reduced Echo Aliasing:} The mix of echoes from transmission and reflection sides in the STARS$\rightarrow$BS link is eliminated. Instead, active sensors on STARS directly receive these echoes, thus facilitating easier and more accurate target location estimations.
    \item \textbf{Low-Complexity Beamforming:} The sensing-at-BS architecture requires consideration of both STARS$\rightarrow$target and STARS$\rightarrow$BS links for beamforming design, leading to high complexity. By contrast, the sensing-at-STARS architecture allows beamforming to be focused solely on enhancing the STARS$\rightarrow$target link, thereby simplifying the beamforming design.
\end{itemize} 
For the sensing-at-STARS architecture, there is a fundamental trade-off between the number of passive STARS elements and the active sensors, which is illustrated in Fig. \ref{fig:element_tradeoff}. It can be observed that, in general, more passive elements are preferred than active sensors because increasing passive elements can enhance the full-space DoFs for both C\&S functions, while increasing active sensors only enhances the reception ability for sensing functions.

\begin{figure}[!t]
    \centering
    \includegraphics[width=0.4\textwidth]{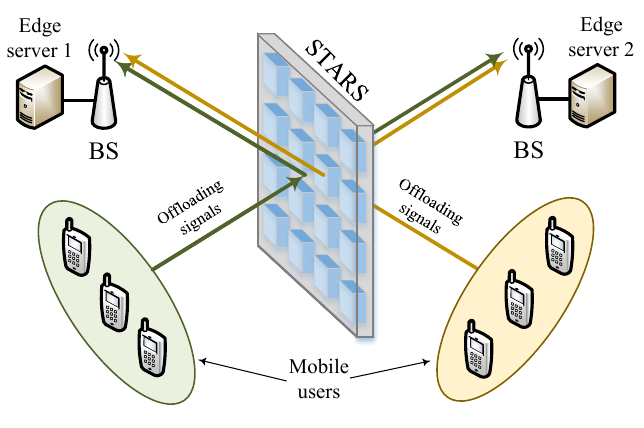}
    \caption{Illustration of full-space MEC enabled by STARS.}
    \label{fig:STAR_MEC}
\end{figure}

\subsection{STARS-MEC}
MEC enables mobile users to offload complicated computational tasks to edge servers, such as machine learning or deep learning tasks, thereby reducing energy consumption and computational latency \cite{8016573, xu2023edge}. RISs have been recognized for their effectiveness in enhancing the offloading link and minimizing offloading delays \cite{bai2020latency, 9279326}. However, conventional RISs, which only offer the reflecting or transmitting capability, are limited to supporting users in a half-space for task offloading to the edge server. Furthermore, in RIS-MEC systems, users are constrained to offload tasks only to the edge servers located on the same side as the RIS. These restrictions inherent in conventional RISs can lead to suboptimal utilization of computational resources and decreased overall computational efficiency. As a remedy, STARS offer the capability for full-space MEC. This advancement enables edge servers to provide computational services to mobile users located on both sides of STARS, as shown in Fig. \ref{fig:STAR_MEC}. Moreover, it allows mobile users to offload tasks to servers situated on either side of STARS, thereby overcoming the spatial constraint inherent in traditional RIS-MEC. From the edge learning perspective, STARS also facilitates decentralized or distributed learning \cite{xu2023edge} by enabling the data exchange and model synchronization among edge servers and users located on different sides of the surface. In \cite{10015822}, the EE gain facilitated by TS-STARS in MEC systems was examined. The study introduced an AO-based approach for STARS beamforming design and time allocation to minimize the total energy consumption of mobile users. The authors of \cite{10121446} explored MS-STARS in MEC systems, where a penalty-based iterative algorithm was proposed for optimizing STARS beamforming to maximize the overall computation rate. To extend the lifespan of mobile devices, \cite{10032506} integrated wireless power transfer into STARS-MEC. This approach initially charges transmission and reflection mobile users via wireless signals from the BS, who then offloaded tasks using the charged energy. In a separate investigation, \cite{10013760} studied a STARS-assisted over-the-air computation system, where the STARS beamforming under the coupled phase-shift model was optimized to minimize the computation mean-squared error. A similar STARS-MEC setup was further studied in \cite{10294010}, where STARS beamforming was optimized to reduce the latency for a group of mobile users offloading tasks to multiple servers located on both sides of STARS.

\begin{figure*}[!t]
    \centering
    \subfigure[Unilateral user clustering.]{
        \includegraphics[width=0.4\textwidth]{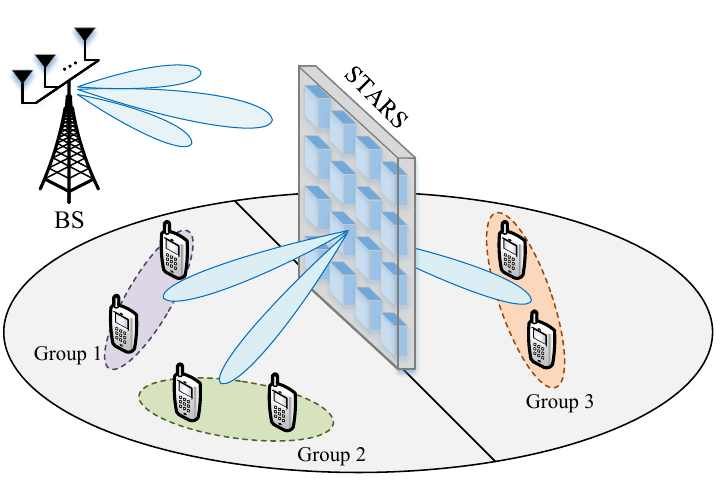}
    }
    \subfigure[Bilateral user clustering.]{
        \includegraphics[width=0.4\textwidth]{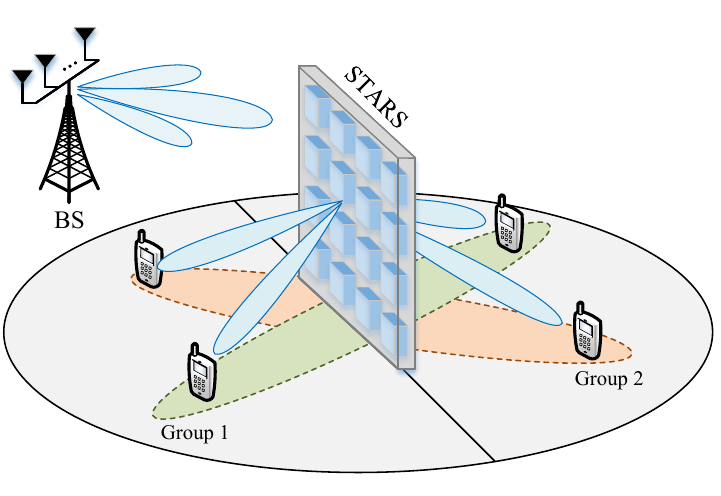}
    }
    \caption{Illustration of two user clustering strategies in downlink STARS-NOMA.}
    \label{fig:STAR_NOMA}
\end{figure*}

\subsection{STARS-NOMA}
NOMA emerges as a significant contender for multiple access in NG wireless networks, enabling multiple users to simultaneously share the same resource block across various domains such as time, frequency, and space \cite{9693417}. NOMA and STARS are complementary to each other. Specifically, NOMA has the potential to significantly boost spectral efficiency and support extensive connectivity in STARS-aided systems. In return, STARS can effectively and intelligently modify the wireless propagation environment, utilizing both transmission and reflection beamforming, to cater to the requirements of NOMA. In the following, we will provide an overview of pertinent research endeavors about the interplay between STARS and NOMA in downlink and uplink systems.

\subsubsection{Downlink Systems} User clustering is critical to fully reap the benefits of NOMA in downlink systems to determine which users share the same resource blocks \cite {9693417}. Traditionally, user clustering is coordinated based on channel characteristics, including channel gain and channel correlation, to ensure that NOMA is effectively implemented within each clustering. Compared to conventional reflecting/transmitting-only RISs, STARS provide additional flexibility for NOMA. On the one hand, in a STARS-NOMA system, user clustering can be carried out within each side of STARS, which is referred to as \emph{unilateral user clustering (UUC)}, as shown in Fig. \ref{fig:STAR_NOMA}(a). In UUC, the channel conditions of users in each cluster can only be adjusted by a single transmission or reflection beamforming matrix. On the other hand, as shown in Fig. \ref{fig:STAR_NOMA}(b), STARS also enable a new user clustering strategy that allows users on different sides to be allocated into one cluster, which is referred to \emph{bilateral user clustering (BUC)}. In BUC, the channel conditions of users in each cluster can be configured by both the transmission and reflection beamforming matrices, thus providing more DoFs for facilitating NOMA. In the following, we review research works on the two user clustering strategies.

$\bullet$ \textbf{Unilateral User Clustering:} In \cite{9863732}, the authors examined a STARS-NOMA system with the fixed UUC. A two-layer iterative algorithm was proposed for optimizing resource allocation, including decoding order, power allocation, BS beamforming, and STARS beamforming, to maximize the sum rate. Further studies in energy-efficient STARS-NOMA designs with UUC were provided in \cite{9956827} and \cite{10052764}. The former introduced a heuristic zero-forcing method for joint optimization of BS and STARS beamforming matrices, while the latter proposed a convex optimization-based approach. To reduce system complexity, the authors of \cite{10138693} explored a hybrid NOMA strategy that dedicates individual time slots to each NOMA cluster, where an iterative algorithm was developed for joint optimization of decoding order, time allocation, and beamforming matrices. Lastly, \cite{10263782} investigated a STARS-assisted cognitive radio NOMA system, where one primary user and two clusters of secondary users are situated on different sides of the STARS. In \cite{10263782}, the outage probability, ergodic rate, and EE for both primary and secondary users were analyzed. 

\begin{table*}[!h]
\caption{Summary of Existing Research Contributions on User Clustering in Downlink STARS-NOMA Systems.}
\label{table_NOMA}
\small
\centering
\resizebox{\textwidth}{!}{
\begin{tabular}{|l|l|l|l|l|}
\hline
\textbf{Schemes}                                                    & \textbf{Characteristics}        & \textbf{Advantages}               & \textbf{Disadvantages}  &\textbf{Ref.}                          \\ \hline
UUC  & \makecell[l]{Independent clustering  on T\&R side}                & \makecell[c]{Low clustering complexity} &  Only T/R BF for each cluster  &  \cite{9863732, 9956827, 10052764, 10138693, 10263782}     \\  \hline
Single-cluster BUC  & \makecell[l]{One cluster for all T\&R users }                         &  No clustering overhead      & High SIC complexity    &  \cite{9956998, 10132045, 10272288, 10284920, 10284719}                         \\ \hline
Multi-cluster BUC  & \makecell[l]{Flexible clustering among T\&R users }                          &  High DoFs for NOMA                                        & High clustering complexity                  &  \cite{9740451, 10223312, 10065555, 10285062}         \\ \hline
\end{tabular}
}
\end{table*}

$\bullet$ \textbf{Bilateral User Clustering:} The simplest two-user case of BUC-based STARS-NOMA systems, involving one transmission user and one reflection user within one NOMA cluster, has garnered considerable research interest to obtain potential design insights \cite{9856598, 9847399, 10280721, 9722712, 10102306, 10272684}. Specifically, the authors of \cite{9856598} divided STARS elements into two groups of equal size for transmission and reflection users and adopted a coherent phase-shift configuration for STARS beamforming. Subsequently, the effect of the proposed approach on the outage probability and ergodic rate was analyzed considering both perfect and imperfect SIC cases. The ES protocol in two-user BUC-based STARS-NOMA systems was further investigated in \cite{9847399}. To mitigate the impact of the energy-splitting ratio on NOMA decoding orders, \cite{9847399} proposed a QoS-based energy-splitting scheme and examined the associated outage probability. Advancing this research, \cite{10280721} introduced a series of optimization-based power allocation and STARS beamforming methods for maximizing the ergodic rate. A comprehensive analysis and comparison of ES, MS, and TS protocols in STARS-NOMA were conducted in \cite{9722712}, focusing on their impacts on the outage probabilities and diversity gains. Considering latency requirements, \cite{10102306} explored the EC and block error rate in two-user BUC-based STARS-NOMA systems. Additionally, \cite{10272684} conceived an FD transmission scheme for BUC-based STARS-NOMA. This approach aimed to enhance communication performance for two cell-edge users by employing a cell-center user as an FD decode-and-forward relay. 

In STARS-NOMA systems with more than two users, the BUC approach becomes much more complex, as the effective channels of NOMA users are influenced by both transmission and reflection beamforming of STARS. A straightforward strategy to circumvent intricate user clustering in such scenarios is to group all transmission and reflection users into a single NOMA cluster. To this end, the authors of \cite{9956998} proposed an index modulation scheme for STARS to support users in such a single NOMA cluster, where the STARS is divided into equally sized sub-surfaces dedicated to specific transmission or reflection users. Expanding on this idea, the authors of \cite{10132045} adapted the index modulation scheme to short-packet transmission scenarios. Diverging from equal partitioning of STARS, the authors of \cite{10272288} introduced a scalable two-stage method, aiming to find a near-optimal STARS element partitioning strategy to maximize achievable rates. To reduce the optimization and channel estimation complexity, the authors of \cite{10284920} proposed a pair of two-timescale transmission schemes for STARS-aided downlink NOMA systems, where the STARS beamforming and power allocation are optimized using the long-term and instantaneous channel information, respectively. Moreover, the authors of \cite{10284719} studied the queue-aware STARS-aided NOMA system considering the stability issues. Apart from the aforementioned BUC approach with one cluster, the multi-cluster-based BUC approach has also been widely investigated. To address the resulting challenging user clustering issue, the authors of \cite{9740451} and \cite{10223312} proposed two matching-theory-based BUC approaches by exploiting the effective channel gain and the possible maximum value of minimum user rate as utility functions, respectively. In \cite{10065555}, the authors introduced a low-complexity optimization-based BUC approach that transforms the complex user clustering problem into a simple linear programming problem, which achieves a near-optimal performance compared to the exhaustive search method. The authors of \cite{10285062} studied a BUC-based STARS-NOMA system with unbalanced user distributions on the transmission and reflection sides. To mitigate the unbalanced effect, a series of phase-shift configuration and energy-splitting strategies were conceived in \cite{10285062}, where the corresponding ergodic rates over Nakagami-$m$ fading channels were analyzed. 

In Table \ref{table_NOMA}, we summarize the representative advantages and disadvantages of the different user clustering approaches in downlink STARS-NOMA systems, where ``T'' and ``R'' represent transmission and reflection, respectively.

\subsubsection{Uplink Systems} Uplink NOMA is generally simpler than downlink NOMA due to the centralization of signal reception. Nonetheless, the use of STARS can still enhance uplink NOMA by dynamically adjusting the decoding order through transmission and reflection beamforming, thus catering to various QoS requirements of NOMA users. For example, to minimize the symbol error rate in uplink NOMA, the authors of \cite{10173573} introduced two effective constellation scaling and rotation schemes for ES-STARS and MS-STARS, respectively. In \cite{9887793} and \cite{10214056}, the authors employed STARS to enhance the communication performance of cell-edge users that share the same resource blocks with cell-center users in uplink NOMA. The impact of FBL on STARS-aided uplink NOMA communications was investigated in \cite{10273229}, where the achievable rate was maximized subject to a maximum decoding error probability. In another study, the authors of \cite{10173617} considered a STARS-aided wireless-powered system, where two users initially harvest energy from the BS, facilitated by ES-STARS or TS-STARS and then utilize the harvested energy to upload information to the BS with NOMA. As a further advance, a STARS-aided backscatter system was studied in \cite{10278099}, where the passive nodes backscatter the modulated signals from the BS to the receiver following a hybrid NOMA-TDMA protocol. The authors of \cite{9815289} explored an integrated framework for uplink communication and over-the-air federated learning in STARS-NOMA, where a successive signal processing scheme was introduced to manage the co-channel interference for communication while also leveraging it for federated learning. The concept of simultaneous downlink and uplink communication in STARS-NOMA systems has also gained growing attention. For instance, the authors of \cite{10224251} studied two-way communication in STARS-NOMA, where the total achievable rate was maximized by jointly optimizing uplink/downlink power allocation and STARS beamforming. Furthermore, the authors of \cite{10153701} incorporated STARS for assisting active downlink communication and uplink backscatter communication with NOMA.

\begin{figure*}[!t]
    \centering
    \subfigure[ES/TS-STARS (Near-field beamfocusing on both sides).]{
        \includegraphics[width=0.7\textwidth]{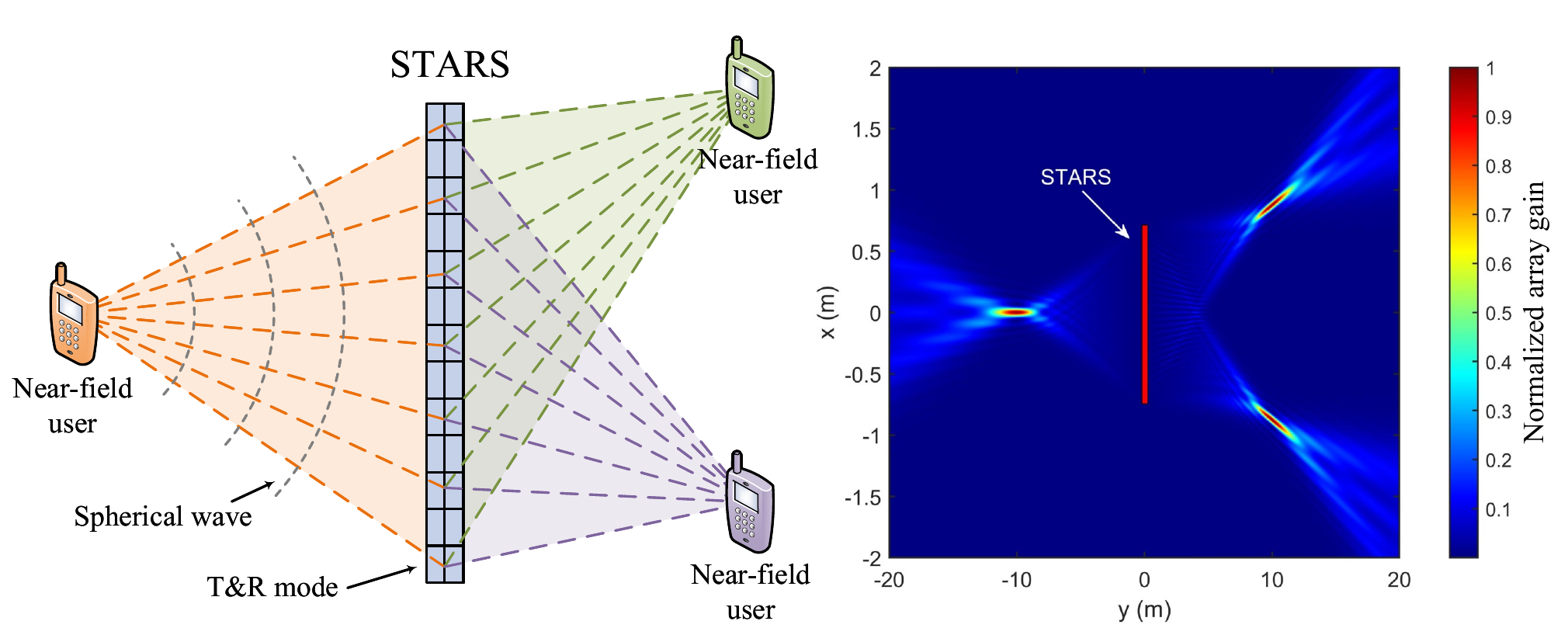}
    }
    \subfigure[MS-STARS (Near-field beamfocusing and far-field beamsteering on two sides).]{
        \includegraphics[width=0.7\textwidth]{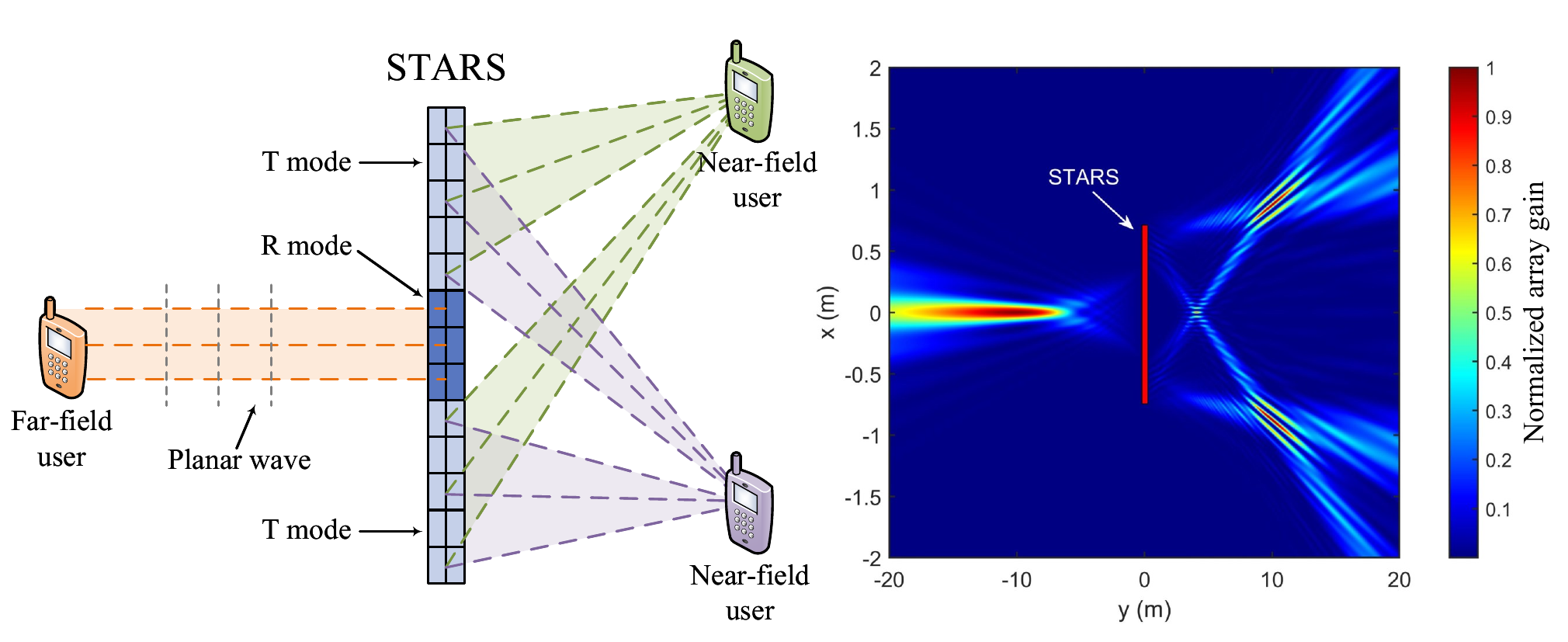}
    }
    \caption{Illustration of STARS-aided NFC and the beamforming patterns achieved using different operating protocols. For the generation of beamforming patterns, we consider a single-antenna BS and a 256-element STARS with 28 GHz carrier frequency and half-wavelength element spacing.}
    \label{fig:STAR_NFC}
\end{figure*}

\subsection{STARS-NFC}
Both far-field and near-field effects are intrinsic EM phenomena associated with radio waves emitted by an antenna or an array of antennas. In the far-field region, wave propagation can typically be represented as planar waves, whereas in the near-field region, it has to be accurately modelled as spherical waves \cite{balanis2016antenna}. The simplicity of the far-field planar-wave model has led to its widespread adoption in earlier generations of wireless networks. However, the emergence of large-scale antenna arrays, such as supermassive MIMO and CF mMIMO \cite{bjornson2019massive}, coupled with the utilization of high-frequency bands like millimeter wave and THz bands \cite{akyildiz2014terahertz}, has made the near-field effect increasingly significant or even dominant in NG wireless networks \cite{10220205}. The extent of the near-field region is proportional to the antenna array's aperture and the signal frequency. For example, following classical Rayleigh distance criteria \cite{balanis2016antenna}, an antenna array with a 1-m aperture and a 28-GHz signal frequency could have a near-field region extending up to 370 m, making NFC design crucial for NG wireless networks.

While the near-field effect leads to a more complex wave-propagation model, it also facilitates a new form of beamforming design, i.e., \emph{beamfocusing} \cite{9738442,10380596}. In contrast to far-field beamsteering, which only differentiates users in the angular domain, near-field beamfocusing can distinguish users in both angular and distance domains, thus greatly enhancing multiple access capabilities but at the cost of higher optimization and channel estimation overhead. Compared to conventional reflecting/transmitting-only RISs, STARS with massive elements not only facilitate full-space near-field beamfocusing through the ES or TS protocol but also enable dynamic switching between far-field beamforming and near-field beamfocusing through the MS protocol. In the following, we will elaborate on the ES/TS-STARS-aided NFC and MS-STARS-aided NFC, respectively. 
\begin{itemize}
    \item \textbf{ES/TS-STARS-aided NFC:} Utilizing the ES/TS protocol can maximize the aperture size of STARS for transmission and/or reflection, thus creating a large near-field region on both sides, as shown in Fig. \ref{fig:STAR_NFC}(a). This setup enables effective near-field beamfocusing across the full space. However, ES/TS-STARS-aided NFC requires a massive number of STARS elements to be optimized for both transmission and reflection as well as complicated channel estimation in both angular and distance domains. To reduce the optimization complexity, the authors of \cite{li2023near} proposed a low-complexity element-wise STARS beamforming design. Furthermore, a unified near-far-field codebook-based STARS beamforming design was introduced in \cite{zhang2024near} to reduce the channel estimation overhead in STARS-NFC.
    
    \item \textbf{MS-STARS-aided NFC:} The MS protocol allows dynamic adjustment of the effective aperture size for transmission and reflection by selectively activating STARS elements in transmission or reflection mode \cite{10192541}, as shown in Fig. \ref{fig:STAR_NFC}(b). This enables switching transmission and reflection users between near-field and far-field regions. Therefore, different forms of beamforming can be realized on different sides of the STARS, thus satisfying specific requirements of transmission and reflection users, such as QoS and channel estimation overhead. 
\end{itemize}

\section{Conclusions}
In this paper, we presented a comprehensive overview of the employment of STARS and their 360° smart radio environment to realize efficient NGMA. First, the basic foundational principles for realizing STARS were discussed together with operating protocols, different categories, and prototypes of STARS. Then, state-of-the-art research contributions on STARS-aided wireless communication were surveyed with a particular focus on performance analysis and STARS beamforming design, which highlighted the benefits of deploying STARS in NG wireless networks. For possible employment of STARS in NGMA, advanced STARS applications for NGMA were discussed in UAV communications, PLS, SWIPT and other scenarios. Furthermore, several promising interplays between STARS and other emerging technologies, including ISAC, MEC, NOMA, and NFC, were put forward. We hope that the discussion presented in this paper can attract more research efforts to unlock the full potential of STARS for NGMA in future wireless networks.

\bibliographystyle{IEEEtran}
\bibliography{mybib}

\begin{thebibliography}{100}
\providecommand{\url}[1]{#1}
\csname url@samestyle\endcsname
\providecommand{\newblock}{\relax}
\providecommand{\bibinfo}[2]{#2}
\providecommand{\BIBentrySTDinterwordspacing}{\spaceskip=0pt\relax}
\providecommand{\BIBentryALTinterwordstretchfactor}{4}
\providecommand{\BIBentryALTinterwordspacing}{\spaceskip=\fontdimen2\font plus
\BIBentryALTinterwordstretchfactor\fontdimen3\font minus
  \fontdimen4\font\relax}
\providecommand{\BIBforeignlanguage}[2]{{%
\expandafter\ifx\csname l@#1\endcsname\relax
\typeout{** WARNING: IEEEtran.bst: No hyphenation pattern has been}%
\typeout{** loaded for the language `#1'. Using the pattern for}%
\typeout{** the default language instead.}%
\else
\language=\csname l@#1\endcsname
\fi
#2}}
\providecommand{\BIBdecl}{\relax}
\BIBdecl

\bibitem{8766143}
Z.~Zhang, Y.~Xiao, Z.~Ma, M.~Xiao, Z.~Ding, X.~Lei, G.~K. Karagiannidis, and
  P.~Fan, ``{6G} wireless networks: Vision, requirements, architecture, and key
  technologies,'' \emph{IEEE Veh. Technol. Mag.}, vol.~14, no.~3, pp. 28--41,
  Sept. 2019.

\bibitem{8869705}
W.~Saad, M.~Bennis, and M.~Chen, ``A vision of {6G} wireless systems:
  Applications, trends, technologies, and open research problems,'' \emph{IEEE
  Network}, vol.~34, no.~3, pp. 134--142, May/June 2020.

\bibitem{10054381}
C.-X. Wang, X.~You, X.~Gao, X.~Zhu, Z.~Li, C.~Zhang, H.~Wang, Y.~Huang,
  Y.~Chen, H.~Haas, J.~S. Thompson, E.~G. Larsson, M.~D. Renzo, W.~Tong,
  P.~Zhu, X.~Shen, H.~V. Poor, and L.~Hanzo, ``On the road to {6G}: Visions,
  requirements, key technologies, and testbeds,'' \emph{IEEE Commun. Surveys
  Tuts.}, vol.~25, no.~2, pp. 905--974, Secondquarter 2023.

\bibitem{9628162}
H.~Guo, J.~Li, J.~Liu, N.~Tian, and N.~Kato, ``A survey on space-air-ground-sea
  integrated network security in {6G},'' \emph{IEEE Commun. Surveys Tuts.},
  vol.~24, no.~1, pp. 53--87, Firstquarter 2022.

\bibitem{8016573}
Y.~Mao, C.~You, J.~Zhang, K.~Huang, and K.~B. Letaief, ``A survey on mobile
  edge computing: The communication perspective,'' \emph{IEEE Commun. Surveys
  Tuts.}, vol.~19, no.~4, pp. 2322--2358, Fourthquarter 2017.

\bibitem{9737357}
F.~Liu, Y.~Cui, C.~Masouros, J.~Xu, T.~X. Han, Y.~C. Eldar, and S.~Buzzi,
  ``Integrated sensing and communications: Toward dual-functional wireless
  networks for 6{G} and beyond,'' \emph{{IEEE} J. Sel. Areas Commun.}, vol.~40,
  no.~6, pp. 1728--1767, Jun. 2022.

\bibitem{10024901}
X.~Mu, Z.~Wang, and Y.~Liu, ``{NOMA} for integrating sensing and communications
  towards 6{G}: A multiple access perspective,'' \emph{{IEEE} Wireless
  Commun.}, Early Access, doi:10.1109/MWC.015.2200559.

\bibitem{8085125}
Y.~Cai, Z.~Qin, F.~Cui, G.~Y. Li, and J.~A. McCann, ``Modulation and multiple
  access for 5{G} networks,'' \emph{IEEE Commun. Surveys Tuts.}, vol.~20,
  no.~1, pp. 629--646, Firstquarter 2018.

\bibitem{9205230}
X.~Chen, D.~W.~K. Ng, W.~Yu, E.~G. Larsson, N.~Al-Dhahir, and R.~Schober,
  ``Massive access for 5{G} and beyond,'' \emph{{IEEE} J. Sel. Areas Commun.},
  vol.~39, no.~3, pp. 615--637, March 2021.

\bibitem{9693417}
Y.~Liu, S.~Zhang, X.~Mu, Z.~Ding, R.~Schober, N.~Al-Dhahir, E.~Hossain, and
  X.~Shen, ``Evolution of {NOMA} toward next generation multiple access
  ({NGMA}) for 6{G},'' \emph{{IEEE} J. Sel. Areas Commun.}, vol.~40, no.~4, pp.
  1037--1071, April 2022.

\bibitem{8114722}
Y.~Liu, Z.~Qin, M.~Elkashlan, Z.~Ding, A.~Nallanathan, and L.~Hanzo,
  ``Nonorthogonal multiple access for 5{G} and beyond,'' \emph{Proc. IEEE},
  vol. 105, no.~12, pp. 2347--2381, Dec. 2017.

\bibitem{9113273}
J.~Zhang, E.~Björnson, M.~Matthaiou, D.~W.~K. Ng, H.~Yang, and D.~J. Love,
  ``Prospective multiple antenna technologies for beyond 5{G},'' \emph{{IEEE}
  J. Sel. Areas Commun.}, vol.~38, no.~8, pp. 1637--1660, Aug. 2020.

\bibitem{9831440}
Y.~Mao, O.~Dizdar, B.~Clerckx, R.~Schober, P.~Popovski, and H.~V. Poor,
  ``Rate-splitting multiple access: Fundamentals, survey, and future research
  trends,'' \emph{IEEE Commun. Surveys Tuts.}, vol.~24, no.~4, pp. 2073--2126,
  Fourthquarter 2022.

\bibitem{9140329}
M.~Di~Renzo, A.~Zappone, M.~Debbah, M.-S. Alouini, C.~Yuen, J.~de~Rosny, and
  S.~Tretyakov, ``Smart radio environments empowered by reconfigurable
  intelligent surfaces: How it works, state of research, and the road ahead,''
  \emph{{IEEE} J. Sel. Areas Commun.}, vol.~38, no.~11, pp. 2450--2525, Nov.
  2020.

\bibitem{9424177}
Y.~Liu, X.~Liu, X.~Mu, T.~Hou, J.~Xu, M.~Di~Renzo, and N.~Al-Dhahir,
  ``Reconfigurable intelligent surfaces: Principles and opportunities,''
  \emph{IEEE Commun. Surveys Tuts.}, vol.~23, no.~3, pp. 1546--1577,
  thirdquarter 2021.

\bibitem{8910627}
Q.~Wu and R.~Zhang, ``Towards smart and reconfigurable environment: Intelligent
  reflecting surface aided wireless network,'' \emph{{IEEE} Commun. Mag.},
  vol.~58, no.~1, pp. 106--112, Jan. 2020.

\bibitem{9475160}
C.~Pan, H.~Ren, K.~Wang, J.~F. Kolb, M.~Elkashlan, M.~Chen, M.~Di~Renzo,
  Y.~Hao, J.~Wang, A.~L. Swindlehurst, X.~You, and L.~Hanzo, ``Reconfigurable
  intelligent surfaces for 6{G} systems: Principles, applications, and research
  directions,'' \emph{{IEEE} Commun. Mag.}, vol.~59, no.~6, pp. 14--20, June
  2021.

\bibitem{9779790}
Z.~Ding, L.~Lv, F.~Fang, O.~A. Dobre, G.~K. Karagiannidis, N.~Al-Dhahir,
  R.~Schober, and H.~V. Poor, ``A state-of-the-art survey on reconfigurable
  intelligent surface-assisted non-orthogonal multiple access networks,''
  \emph{Proc. IEEE}, vol. 110, no.~9, pp. 1358--1379, Sept. 2022.

\bibitem{9437234}
J.~Xu, Y.~Liu, X.~Mu, and O.~A. Dobre, ``{STAR-RIS}s: Simultaneous transmitting
  and reflecting reconfigurable intelligent surfaces,'' \emph{{IEEE} Commun.
  Lett.}, vol.~25, no.~9, pp. 3134--3138, Sept. 2021.

\bibitem{9690478}
Y.~Liu, X.~Mu, J.~Xu, R.~Schober, Y.~Hao, H.~V. Poor, and L.~Hanzo, ``{STAR}:
  Simultaneous transmission and reflection for 360° coverage by intelligent
  surfaces,'' \emph{{IEEE} Wireless Commun.}, vol.~28, no.~6, pp. 102--109,
  Dec. 2021.

\bibitem{xu_vtmag}
J.~Xu, Y.~Liu, X.~Mu, J.~T. Zhou, L.~Song, H.~V. Poor, and L.~Hanzo,
  ``Simultaneously transmitting and reflecting intelligent omni-surfaces:
  Modeling and implementation,'' \emph{IEEE Veh. Technol. Mag.}, vol.~17,
  no.~2, pp. 46--54, June 2022.

\bibitem{zhu2014dynamic}
B.~O. Zhu, K.~Chen, N.~Jia, L.~Sun, J.~Zhao, T.~Jiang, and Y.~Feng, ``Dynamic
  control of electromagnetic wave propagation with the equivalent principle
  inspired tunable metasurface,'' \emph{Sci. Rep.}, vol.~4, no.~1, pp. 1--7,
  May 2014.

\bibitem{xu_coupled}
J.~Xu, Y.~Liu, X.~Mu, R.~Schober, and H.~V. Poor, ``{STAR-RISs}: A correlated
  {T\&R} phase-shift model and practical phase-shift configuration
  strategies,'' \emph{IEEE J. Sel.Top. Signal Process.}, vol.~16, no.~5, pp.
  1097--1111, Aug. 2022.

\bibitem{9570143}
X.~Mu, Y.~Liu, L.~Guo, J.~Lin, and R.~Schober, ``Simultaneously transmitting
  and reflecting ({STAR}) {RIS} aided wireless communications,'' \emph{{IEEE}
  Trans. Wireless Commun.}, vol.~21, no.~5, pp. 3083--3098, May 2022.

\bibitem{Zhang_2020}
X.~Zhang, D.~Tang, L.~Zhou, G.~Liang, D.~Feng, and Y.~Guo, ``A quasi-continuous
  all-dielectric metasurface for broadband and high-efficiency holographic
  images,'' \emph{J. Phys. D: Appl. Phys.}, vol.~53, no.~46, p. 465105, Aug.
  2020.

\bibitem{10192541}
J.~Xu, X.~Mu, and Y.~Liu, ``Exploiting {STAR-RIS}s in near-field
  communications,'' \emph{{IEEE} Trans. Wireless Commun.}, vol.~23, no.~3, pp.
  2181--2196, March 2024.

\bibitem{xu_active}
J.~Xu, J.~Zuo, J.~T. Zhou, and Y.~Liu, ``Active simultaneously transmitting and
  reflecting ({STAR})-{RIS}s: Modeling and analysis,'' \emph{{IEEE} Commun.
  Lett.}, vol.~27, no.~9, pp. 2466--2470, Sept. 2023.

\bibitem{9961851}
H.~Luo, L.~Lv, Q.~Wu, Z.~Ding, N.~Al-Dhahir, and J.~Chen, ``Beamforming design
  for active {IOS} aided {NOMA} networks,'' \emph{{IEEE} Wireless Commun.
  Lett.}, vol.~12, no.~2, pp. 282--286, Feb. 2023.

\bibitem{10264149}
X.~Ma, X.~Lei, P.~T. Mathiopoulos, and D.~B.~d. Costa, ``Active {STAR-RIS}
  aided cell-free massive {MIMO}: A performance study,'' \emph{{IEEE} Trans.
  Veh. Technol.}, vol.~73, no.~2, pp. 2936--2941, Feb. 2024.

\bibitem{10153967}
Y.~Ma, M.~Li, Y.~Liu, Q.~Wu, and Q.~Liu, ``Optimization for reflection and
  transmission dual-functional active {RIS}-assisted systems,'' \emph{{IEEE}
  Trans. Commun.}, vol.~71, no.~9, pp. 5534--5548, Sept. 2023.

\bibitem{10227341}
C.~Zhou, B.~Lyu, S.~Gong, and C.~You, ``Active {STAR-RIS}-assisted symbiotic
  radio communications under hardware impairments,'' \emph{{IEEE} Commun.
  Lett.}, vol.~27, no.~10, pp. 2797--2801, Oct. 2023.

\bibitem{PhysRevLett}
F.~Monticone, N.~M. Estakhri, and A.~Al\`u, ``Full control of nanoscale optical
  transmission with a composite metascreen,'' \emph{Phys. Rev. Lett.}, vol.
  110, p. 203903, May 2013.

\bibitem{ntt}
\BIBentryALTinterwordspacing
``{DOCOMO} conducts world's first successful trial of transparent dynamic
  metasurface''. [Online]. Available:
  \url{www.docomo.ne.jp/english/info/media_center/pr/2020/0117_00.html}
\BIBentrySTDinterwordspacing

\bibitem{9895224}
S.~Zeng, H.~Zhang, B.~Di, Y.~Liu, M.~D. Renzo, Z.~Han, H.~V. Poor, and L.~Song,
  ``Intelligent omni-surfaces: Reflection-refraction circuit model,
  full-dimensional beamforming, and system implementation,'' \emph{{IEEE}
  Trans. Commun.}, vol.~70, no.~11, pp. 7711--7727, Nov. 2022.

\bibitem{10288376}
Y.~Liu, J.~Kelly, M.~Holm, S.~Gopal, S.~R. Aghdam, and Y.~Liu, ``Unit cell
  design for intelligent reflecting and refracting surface ({IR$^2$S}) with
  independent electronic control capability,'' \emph{IEEE Antennas Wirel.
  Propag. Lett.}, vol.~23, no.~1, pp. 414--418, Jan. 2024.

\bibitem{10177915}
M.~Xiang, Y.~Xiao, J.~Deng, S.~Xu, and F.~Yang, ``Simultaneous transmitting and
  reflecting reconfigurable array ({STAR-RA}) with independent beams,''
  \emph{IEEE Trans. Antennas Propag.}, vol.~71, no.~10, pp. 8338--8343, Oct.
  2023.

\bibitem{10156858}
Q.~Li, M.~El-Hajjar, Y.~Sun, I.~Hemadeh, A.~Shojaeifard, Y.~Liu, and L.~Hanzo,
  ``Achievble rate analysis of the {STAR-RIS}-aided {NOMA} uplink in the face
  of imperfect {CSI} and hardware impairments,'' \emph{{IEEE} Trans. Commun.},
  vol.~71, no.~10, pp. 6100--6114, Oct. 2023.

\bibitem{10466748}
Q.~Li, M.~El-Hajjar, and L.~Hanzo, ``Ergodic spectral efficiency analysis of
  intelligent omni-surface aided systems suffering from imperfect {CSI} and
  hardware impairments,'' \emph{{IEEE} Trans. Commun.}, Early Access, doi:
  10.1109/TCOMM.2024.3376592.

\bibitem{Qingchao}
Q.~Li, M.~El-Hajjar, Y.~Sun, I.~Hemadeh, Y.~Tsai, A.~Shojaeifard, and L.~Hanzo,
  ``Achievable rate analysis of intelligent omni-surface assisted {NOMA}
  holographic {MIMO} systems,'' \emph{arXiv preprint arXiv:2405.01136}, 2024.

\bibitem{9815097}
H.~Liu, G.~Li, X.~Li, Y.~Liu, G.~Huang, and Z.~Ding, ``Effective capacity
  analysis of {STAR-RIS}-assisted {NOMA} networks,'' \emph{{IEEE} Wireless
  Commun. Lett.}, vol.~11, no.~9, pp. 1930--1934, Sep. 2022.

\bibitem{9869706}
J.~Chen and X.~Yu, ``Ergodic rate analysis and phase design of {STAR-RIS} aided
  {NOMA} with statistical {CSI},'' \emph{{IEEE} Commun. Lett.}, vol.~26,
  no.~12, pp. 2889--2893, Dec. 2022.

\bibitem{10175074}
A.~Papazafeiropoulos, L.-N. Tran, Z.~Abdullah, P.~Kourtessis, and
  S.~Chatzinotas, ``Achievable rate of a {STAR-RIS} assisted massive {MIMO}
  system under spatially-correlated channels,'' \emph{{IEEE} Trans. Wireless
  Commun.}, vol.~23, no.~2, pp. 1550--1564, Feb. 2024.

\bibitem{9843866}
B.~Zhao, C.~Zhang, W.~Yi, and Y.~Liu, ``Ergodic rate analysis of {STAR-RIS}
  aided {NOMA} systems,'' \emph{{IEEE} Commun. Lett.}, vol.~26, no.~10, pp.
  2297--2301, Oct. 2022.

\bibitem{10373089}
H.~Ge, A.~Papazafeiropoulos, and T.~Ratnarajah, ``Impact of phase noise in
  downlink {STAR-RIS-Aided} massive {MIMO} systems,'' \emph{{IEEE} Commun.
  Lett.}, vol.~28, no.~2, pp. 392--396, Feb. 2024.

\bibitem{10297571}
A.~Papazafeiropoulos, H.~Q. Ngo, P.~Kourtessis, and S.~Chatzinotas,
  ``{STAR-RIS} assisted cell-free massive {MIMO} system under
  spatially-correlated channels,'' \emph{{IEEE} Trans. Veh. Technol.}, vol.~73,
  no.~3, pp. 3932--3948, March 2024.

\bibitem{9864148}
T.~H. Nguyen and T.~T. Nguyen, ``On performance of {STAR-RIS}-enabled multiple
  two-way full-duplex {D2D} communication systems,'' \emph{IEEE Access},
  vol.~10, pp. 89\,063--89\,071, 2022.

\bibitem{10345673}
X.~Sheng, X.~Li, G.~Chen, G.~Huang, C.~Han, and Z.~Ding, ``Performance analysis
  of {STAR-RIS} assisted secure cognitive {NOMA-HARQ} networks,'' \emph{{IEEE}
  Wireless Commun. Lett.}, vol.~13, no.~3, pp. 696--700, March 2024.

\bibitem{9935303}
J.~Xu, X.~Mu, J.~T. Zhou, and Y.~Liu, ``Simultaneously transmitting and
  reflecting {(STAR)-RIS}s: Are they applicable to dual-sided incidence?''
  \emph{{IEEE} Wireless Commun. Lett.}, vol.~12, no.~1, pp. 129--133, Jan.
  2023.

\bibitem{9786058}
A.~Papazafeiropoulos, Z.~Abdullah, P.~Kourtessis, S.~Kisseleff, and
  I.~Krikidis, ``Coverage probability of {STAR-RIS}-assisted massive {MIMO}
  systems with correlation and phase errors,'' \emph{{IEEE} Wireless Commun.
  Lett.}, vol.~11, no.~8, pp. 1738--1742, Aug. 2022.

\bibitem{9462949}
C.~Wu, Y.~Liu, X.~Mu, X.~Gu, and O.~A. Dobre, ``Coverage characterization of
  {STAR-RIS} networks: {NOMA} and {OMA},'' \emph{{IEEE} Commun. Lett.},
  vol.~25, no.~9, pp. 3036--3040, Sept. 2021.

\bibitem{ghadi2023analytical}
F.~R. Ghadi, F.~J. Lopez-Martinez, and K.-K. Wong, ``Analytical
  characterization of coverage regions for {STAR-RIS}-aided {NOMA/OMA}
  communication systems,'' \emph{{IEEE} Commun. Lett.}, vol.~27, no.~11, pp.
  3063--3067, Nov. 2023.

\bibitem{9808307}
Z.~Xie, W.~Yi, X.~Wu, Y.~Liu, and A.~Nallanathan, ``{STAR-RIS} aided {NOMA} in
  multicell networks: A general analytical framework with {G}amma distributed
  channel modeling,'' \emph{{IEEE} Trans. Commun.}, vol.~70, no.~8, pp.
  5629--5644, Aug. 2022.

\bibitem{9786807}
M.~Aldababsa, A.~Khaleel, and E.~Basar, ``{STAR-RIS-NOMA} networks: An error
  performance perspective,'' \emph{{IEEE} Commun. Lett.}, vol.~26, no.~8, pp.
  1784--1788, Aug. 2022.

\bibitem{10049460}
F.~Karim, S.~K. Singh, K.~Singh, S.~Prakriya, and M.~F. Flanagan, ``On the
  performance of {STAR-RIS}-aided {NOMA} at finite blocklength,'' \emph{{IEEE}
  Wireless Commun. Lett.}, vol.~12, no.~5, pp. 868--872, May 2023.

\bibitem{9920228}
Y.~Lin, Y.~Shen, and A.~Li, ``Simultaneous transmission and reflection
  beamforming design for {RIS}-aided {MU-MISO},'' \emph{{IEEE} Trans. Veh.
  Technol.}, vol.~72, no.~3, pp. 4040--4045, March 2023.

\bibitem{10130543}
L.~Xue, K.~Wang, Z.~Yang, and M.~Peng, ``Max-min energy-efficiency fair
  optimization in {STAR-RIS} assisted communication system,'' \emph{IEEE
  Access}, vol.~11, pp. 51\,106--51\,116, 2023.

\bibitem{9785636}
Y.~Wang, P.~Guan, H.~Yu, and Y.~Zhao, ``Transmit power optimization of
  simultaneous transmission and reflection {RIS} assisted full-duplex
  communications,'' \emph{IEEE Access}, vol.~10, pp. 61\,192--61\,200, 2022.

\bibitem{9838767}
Y.~Liu, X.~Mu, R.~Schober, and H.~V. Poor, ``Simultaneously transmitting and
  reflecting ({STAR})-{RIS}s: A coupled phase-shift model,'' in \emph{Proc.
  {IEEE} Int. Conf. Commun. ({ICC})}, 2022, pp. 2840--2845.

\bibitem{10224271}
C.~Zhou, B.~Lyu, Y.~Feng, and D.~T. Hoang, ``Transmit power minimization for
  {STAR-RIS} empowered symbiotic radio communications,'' \emph{IEEE Trans.
  Cogn. Commun. Netw.}, vol.~9, no.~6, pp. 1641--1656, Dec. 2023.

\bibitem{9629335}
H.~Niu, Z.~Chu, F.~Zhou, P.~Xiao, and N.~Al-Dhahir, ``Weighted sum rate
  optimization for {STAR-RIS}-assisted {MIMO} system,'' \emph{{IEEE} Trans.
  Veh. Technol.}, vol.~71, no.~2, pp. 2122--2127, Feb. 2022.

\bibitem{9751144}
H.~Niu and X.~Liang, ``Weighted sum-rate maximization for star-riss-aided
  networks with coupled phase-shifters,'' \emph{IEEE Syst. J.}, vol.~17, no.~1,
  pp. 1083--1086, March 2023.

\bibitem{10325546}
S.~Huang, W.~Wang, R.~Jiang, X.~Wang, Z.~Fei, C.~Huang, J.~Li, S.~Ren, X.~Li,
  and H.~Dang, ``Average sum-rate maximization for coupled phase-shift
  {STAR-RIS} enhanced multi-user {MISO-OFDM} system,'' \emph{{IEEE} Trans.
  Commun.}, vol.~72, no.~3, pp. 1457--1473, March 2024.

\bibitem{9935266}
Z.~Wang, X.~Mu, Y.~Liu, and R.~Schober, ``Coupled phase-shift {STAR-RIS}s: A
  general optimization framework,'' \emph{{IEEE} Wireless Commun. Lett.},
  vol.~12, no.~2, pp. 207--211, Feb. 2023.

\bibitem{10093070}
A.~Papazafeiropoulos, A.~M. Elbir, P.~Kourtessis, I.~Krikidis, and
  S.~Chatzinotas, ``Cooperative {RIS} and {STAR-RIS} assisted m{MIMO}
  communication: Analysis and optimization,'' \emph{{IEEE} Trans. Veh.
  Technol.}, vol.~72, no.~9, pp. 11\,975--11\,989, Sept. 2023.

\bibitem{10050140}
Q.~Gao, Y.~Liu, X.~Mu, M.~Jia, D.~Li, and L.~Hanzo, ``Joint location and
  beamforming design for {STAR-RIS} assisted {NOMA} systems,'' \emph{{IEEE}
  Trans. Commun.}, vol.~71, no.~4, pp. 2532--2546, Apr. 2023.

\bibitem{10254537}
Y.~Pan, Z.~Qin, J.-B. Wang, Y.~Chen, H.~Yu, and A.~Tang, ``Joint deployment and
  beamforming design for {STAR-RIS} aided communication,'' \emph{{IEEE} Commun.
  Lett.}, vol.~27, no.~11, pp. 3083--3087, Nov. 2023.

\bibitem{9837935}
R.~Zhong, Y.~Liu, X.~Mu, Y.~Chen, X.~Wang, and L.~Hanzo, ``Hybrid reinforcement
  learning for {STAR-RIS}s: A coupled phase-shift model based beamformer,''
  \emph{{IEEE} J. Sel. Areas Commun.}, vol.~40, no.~9, pp. 2556--2569, Sept.
  2022.

\bibitem{9964251}
Y.~Guo, F.~Fang, D.~Cai, and Z.~Ding, ``Energy-efficient design for a {NOMA}
  assisted {STAR-RIS} network with deep reinforcement learning,'' \emph{{IEEE}
  Trans. Veh. Technol.}, vol.~72, no.~4, pp. 5424--5428, April 2023.

\bibitem{10306287}
P.~S. Aung, L.~X. Nguyen, Y.~K. Tun, Z.~Han, and C.~S. Hong, ``Deep
  reinforcement learning based joint spectrum allocation and configuration
  design for {STAR-RIS}-assisted {V2X} communications,'' \emph{IEEE Internet
  Things J.}, vol.~11, no.~7, pp. 11\,298--11\,311, April 2024.

\bibitem{10187159}
X.~Gao, W.~Yi, Y.~Liu, J.~Zhang, and P.~Zhang, ``{DRL} enabled coverage and
  capacity optimization in {STAR-RIS}-assisted networks,'' \emph{{IEEE} Trans.
  Commun.}, vol.~71, no.~11, pp. 6616--6632, Nov. 2023.

\bibitem{10049110}
R.~Zhong, X.~Mu, Y.~Liu, Y.~Chen, J.~Zhang, and P.~Zhang, ``{STAR-RIS}s
  assisted {NOMA} networks: A distributed learning approach,'' \emph{IEEE J.
  Sel. Top. Signal Process.}, vol.~17, no.~1, pp. 264--278, Jan. 2023.

\bibitem{10172151}
H.~Du, R.~Zhang, D.~Niyato, J.~Kang, Z.~Xiong, D.~I. Kim, X.~S. Shen, and H.~V.
  Poor, ``Exploring collaborative distributed diffusion-based {AI}-generated
  content ({AIGC}) in wireless networks,'' \emph{IEEE Network}, Early Access,
  doi: 10.1109/MNET.006.2300223.

\bibitem{GAI}
H.~Du, R.~Zhang, Y.~Liu, J.~Wang, Y.~Lin, Z.~Li, D.~Niyato, J.~Kang, Z.~Xiong,
  S.~Cui, B.~Ai, H.~Zhou, and D.~I. Kim, ``Enhancing deep reinforcement
  learning: A tutorial on generative diffusion models in network
  optimization,'' \emph{arXiv preprint arXiv:2308.05384}, 2023.

\bibitem{8918497}
Y.~Zeng, Q.~Wu, and R.~Zhang, ``Accessing from the sky: A tutorial on {UAV}
  communications for 5{G} and beyond,'' \emph{Proc. IEEE}, vol. 107, no.~12,
  pp. 2327--2375, Sept. 2019.

\bibitem{10354335}
Y.~Peng, J.~Tang, Q.~Yang, Z.~Han, and J.~Ma, ``Joint power allocation
  algorithm for {UAV}-borne simultaneous transmitting and reflecting
  reconfigurable intelligent surface-assisted non-orthogonal multiple access
  system,'' \emph{IEEE Access}, vol.~11, pp. 140\,506--140\,518, 2023.

\bibitem{10376206}
L.~Guo, J.~Jia, J.~Chen, and X.~Wang, ``Secure communication optimization in
  {NOMA} systems with {UAV}-mounted {STAR-RIS},'' \emph{IEEE Trans. Inf.
  Forensics Secur.}, vol.~19, pp. 2300--2314, Dec. 2024.

\bibitem{STAR_UAV_3D}
X.~Wang, Z.~Lin, F.~Lin, and P.~Xiao, ``Quantum sensing based joint {3D} beam
  training for {UAV}-mounted {STAR-RIS} aided terahertz multi-user massive
  {MIMO} systems,'' \emph{arXiv preprint arXiv:2212.07731}, 2022.

\bibitem{10453937}
T.~Kim, M.~Jung, and H.~Son, ``Joint user scheduling and phase shift
  optimization for {STAR-RIS}-assisted multicast satellite communications,''
  \emph{IEEE Trans. Aerosp. Electron. Syst.}, pp. 1--10, Early Access, doi:
  10.1109/TAES.2024.3371398.

\bibitem{STAR_UAV_MEC1}
P.~S. Aung, L.~X. Nguyen, Y.~K. Tun, Z.~Han, and C.~S. Hong, ``Aerial
  {STAR-RIS} empowered {MEC}: A {DRL} approach for energy minimization,''
  \emph{{IEEE} Wireless Commun. Lett.}, vol.~13, no.~5, pp. 1409--1413, May
  2024.

\bibitem{STAR_UAV_MEC2}
H.~Xiao, X.~Hu, P.~Mu, W.~Zhang, W.~Wang, K.~Wong, and K.~Yang, ``{STAR-RIS}
  enhanced {UAV}-enabled {MEC} networks with bi-directional task offloading,''
  \emph{arXiv preprint arXiv:2401.05725}, 2023.

\bibitem{9849460}
J.~Zhao, Y.~Zhu, X.~Mu, K.~Cai, Y.~Liu, and L.~Hanzo, ``Simultaneously
  transmitting and reflecting reconfigurable intelligent surface ({STAR-RIS})
  assisted {UAV} communications,'' \emph{{IEEE} J. Sel. Areas Commun.},
  vol.~40, no.~10, pp. 3041--3056, Oct. 2022.

\bibitem{9878137}
Q.~Zhang, Y.~Zhao, H.~Li, S.~Hou, and Z.~Song, ``Joint optimization of
  {STAR-RIS} assisted {UAV} communication systems,'' \emph{{IEEE} Wireless
  Commun. Lett.}, vol.~11, no.~11, pp. 2390--2394, Nov. 2022.

\bibitem{10286085}
P.~Zhu, L.~Qin, J.~Wang, Y.~Li, X.~Li, and W.~Xie, ``Optimized trajectory and
  passive beamforming for {STAR-RIS}-assisted {UAV}-empowered {O2I} {WPCN},''
  \emph{{IEEE} Wireless Commun. Lett.}, vol.~13, no.~1, pp. 163--167, Jan.
  2024.

\bibitem{10083240}
Y.~Su, X.~Pang, W.~Lu, N.~Zhao, X.~Wang, and A.~Nallanathan, ``Joint location
  and beamforming optimization for {STAR-RIS} aided {NOMA-UAV} networks,''
  \emph{{IEEE} Trans. Veh. Technol.}, vol.~72, no.~8, pp. 11\,023--11\,028,
  Aug. 2023.

\bibitem{10320337}
J.~Lei, T.~Zhang, X.~Mu, and Y.~Liu, ``{NOMA} for {STAR-RIS} assisted {UAV}
  networks,'' \emph{{IEEE} Trans. Commun.}, vol.~72, no.~3, pp. 1732--1745,
  March 2024.

\bibitem{9525400}
H.~Niu, Z.~Chu, F.~Zhou, and Z.~Zhu, ``Simultaneous transmission and reflection
  reconfigurable intelligent surface assisted secrecy {MISO} networks,''
  \emph{{IEEE} Commun. Lett.}, vol.~25, no.~11, pp. 3498--3502, Nov. 2021.

\bibitem{9774882}
W.~Wang, W.~Ni, H.~Tian, Z.~Yang, C.~Huang, and K.-K. Wong, ``Safeguarding
  {NOMA} networks via reconfigurable dual-functional surface under imperfect
  {CSI},'' \emph{IEEE J. Sel. Top. Signal Process.}, vol.~16, no.~5, pp.
  950--966, Aug. 2022.

\bibitem{10005206}
H.~Jia, L.~Ma, and S.~Valaee, ``{STAR-RIS} enabled downlink secure {NOMA}
  network under imperfect {CSI} of eavesdroppers,'' \emph{{IEEE} Commun.
  Lett.}, vol.~27, no.~3, pp. 802--806, March 2023.

\bibitem{10040906}
Z.~Zhang, Z.~Wang, Y.~Liu, B.~He, L.~Lv, and J.~Chen, ``Security enhancement
  for coupled phase-shift {STAR-RIS} networks,'' \emph{{IEEE} Trans. Veh.
  Technol.}, vol.~72, no.~6, pp. 8210--8215, June 2023.

\bibitem{10304291}
T.~Zhou, K.~Xu, G.~Hu, X.~Xia, W.~Xie, and C.~Li, ``Robust beamforming design
  for {STAR-RIS}-assisted anti-jamming and secure transmission,'' \emph{IEEE
  Trans. Green Commun. Netw.}, vol.~8, no.~1, pp. 345--361, March 2024.

\bibitem{10002889}
G.~Hu, Z.~Li, J.~Si, K.~Xu, Y.~Cai, D.~Xu, and N.~Al-Dhahir, ``Analysis and
  optimization of {STAR-RIS}-assisted proactive eavesdropping with statistical
  {CSI},'' \emph{{IEEE} Trans. Veh. Technol.}, vol.~72, no.~5, pp. 6850--6855,
  May 2023.

\bibitem{10315044}
Z.~Xie, Y.~Liu, W.~Yi, X.~Wu, and A.~Nallanathan, ``Physical layer security for
  {STAR-RIS-NOMA}: A stochastic geometry approach,'' \emph{{IEEE} Trans.
  Wireless Commun.}, Early Access, doi: 10.1109/TWC.2023.3329871.

\bibitem{10304608}
G.~Zhu, X.~Mu, L.~Guo, A.~Huang, and S.~Xu, ``Robust resource allocation for
  {STAR-RIS} assisted {SWIPT} systems,'' \emph{{IEEE} Trans. Wireless Commun.},
  Early Access, doi:10.1109/TWC.2023.3327502.

\bibitem{8214104}
T.~D. Ponnimbaduge~Perera, D.~N.~K. Jayakody, S.~K. Sharma, S.~Chatzinotas, and
  J.~Li, ``Simultaneous wireless information and power transfer ({SWIPT}):
  Recent advances and future challenges,'' \emph{{IEEE} Commun. Surv. Tut.},
  vol.~20, no.~1, pp. 264--302, Firstquarter 2018.

\bibitem{10283600}
J.~Yaswanth, M.~Katwe, K.~Singh, O.~Taghizadeh, A.~Schmeink, and C.~Pan,
  ``Joint beamforming design for {STAR-RIS}-aided {MU-MIMO} system with
  {SWIPT},'' in \emph{Proc. IEEE Int. Conf. Commun. Workshops (ICC Workshops)},
  2023, pp. 574--579.

\bibitem{10278936}
H.~Zhang, J.~Nie, Y.~Yu, Z.~Xiong, W.~Jiang, and D.~Niyato, ``Performance
  analysis for {STAR-RIS} assisted {SWIPT} system over rayleigh fading
  channel,'' in \emph{Proc. {IEEE} Int. Conf. Commun. ({ICC})}, 2023, pp.
  1468--1473.

\bibitem{10086660}
W.~Du, Z.~Chu, G.~Chen, P.~Xiao, Y.~Xiao, X.~Wu, and W.~Hao, ``{STAR-RIS}
  assisted wireless powered {I}o{T} networks,'' \emph{{IEEE} Trans. Veh.
  Technol.}, vol.~72, no.~8, pp. 10\,644--10\,658, Aug. 2023.

\bibitem{10333826}
J.~Jiang, B.~Lyu, P.~Chen, and Z.~Yang, ``Active {STAR-RIS} assisted wireless
  information and power transfer systems,'' in \emph{Proc. IEEE 98th Veh.
  Technol. Conf. (VTC-Fall)}, 2023, pp. 1--7.

\bibitem{10133914}
Z.~Wang, X.~Mu, J.~Xu, and Y.~Liu, ``Simultaneously transmitting and reflecting
  surface ({STARS}) for {T}erahertz communications,'' \emph{IEEE J. Sel. Top.
  Signal Process.}, vol.~17, no.~4, pp. 861--877, July 2023.

\bibitem{STAR_THZ}
W.~Yan, W.~Hao, G.~Sun, C.~Huang, and Q.~Wu, ``Wideband beamforming for
  {STAR-RIS}-assisted {TH}z communications with three-side beam split,''
  \emph{arXiv preprint arXiv:2310.13933}, 2023.

\bibitem{6146494}
D.~Lee, H.~Seo, B.~Clerckx, E.~Hardouin, D.~Mazzarese, S.~Nagata, and
  K.~Sayana, ``Coordinated multipoint transmission and reception in
  {LTE}-advanced: deployment scenarios and operational challenges,''
  \emph{{IEEE} Commun. Mag.}, vol.~50, no.~2, pp. 148--155, Feb. 2012.

\bibitem{9622133}
T.~Hou, J.~Wang, Y.~Liu, X.~Sun, A.~Li, and B.~Ai, ``A joint design for
  {STAR-RIS} enhanced {NOMA-CoMP} networks: A
  simultaneous-signal-enhancement-and-cancellation-based ({SSECB}) design,''
  \emph{{IEEE} Trans. Veh. Technol.}, vol.~71, no.~1, pp. 1043--1048, Jan.
  2022.

\bibitem{9687468}
Z.~Feng, Z.~Wei, X.~Chen, H.~Yang, Q.~Zhang, and P.~Zhang, ``Joint
  communication, sensing, and computation enabled 6{G} intelligent machine
  system,'' \emph{IEEE Network}, vol.~35, no.~6, pp. 34--42, Nov/Dec 2021.

\bibitem{10220205}
Y.~Liu, Z.~Wang, J.~Xu, C.~Ouyang, X.~Mu, and R.~Schober, ``Near-field
  communications: A tutorial review,'' \emph{IEEE Open J. Commun. Soc.},
  vol.~4, pp. 1999--2049, Aug. 2023.

\bibitem{9732186}
S.~Buzzi, E.~Grossi, M.~Lops, and L.~Venturino, ``Foundations of {MIMO} radar
  detection aided by reconfigurable intelligent surfaces,'' \emph{{IEEE} Trans.
  Signal Process.}, vol.~70, pp. 1749--1763, Mar. 2022.

\bibitem{10155669}
Z.~Zhang, Y.~Liu, Z.~Wang, and J.~Chen, ``{STARS-ISAC}: How many sensors do we
  need?'' \emph{{IEEE} Trans. Wireless Commun.}, vol.~23, no.~2, pp.
  1085--1099, Feb. 2024.

\bibitem{wang2023dual}
B.~Wang, H.~Li, S.~Shen, Z.~Cheng, and B.~Clerckx, ``A dual-function
  radar-communication system empowered by beyond diagonal reconfigurable
  intelligent surface,'' \emph{arXiv preprint arXiv:2301.03286}, 2023.

\bibitem{10050406}
Z.~Wang, X.~Mu, and Y.~Liu, ``{STARS} enabled integrated sensing and
  communications,'' \emph{{IEEE} Trans. Wireless Commun.}, vol.~22, no.~10, pp.
  6750--6765, Oct. 2023.

\bibitem{10178069}
Z.~Liu, X.~Li, H.~Ji, H.~Zhang, and V.~C.~M. Leung, ``Toward
  {STAR-RIS}-empowered integrated sensing and communications: Joint active and
  passive beamforming design,'' \emph{{IEEE} Trans. Veh. Technol.}, vol.~72,
  no.~12, pp. 15\,991--16\,005, Dec. 2023.

\bibitem{10311519}
Y.~Wang, Z.~Yang, J.~Cui, P.~Xu, G.~Chen, T.~Q.~S. Quek, and R.~Tafazolli,
  ``Optimizing the fairness of {STAR-RIS} and {NOMA} assisted integrated
  sensing and communication systems,'' \emph{{IEEE} Trans. Wireless Commun.},
  Early Access, doi:10.1109/TWC.2023.3328872.

\bibitem{10188900}
C.~Wang, C.-C. Wang, Z.~Li, D.~W.~K. Ng, K.-K. Wong, N.~Al-Dhahir, and
  D.~Niyato, ``{STAR-RIS}-enabled secure dual-functional radar-communications:
  Joint waveform and reflective beamforming optimization,'' \emph{IEEE Trans.
  Inf. Forensics Secur.}, vol.~18, pp. 4577--4592, Jul. 2023.

\bibitem{10238433}
W.~Sun, S.~Sun, X.~Su, and R.~Liu, ``Security-ensured integrated sensing and
  communication ({ISAC}) systems enabled by phase-coupled intelligent
  omni-surfaces ({IOS}),'' \emph{{IEEE} Trans. Wireless Commun.}, vol.~23,
  no.~4, pp. 3480--3492, April 2024.

\bibitem{10226306}
K.~Meng, Q.~Wu, W.~Chen, and D.~Li, ``Sensing-assisted communication in
  vehicular networks with intelligent surface,'' \emph{{IEEE} Trans. Veh.
  Technol.}, vol.~73, no.~1, pp. 876--893, Jan. 2024.

\bibitem{9724202}
X.~Shao, C.~You, W.~Ma, X.~Chen, and R.~Zhang, ``Target sensing with
  intelligent reflecting surface: Architecture and performance,'' \emph{{IEEE}
  J. Sel. Areas Commun.}, vol.~40, no.~7, pp. 2070--2084, Jul. 2022.

\bibitem{xu2023edge}
W.~Xu, Z.~Yang, D.~W.~K. Ng, M.~Levorato, Y.~C. Eldar, and M.~Debbah, ``Edge
  learning for {B5G} networks with distributed signal processing: Semantic
  communication, edge computing, and wireless sensing,'' \emph{{IEEE} J. Sel.
  Topics Signal Process.}, vol.~17, no.~1, pp. 9--39, Jan. 2023.

\bibitem{bai2020latency}
T.~Bai, C.~Pan, Y.~Deng, M.~Elkashlan, A.~Nallanathan, and L.~Hanzo, ``Latency
  minimization for intelligent reflecting surface aided mobile edge
  computing,'' \emph{{IEEE} J. Sel. Areas Commun.}, vol.~38, no.~11, pp.
  2666--2682, Nov. 2020.

\bibitem{9279326}
F.~Zhou, C.~You, and R.~Zhang, ``Delay-optimal scheduling for irs-aided mobile
  edge computing,'' \emph{{IEEE} Wireless Commun. Lett.}, vol.~10, no.~4, pp.
  740--744, Apr. 2021.

\bibitem{10015822}
Q.~Zhang, Y.~Wang, H.~Li, S.~Hou, and Z.~Song, ``Resource allocation for energy
  efficient {STAR-RIS} aided {MEC} systems,'' \emph{{IEEE} Wireless Commun.
  Lett.}, vol.~12, no.~4, pp. 610--614, Apr. 2023.

\bibitem{10121446}
Z.~Liu, Z.~Li, M.~Wen, Y.~Gong, and Y.-C. Wu, ``{STAR-RIS}-aided mobile edge
  computing: Computation rate maximization with binary amplitude
  coefficients,'' \emph{{IEEE} Trans. Commun.}, vol.~71, no.~7, pp. 4313--4327,
  Jul. 2023.

\bibitem{10032506}
X.~Qin, Z.~Song, T.~Hou, W.~Yu, J.~Wang, and X.~Sun, ``Joint resource
  allocation and configuration design for {STAR-RIS}-enhanced wireless-powered
  {MEC},'' \emph{{IEEE} Trans. Commun.}, vol.~71, no.~4, pp. 2381--2395, Apr.
  2023.

\bibitem{10013760}
X.~Zhai, G.~Han, Y.~Cai, Y.~Liu, and L.~Hanzo, ``Simultaneously transmitting
  and reflecting ({STAR}) {RIS} assisted over-the-air computation systems,''
  \emph{{IEEE} Trans. Commun.}, vol.~71, no.~3, pp. 1309--1322, Mar. 2023.

\bibitem{10294010}
A.~A. Al-Habob, O.~Waqar, and H.~Tabassum, ``Latency minimization in
  phase-coupled {STAR-RIS} assisted multi-{MEC} server systems,'' in
  \emph{Proc. IEEE Annu. Int. Symp. Pers., Indoor Mobile Radio Commun.
  (PIMRC)}, Toronto, ON, Canada, Sep. 2023, pp. 1--7.

\bibitem{9863732}
J.~Zuo, Y.~Liu, Z.~Ding, L.~Song, and H.~V. Poor, ``Joint design for
  simultaneously transmitting and reflecting ({STAR}) {RIS} assisted {NOMA}
  systems,'' \emph{{IEEE} Trans. Wireless Commun.}, vol.~22, no.~1, pp.
  611--626, Jan. 2023.

\bibitem{9956827}
F.~Fang, B.~Wu, S.~Fu, Z.~Ding, and X.~Wang, ``Energy-efficient design of
  {STAR-RIS} aided {MIMO-NOMA} networks,'' \emph{{IEEE} Trans. Commun.},
  vol.~71, no.~1, pp. 498--511, Jan. 2023.

\bibitem{10052764}
T.~Wang, F.~Fang, and Z.~Ding, ``Joint phase shift and beamforming design in a
  multi-user {MISO STAR-RIS} assisted downlink {NOMA} network,'' \emph{{IEEE}
  Trans. Veh. Technol.}, vol.~72, no.~7, pp. 9031--9043, Jul. 2023.

\bibitem{10138693}
M.~Asif, A.~Ihsan, W.~U. Khan, Z.~Ali, S.~Zhang, and S.~X. Wu,
  ``Energy-efficient beamforming and resource optimization for {STAR-IRS}
  enabled hybrid-{NOMA} 6{G} communications,'' \emph{IEEE Trans. Green Commun.
  Netw.}, vol.~7, no.~3, pp. 1356--1368, Sep. 2023.

\bibitem{10263782}
X.~Li, X.~Gao, L.~Yang, H.~Liu, J.~Wang, and K.~M. Rabie, ``Performance
  analysis of {STAR-RIS-CR-NOMA} based consumer {IoT} networks for resilient
  industry 5.0,'' \emph{IEEE Trans. Consum. Electron.}, vol.~70, no.~1, pp.
  1380--1391, Feb. 2024.

\bibitem{9956998}
J.~Zhu, P.~Gao, G.~Chen, P.~Xiao, and A.~Quddus, ``Index modulation for
  {STAR-RIS} assisted {NOMA} system,'' \emph{{IEEE} Commun. Lett.}, vol.~27,
  no.~2, pp. 716--720, Feb. 2023.

\bibitem{10132045}
T.-H. Vu, T.-V. Nguyen, Q.-V. Pham, D.~Benevides~da Costa, and S.~Kim,
  ``{STAR-RIS}-enabled short-packet {NOMA} systems,'' \emph{{IEEE} Trans. Veh.
  Technol.}, vol.~72, no.~10, pp. 13\,764--13\,769, Oct. 2023.

\bibitem{10272288}
M.~Aldababsa, A.~Khaleel, and E.~Basar, ``Simultaneous transmitting and
  reflecting reconfigurable intelligent surfaces-empowered {NOMA} networks,''
  \emph{IEEE Syst. J.}, vol.~17, no.~4, pp. 5441--5451, Dec. 2023.

\bibitem{10284920}
C.~Wu, C.~You, Y.~Liu, S.~Han, and M.~Di~Renzo, ``Two-timescale design for
  {STAR-RIS} aided {NOMA} systems,'' \emph{{IEEE} Trans. Commun.}, vol.~72,
  no.~1, pp. 585--600, Jan. 2024.

\bibitem{10284719}
N.~Zhang, Y.~Liu, X.~Mu, W.~Wang, and A.~Huang, ``Queue-aware {STAR-RIS}
  assisted {NOMA} communication systems,'' \emph{{IEEE} Trans. Wireless
  Commun.}, vol.~23, no.~5, pp. 4786--4801, May 2024.

\bibitem{9740451}
C.~Wu, X.~Mu, Y.~Liu, X.~Gu, and X.~Wang, ``Resource allocation in
  {STAR-RIS}-aided networks: {OMA} and {NOMA},'' \emph{{IEEE} Trans. Wireless
  Commun.}, vol.~21, no.~9, pp. 7653--7667, Sep. 2022.

\bibitem{10223312}
J.~Lei, T.~Zhang, and Y.~Liu, ``Hybrid {NOMA} for {STAR-RIS} enhanced
  communication,'' \emph{{IEEE} Trans. Veh. Technol.}, vol.~73, no.~1, pp.
  1497--1502, Jan. 2024.

\bibitem{10065555}
M.~F.~U. Abrar, M.~Talha, R.~I. Ansari, S.~A. Hassan, and H.~Jung,
  ``Optimization of {STAR-RIS}-assisted hybrid {NOMA} {mmWave} communication,''
  \emph{{IEEE} Trans. Veh. Technol.}, vol.~72, no.~8, pp. 10\,146--10\,161,
  Aug. 2023.

\bibitem{10285062}
H.~Wen, A.~M. Tota~Khel, and K.~A. Hamdi, ``Phase shift configuration
  strategies for unbalanced {T\&R} users in {STAR-RIS}-aided {NOMA},''
  \emph{{IEEE} Commun. Lett.}, vol.~27, no.~12, pp. 3404--3408, Dec. 2023.

\bibitem{9856598}
X.~Yue, J.~Xie, Y.~Liu, Z.~Han, R.~Liu, and Z.~Ding, ``Simultaneously
  transmitting and reflecting reconfigurable intelligent surface assisted
  {NOMA} networks,'' \emph{{IEEE} Trans. Wireless Commun.}, vol.~22, no.~1, pp.
  189--204, Jan. 2023.

\bibitem{9847399}
S.~Yang, J.~Zhang, W.~Xia, H.~Gao, and H.~Zhu, ``Joint power and discrete
  amplitude allocation for {STAR-RIS}-aided {NOMA} system,'' \emph{{IEEE}
  Trans. Veh. Technol.}, vol.~71, no.~12, pp. 13\,382--13\,386, Dec. 2022.

\bibitem{10280721}
W.~Xu, J.~Chen, and X.~Yu, ``Joint design of power allocation and amplitude
  coefficients for ergodic rate optimization in {STAR-RIS}-aided {NOMA}
  system,'' \emph{IEEE Syst. J.}, vol.~17, no.~4, pp. 5452--5463, Dec. 2023.

\bibitem{9722712}
C.~Zhang, W.~Yi, Y.~Liu, Z.~Ding, and L.~Song, ``{STAR-IOS} aided {NOMA}
  networks: Channel model approximation and performance analysis,''
  \emph{{IEEE} Trans. Wireless Commun.}, vol.~21, no.~9, pp. 6861--6876, Sep.
  2022.

\bibitem{10102306}
J.~Xu, L.~Yuan, N.~Yang, N.~Yang, and Y.~Guo, ``Performance analysis of
  {STAR-IRS} aided {NOMA} short-packet communications with statistical {CSI},''
  \emph{{IEEE} Trans. Veh. Technol.}, vol.~72, no.~9, pp. 12\,385--12\,390,
  Sep. 2023.

\bibitem{10272684}
Q.~Wang, X.~Pang, C.~Wu, L.~Xu, N.~Zhao, and F.~R. Yu, ``Transmit power
  minimization for {STAR-RIS} aided {FD-NOMA} networks,'' \emph{{IEEE} Trans.
  Veh. Technol.}, vol.~73, no.~3, pp. 4389--4394, March 2024.

\bibitem{10173573}
B.~Y.~D. Rito and K.~H. Li, ``{SER}-effective constellation scaling and
  rotation in {STAR-RIS}-assisted uplink {NOMA},'' \emph{{IEEE} Commun. Lett.},
  vol.~27, no.~9, pp. 2506--2510, Sep. 2023.

\bibitem{9887793}
H.~Ma, H.~Wang, H.~Li, and Y.~Feng, ``Transmit power minimization for
  {STAR-RIS}-empowered uplink {NOMA} system,'' \emph{{IEEE} Wireless Commun.
  Lett.}, vol.~11, no.~11, pp. 2430--2434, Nov. 2022.

\bibitem{10214056}
S.~Fu, H.~Wang, H.~Zhao, and H.~Ma, ``On {STAR-RIS}-aided {NOMA} with
  multi-group detection,'' \emph{{IEEE} Wireless Commun. Lett.}, vol.~12,
  no.~11, pp. 1971--1975, Nov. 2023.

\bibitem{10273229}
S.~Lv, X.~Xu, S.~Han, Y.~Liu, P.~Zhang, and A.~Nallanathan, ``{STAR-RIS}
  enhanced finite blocklength transmission for uplink {NOMA} networks,''
  \emph{{IEEE} Trans. Commun.}, vol.~72, no.~1, pp. 273--287, Jan. 2024.

\bibitem{10173617}
K.~Xie, G.~Cai, G.~Kaddoum, and J.~He, ``Performance analysis and resource
  allocation of {STAR-RIS}-aided wireless-powered {NOMA} system,'' \emph{{IEEE}
  Trans. Commun.}, vol.~71, no.~10, pp. 5740--5755, Oct. 2023.

\bibitem{10278099}
S.~Basharat, S.~A. Hassan, H.~Jung, A.~Mahmood, Z.~Ding, and M.~Gidlund, ``On
  the statistical channel distribution and effective capacity analysis of
  {STAR-RIS}-assisted {BAC-NOMA} systems,'' \emph{{IEEE} Trans. Wireless
  Commun.}, vol.~23, no.~5, pp. 4675--4690, May 2024.

\bibitem{9815289}
W.~Ni, Y.~Liu, Y.~C. Eldar, Z.~Yang, and H.~Tian, ``{STAR-RIS} integrated
  nonorthogonal multiple access and over-the-air federated learning: Framework,
  analysis, and optimization,'' \emph{IEEE Internet Things J.}, vol.~9, no.~18,
  pp. 17\,136--17\,156, Sep. 2022.

\bibitem{10224251}
P.~Wang, H.~Wang, H.~Ma, Z.~Shi, and Y.~Fu, ``Sum rate maximization for
  {STAR-RIS} aided {NOMA} system with two-way communication,'' \emph{{IEEE}
  Commun. Lett.}, vol.~27, no.~10, pp. 2857--2861, Oct. 2023.

\bibitem{10153701}
A.~Huang, X.~Mu, and L.~Guo, ``{STAR-RIS} assisted downlink active and uplink
  backscatter communications with {NOMA},'' \emph{{IEEE} Trans. Veh. Technol.},
  vol.~72, no.~11, pp. 14\,516--14\,530, Nov. 2023.

\bibitem{balanis2016antenna}
C.~A. Balanis, \emph{Antenna Theory: Analysis and Design}.\hskip 1em plus 0.5em
  minus 0.4em\relax Hoboken, NJ, USA: Wiley, 2016.

\bibitem{bjornson2019massive}
E.~Bj{\"o}rnson, L.~Sanguinetti, H.~Wymeersch, J.~Hoydis, and T.~L. Marzetta,
  ``Massive {MIMO} is a reality—what is next?: Five promising research
  directions for antenna arrays,'' \emph{Digit. Signal Process.}, vol.~94, pp.
  3--20, Nov. 2019.

\bibitem{akyildiz2014terahertz}
I.~F. Akyildiz, J.~M. Jornet, and C.~Han, ``Terahertz band: Next frontier for
  wireless communications,'' \emph{Phys. Commun.}, vol.~12, pp. 16--32, Sep.
  2014.

\bibitem{9738442}
H.~Zhang, N.~Shlezinger, F.~Guidi, D.~Dardari, M.~F. Imani, and Y.~C. Eldar,
  ``Beam focusing for near-field multiuser {MIMO} communications,''
  \emph{{IEEE} Trans. Wireless Commun.}, vol.~21, no.~9, pp. 7476--7490, Sep.
  2022.

\bibitem{10380596}
X.~Mu, J.~Xu, Y.~Liu, and L.~Hanzo, ``Reconfigurable intelligent surface-aided
  near-field communications for 6{G}: Opportunities and challenges,''
  \emph{IEEE Veh. Technol. Mag.}, vol.~19, no.~1, pp. 65--74, March 2024.

\bibitem{li2023near}
H.~Li, Y.~Liu, X.~Mu, Y.~Chen, Z.~Pan, and Y.~C. Eldar, ``Near-field
  beamforming for {STAR-RIS} networks,'' \emph{arXiv preprint
  arXiv:2306.14587}, 2023.

\bibitem{zhang2024near}
S.~Zhang, Y.~Zhang, and B.~Di, ``Near-far field codebook design for {IOS}-aided
  multi-user communications,'' \emph{arXiv preprint arXiv:2401.08165}, 2024.

\end{thebibliography}

\end{document}